\documentclass{ieeeaccess}
\usepackage{cite}
\usepackage{textcomp}
\usepackage{colortbl}
\usepackage[table]{xcolor}
\usepackage{float} 
\usepackage{amsmath,amsfonts,amssymb,mathrsfs,amsthm,amsbsy,graphicx,bm,balance,mathtools,xfrac,nccmath,siunitx,cite,color,soul}

\usepackage{tikz}
\usepackage{threeparttable}

\NewSpotColorSpace{PANTONE}
\AddSpotColor{PANTONE} {PANTONE3015C} {PANTONE\SpotSpace 3015\SpotSpace C} {1 0.3 0 0.2}
\SetPageColorSpace{PANTONE}%

\usepackage{multirow}

\newtheorem{rem}{Remark}

\usepackage[font=scriptsize,labelfont=bf]{caption}

\usepackage{pgfplots}
\pgfplotsset{compat=newest}
\usetikzlibrary{plotmarks}
\usetikzlibrary{arrows.meta}
\usepgfplotslibrary {patchplots}
\usepackage{grffile}

\usepackage{pifont}
\newcommand{\cmark}{\ding{51}}
\newcommand{\xmark}{\ding{55}}

\usepackage{algpseudocode}
\usepackage{algorithm}

\usepackage[thinlines]{easytable}
\newcommand{\bc}{\text{BackCom}\xspace}
\newcommand{\mbc}{\text{MoBC}\xspace} 
\newcommand{\bbc}{\text{BiBC}\xspace}
\newcommand{\abc}{\text{AmBC}\xspace} 
\usepackage[utf8]{inputenc}
\usepackage{booktabs, caption, makecell}

\usepackage{threeparttable}

\usepackage{url}
\usepackage{array}
\usepackage{makecell}

 \usepackage[utf8]{inputenc}
\usepackage{booktabs, subfigure, caption, makecell,xspace}

\usepackage{threeparttable}

\definecolor{ambe}{rgb}{1.0, 0.49, 0.0}

 \theoremstyle{definition}
 \makeatletter
\newcommand{\printfnsymbol}[1]{
  \textsuperscript{\@fnsymbol{#1}}
}

\begin{document}
\history{Date of publication xxxx 00, 0000, date of current version xxxx 00, 0000.}
\doi{xxx}

\title{Signal Detection in Ambient Backscatter Systems: Fundamentals, Methods,  and Trends}
\author{\uppercase{Shayan Zargari}\authorrefmark{1}, \IEEEmembership{Student Member, IEEE}, \uppercase{Azar Hakimi}\authorrefmark{1}, \IEEEmembership{Student Member, IEEE},
\uppercase{Fatemeh Rezaei}\authorrefmark{1}, \IEEEmembership{Member, IEEE}, \uppercase{Chintha Tellambura}\authorrefmark{1},
\IEEEmembership{Fellow, IEEE}, and \uppercase{Amine Maaref}\authorrefmark{2}, \IEEEmembership{Senior Member, IEEE} }
\address[1]{Department of Electrical and Computer Engineering, University of Alberta, Edmonton, AB, T6G 1H9, Canada (e-mail: \{zargari, hakimina, rezaeidi, ct4\}@ualberta.ca)}
\address[2]{Huawei Canada, 303 Terry Fox Drive, Suite 400, Ottawa, Ontario K2K 3J1 (e-mail: Amine.Maaref@huawei.com).}
\tfootnote{``This work was supported in part by Huawei Technologies Canada Company, Ltd.''}

\markboth
{Shayan Zargari \headeretal: Signal Detection in Ambient Backscatter Systems: Fundamentals, Methods,  and Trends}
{Shayan Zargari \headeretal: Signal Detection in Ambient Backscatter Systems: Fundamentals, Methods,  and Trends}

\corresp{Corresponding author: Shayan Zargari (e-mail: zargari@ualberta.ca).}

\begin{abstract}  
Internet-of-Things (IoT) is rapidly growing in wireless technology, aiming to connect vast numbers of devices to gather and distribute vital information. Despite individual devices having low energy consumption, the cumulative demand results in significant energy usage. Consequently, the concept of ultra-low-power tags gains appeal. Such tags communicate by reflecting rather than generating the radio frequency (RF) signals by themselves. Thus, these backscatter tags can be low-cost and battery-free. The RF signals can be ambient sources such as wireless-fidelity (Wi-Fi), cellular, or television (TV) signals, or the system can generate them externally. Backscatter channel characteristics are different from conventional point-to-point or cooperative relay channels. These systems are also affected by a strong interference link between the RF source and the tag besides the direct and backscattering links, making signal detection challenging.
This paper provides an overview of the fundamentals, challenges, and ongoing research in signal detection for \abc networks. It delves into various detection methods, discussing their advantages and drawbacks. The paper's emphasis on signal detection sets it apart and positions it as a valuable resource for IoT and wireless communication professionals and researchers.
\end{abstract}

 
\maketitle

\IEEEdisplaynontitleabstractindextext

\IEEEpeerreviewmaketitle

\begin{figure}[t]
\centering
	\includegraphics[width=3.5in]{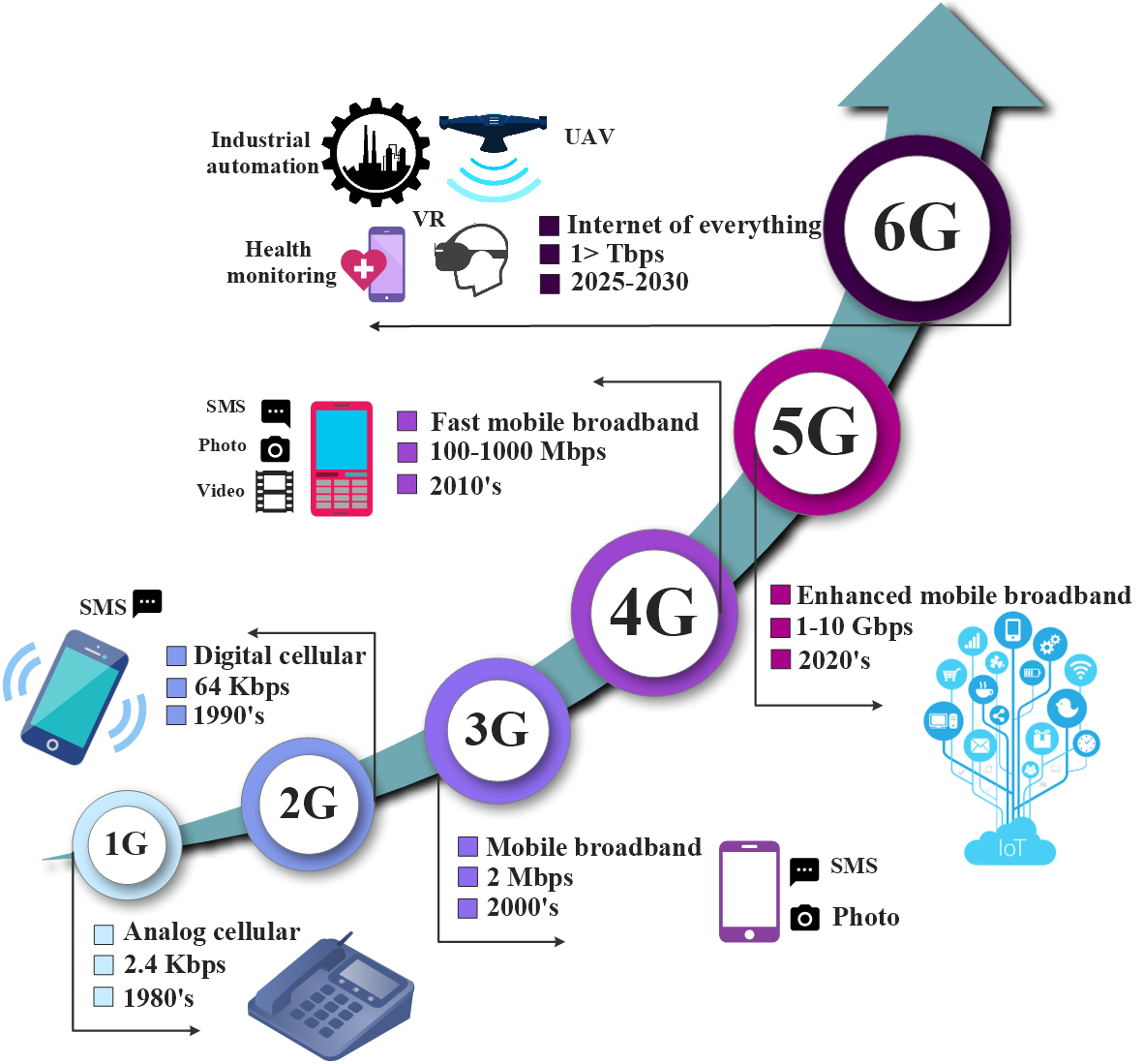}
	\caption{Evolution of mobile wireless systems.}\label{mobile_evolution}
\end{figure}

\IEEEraisesectionheading{\section{Introduction}\label{sec:introduction}}
Wireless communication systems have evolved since the first generation (1G) in the 1980s, with each subsequent generation bringing improvements in quality of service (QoS), services, and features. Fig. \ref{mobile_evolution} shows advancements across generations. While the first to four generations focused on human-type communication (HTC), fifth generation (5G) and future networks support both HTC and machine-type communication (MTC) \cite{3GPP2017study}. MTC enables wireless machine communication, enabling the development of Internet-of-Things (IoT) networks \cite{stoyanova2020survey,al2015internet}.

IoT connects information-sensing devices, enabling interconnectivity between people, machines, and various things. The number of connected devices surpassed $20$ billion in 2017 and is projected to exceed $75$ billion by $2025$, driven by the expansion of IoT networks in industrial, parcel monitoring, intelligent farming, smart home, and other applications \cite{Wei_Zhou,Kinza_Shafique}. With 5G, IoT networks can leverage services like ultra-reliable low-latency communication (URLLC), massive machine-type communication (mMTC), and enhanced mobile broadband (eMBB), enabling low latency and low-power wireless communication for a larger number of devices/sensors compared to previous generations \cite{Maria_Stoyanova}.

\begin{figure}[t]
\centering
	\includegraphics[width=3.5in]{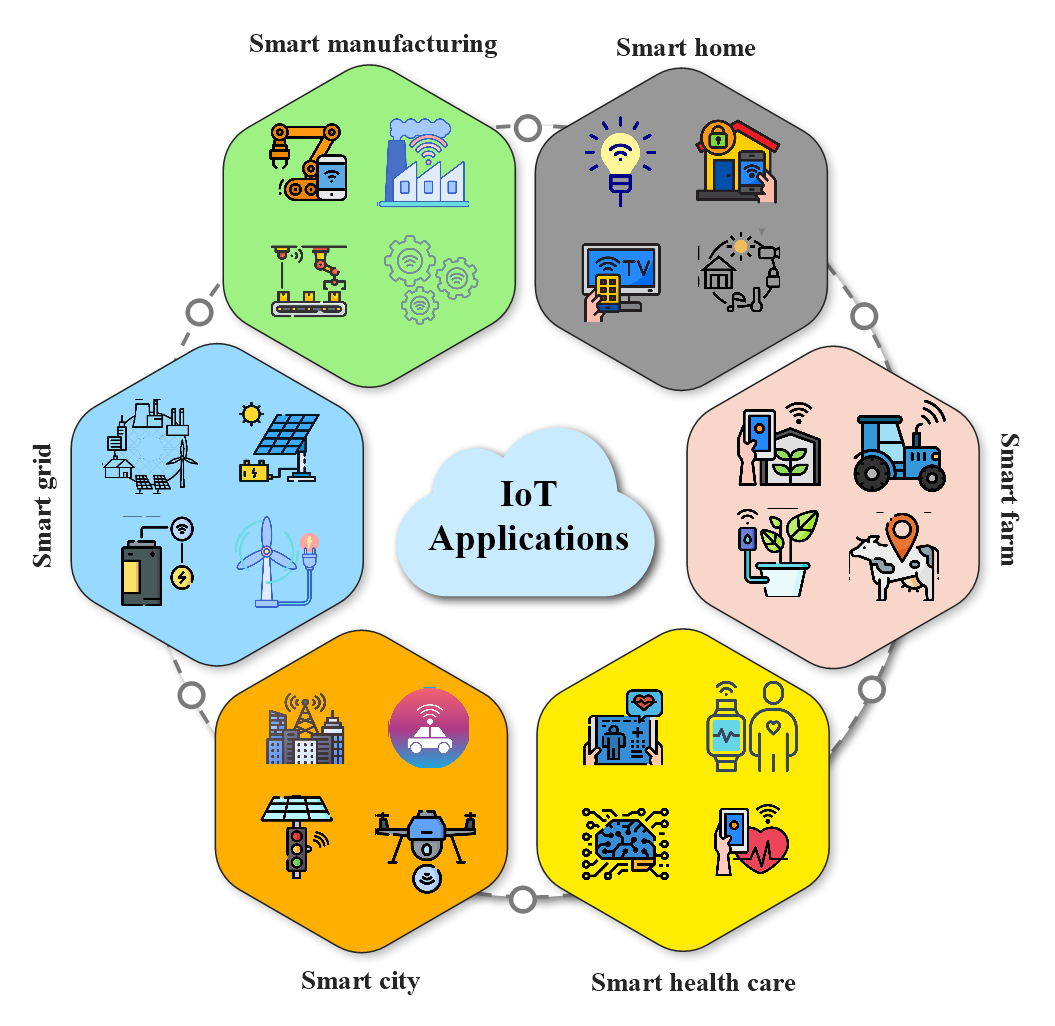}
	\caption{Emerging IoT application scenarios in 6G.}\label{IoT_application}
\end{figure}

The rapid evolution of intelligent IoT networks, surpassing the capabilities of 5G,  is driving the need for sixth-generation (6G) wireless systems,\cite{Amitabha_Ghosh}. 6G will enhance existing IoT networks by introducing new services and technologies, improving user experience and service quality. These advancements include high throughput, ultra-low latency communications, and massive/autonomous networks \cite{Braud2021,Dinh_Nguyen}. Building a super-smart society is a key prospect of 6G, leveraging intelligent mobile devices, autonomous vehicles, and other technologies (Fig. \ref{IoT_application}). This vision entails embedding millions of sensors into cities, automobiles, homes, industries, and other environments \cite{Chowdhury2020, Lopez2021}.

The proliferation of IoT devices necessitates access to the internet, however, it might be difficult to power them continually for monitoring, controlling, and other purposes. These include the need for an uninterrupted energy supply for sensors and controlling/computing devices. The disadvantages of battery-operated devices include high prices and ineffective battery replacement, particularly in hazardous areas where several sensors are employed. Consequently, the emergence of passive IoT, also known as battery-free IoT, is addressing these issues \cite{Qualcomm2022, Huawei, SA1}.

Passive IoT devices rely on energy harvesting (EH) instead of batteries, drawing energy from sources like solar, motion/vibration, ambient radio-frequency (RF), or RF signals generated by radio frequency identification (RFID) readers \cite{Qualcomm2022, Huawei}. To operate with EH, these devices must function with ultra-low power levels ranging  \qty{1}{\uW}$\sim$\qty{100}{\uW}, as discussed by the 3rd Generation Partnership Project (3GPP) \cite{Huawei, Huawei_ambient}. Passive modulation enables this low-power operation, where backscatter communication (\bc) tags reflect incident RF signals and vary the reflection coefficient by switching the impedance presented to the antenna \cite{Xu2018,Nguyen_Van}. This modulation process, known as load modulation \cite{Lu2018}, allows tags to harvest energy from incident RF signals, such as those from cellular base stations (BSs), television (TV) towers, and wireless-fidelity (Wi-Fi) access points (APs), for internal operations. Tags, with their passive modulation, do not require active RF components, making them cost-effective and having ultra-low energy requirements (a few {\qty{}{\nano \watt}} {–} {\qty{}{\uW}}) due to their simple RF design \cite{HoangBook2020}. Indeed, tags require less than 10$\mu$W of power, which is notably lower than the power consumption of active radio devices \cite{Nguyen_Van}. Active radio devices generally require more power due to their need to generate their own signals, unlike tags that merely reflect existing signals. Consequently, \bc  is an emerging technology for green IoT, enabling joint sensing and data transmission with power consumption at the micro-watt level \cite{Usman_Saleh_Toro}.

\begin{table*}[t]
 \centering
\caption{List of acronyms and their full forms}
\label{tab:acronyms}
\begin{tabular}{|l|l|l|l|}
\hline
\rowcolor[HTML]{EFEFEF}
\textbf{Acronyms} & \textbf{Full form} & \textbf{Acronyms} & \textbf{Full form} \\ \hline
1G & First generation & 5G & Fifth generation \\ \hline
6G & Sixth generation & ADC & Analog-to-digital converter \\ \hline
\abc & Ambient backscatter communication & AoA & Angle of arrival \\ \hline
AP & Access point & ASK & Amplitude shift keying \\ \hline
AWGN & Additive white Gaussian noise & BackComm & Backscatter communication \\ \hline
\bbc & Bistatic backscatter communication & BER & Bit error rate \\ \hline
BFSK & Binary frequency shift keying & BPSK & Binary phase shift keying \\ \hline
BS & Base station & CE & Channel estimation \\ \hline
CFO & Carrier frequency offset & CLS & Clustering with labeled signals \\ \hline
CLUS & Clustering with labeled and unlabeled signals & CNN & Convolutional neural network \\ \hline
CP & Cyclic prefix & CR & Cognitive radio \\ \hline
CSI & Channel state information & DC & Direct current \\ \hline
DFT & Discrete Fourier transformation & DNN & Deep neural network \\ \hline
DRL & Deep reinforcement learning & DSK & Delay-shift keying \\ \hline
DPSK & Differential phase shift keying & ED & Energy detector \\ \hline
EGC & Equal gain combining & EH & Energy harvesting \\ \hline
eMBB & Enhanced mobile broadband & FDA & Frequency diverse array \\ \hline
FM & Frequency modulation & FSK & Frequency shift keying \\ \hline
GAMP & Generalized approximate message passing & GESM & Generalized spatial modulation \\ \hline
GLRT & Generalized likelihood ratio test & GMM & Gaussian mixture model \\ \hline
HSR & High-speed rail & HTC & Human-type communication \\ \hline
IC & Integrated circuit & IoT & Internet-of-Things \\ \hline
INR & Interference to noise ratio & IRS & Intelligent reflecting surface \\ \hline
ISI & Inter-symbol interference & KNN & K-nearest neighbors \\ \hline
LDPC & Low-density parity-check code & LMMSE & Linear minimum mean-square error \\ \hline
LS & Least square & MAC & Multiple access channel \\ \hline
MAP & Maximum a-posteriori probability & MIMO & Multiple-input multiple-output \\ \hline
ML & Maximum likelihood & mMTC & Massive machine-type communication \\ \hline
MoBC & Monostatic backscatter communication & $M$-PSK & $M$-ary phase shift keying \\ \hline
MRC & Maximum ratio combining & NP & Neyman-Pearson \\ \hline
NRZ & Non-return to zero & OFDM & Orthogonal frequency division multiplexing \\ \hline
OOK & On-off keying & OSTBC & Orthogonal space-time block code \\ \hline
PA & Phase array & PAM & Pulse amplitude modulation \\ \hline
PDF & Probability density function & QAM & Quadrature amplitude modulation \\ \hline
QPSK & Quadrature phase-shift keying & RF & Radio frequency \\ \hline
RFID & Radio frequency identification & ROC & Receiver operating characteristic \\ \hline
RPM & Reflection pattern modulation & SER & Symbol error rate \\ \hline
SIC & Successive interference cancellation & SINR & Signal-to-interference-to-noise ratio \\ \hline
SISO & Single-input single-output & SM & Spatial modulation \\ \hline
SNR & Signal-to-noise ratio & SR & Symbiotic radio \\ \hline
SVD & Singular value decomposition & SVM & Support vector machine \\ \hline
TDMA & Time division multiple access & TV & Television \\ \hline
UMPT & Uniformly most powerful test & URLLC & Ultra-reliable low-latency communication \\ \hline
Wi-Fi & Wireless-fidelity & ZF & Zero-forcing \\ \hline
\end{tabular}
\end{table*}

However, signal detection of \abc,  i.e.,  the process of extracting tag data by the reader, presents formidable challenges that distinguish it from the detection of conventional wireless signals. Low reliability of ambient backscatter systems (\abc) where the tags coexist with other communication systems and harvest energy from ambient RF sources (Section \ref{Backscatter}), is a significant challenge. For instance,  a single-antenna reader with an energy detector (ED)  can achieve detection probability from $0.6$ to $0.7$, given an  SNR of $0$ dB \cite{zargari2023improved}. Signal detection is particularly challenging due to multiple factors, including strict energy constraints, limited hardware capabilities, passive modulation techniques, and the unique characteristics of the communication channel (see Section \ref{General_AmBC System_Model}). Additionally, the presence of unknown ambient RF sources with high interfering power exacerbates the detection challenges. Hence, while signal detection methods are well-established in conventional communication systems, they cannot be directly adopted in \abc.  These challenges have prompted intensive research and development efforts focused on exploring efficient and reliable symbol detection methods tailored explicitly for \abc systems \cite{ Chen_Chen_2, William_Barott, Yunkai_Hu, Georgios_Vougioukas_1}.

Before delving into the specifics of \abc, we give a brief overview of \bc in general. 

\begin{table*}[h]
\rowcolors{2}{}{teal!10}
\caption{{Summary of related works.}    } \label{summary}
\begin{center}
\begin{threeparttable}
\begin{tabular}{|l|p{5cm}|lllll|}
\hline
\multirow{2}{*}{\textbf{Reference}} & \multirow{2}{*}{\textbf{Objective}}       & \multicolumn{5}{c|}{\textbf{Contribution}}   \\ \cline{3-7} 
& & \multicolumn{1}{l|}{(a)} & \multicolumn{1}{l|}{(b)} & \multicolumn{1}{l|}{(c)} & \multicolumn{1}{l|}{(d)} & \multicolumn{1}{l|}{(e)} \\ \hline \hline
\cite{Nguyen_Van}    &   Survey on fundamentals of  \bc    & \multicolumn{1}{l|}{\cmark }     & \multicolumn{1}{l|}{\xmark  }     & \multicolumn{1}{l|}{\xmark }     & \multicolumn{1}{l|}{\xmark }    & \multicolumn{1}{l|}{\cmark }   
\\ \hline

\cite{Usman_Saleh_Toro}  & Comprehensive investigation of signal processing issues in \bc       & \multicolumn{1}{l|}{\cmark }     & \multicolumn{1}{l|}{\xmark  }     & \multicolumn{1}{l|}{\xmark }    & \multicolumn{1}{l|}{\xmark }    & \multicolumn{1}{l|}{\cmark }       \\ \hline

\cite{Diluka2022}   & Feasibility study and comprehensive survey       & \multicolumn{1}{l|}{\cmark  }     & \multicolumn{1}{l|}{\xmark}     &   \multicolumn{1}{l|}{\xmark }  & \multicolumn{1}{l|}{\xmark }    & \multicolumn{1}{l|}{\cmark }     \\ \hline
\cite{Rezaei2023}   &    Survey on potential BackCom coding methods     & \multicolumn{1}{l|}{\cmark  }     & \multicolumn{1}{l|}{\xmark }     & \multicolumn{1}{l|}{\xmark }     & \multicolumn{1}{l|}{\xmark }    & \multicolumn{1}{l|}{\cmark }   
\\ \hline

\cite{Fatemeh_Rezaei}  & Survey on wireless-powered network with \bc        & \multicolumn{1}{l|}{\cmark }     & \multicolumn{1}{l|}{\xmark  }     & \multicolumn{1}{l|}{\xmark }    & \multicolumn{1}{l|}{\cmark }    & \multicolumn{1}{l|}{\cmark }       \\ \hline

\cite{Song2022}  & Designing antennas and integrating RF systems for \bc       & \multicolumn{1}{l|}{\cmark }     & \multicolumn{1}{l|}{\xmark  }     & \multicolumn{1}{l|}{\xmark }    & \multicolumn{1}{l|}{\xmark }    & \multicolumn{1}{l|}{\cmark }       \\ \hline
\textbf{This paper}     &  Comprehensive survey on \bc signal detection Schemes   & \multicolumn{1}{l|}{\cmark  }     & \multicolumn{1}{l|}{\cmark  }     &  \multicolumn{1}{l|}{\cmark }   & \multicolumn{1}{l|}{\cmark  }    & \multicolumn{1}{l|}{\cmark }   
\\ \hline
\end{tabular}
\begin{tablenotes}\footnotesize
\item (a) Fundamentals of \bc  \,  (b) Detection scheme for \bc  \, (c) 
 Channel estimation in \bc  \, (d) Multi-antenna analysis  \,  (e) Challenges \& applications
\end{tablenotes}
\end{threeparttable}
\end{center}
\end{table*}

\subsection{Backscatter Communication}\label{Backscatter}

The key principle is that tags reflect a radio frequency (RF) signal to communicate with a reader \cite{HoangBook2020}. The tag comprises a simple low-power integrated circuit (IC) designed to operate within an ultra-low power range, typically between {\qty{1}{\uW}} and {\qty{100}{\uW}}. Notably, its cost is impressively economical, typically ranging from {\num{0.01}}\$$\sim${\num{0.5}}\$  per device. Furthermore, it boasts an incredibly compact and printable form factor, making it ideal for various applications. Moreover, the tag has an exceptionally long lifespan, lasting for more than {\num{10}} years without degradation \cite{Huawei, Huawei_ambient}. EH can power tags, which save batteries
and associated costs (see Section {\ref{tag_reflection}}).

\begin{figure*}[t]
\centering
	\includegraphics[width=6.5in]{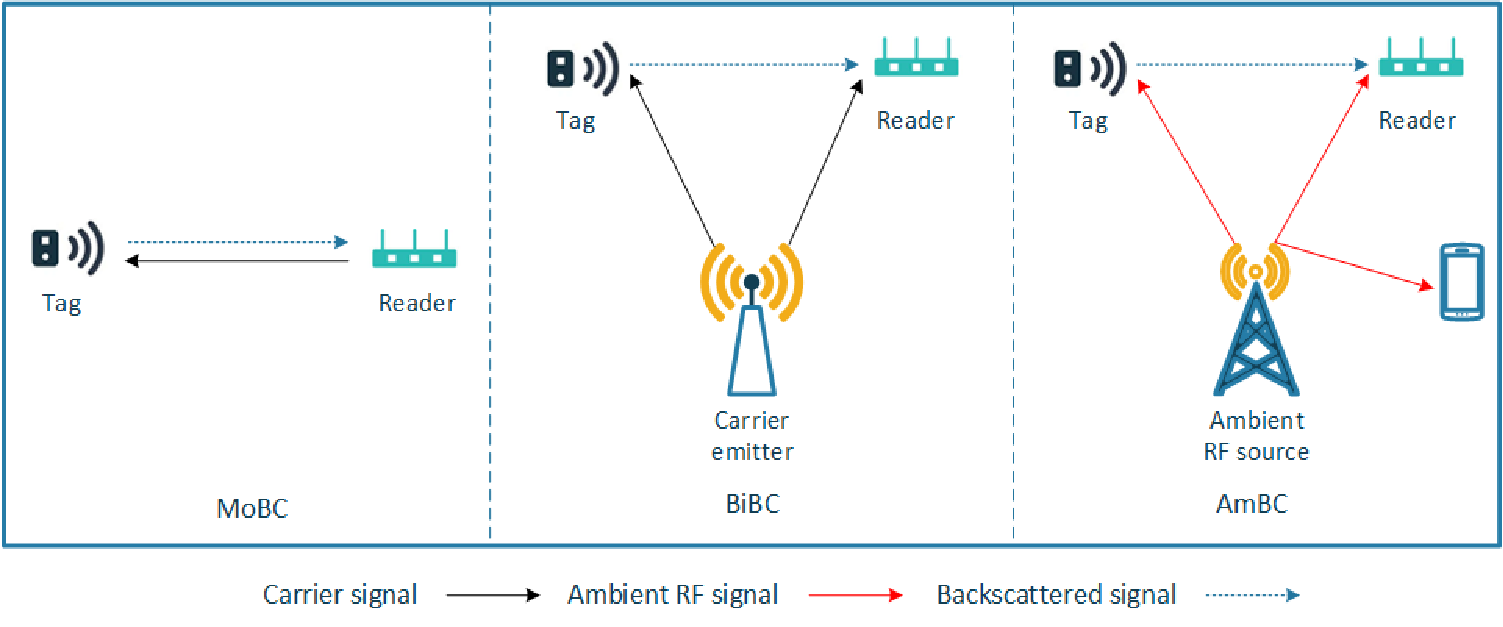}
	\caption{Paradigms for backscatter communications.}\label{configurations}
\end{figure*}

Depending on the type of RF source which generates the RF signal and empowers the tag, there are three types of {\bc} networks: Monostatic, bistatic, and ambient (Fig.{~\ref{configurations}}) \cite{Giovanni_Vannucci,Aggelos_Bletsas,griffin2009complete}. In monostatic backscatter communication ({\mbc}),  the reader and emitter are co-located, whereas, the bistatic backscatter communication ({\bbc}) dislocates the two functions of the reader; it deploys single or multiple dedicated RF emitters to energize the tags and enable backscatter modulation.  In contrast, {\abc} harnesses the existing ambient legacy RF sources such as cellular BSs, TV towers, and Wi-Fi access points,  instead of relying on dedicated ones. This innovative approach not only mitigates the need for new spectrum allocation but also enhances spectrum resource utilization.  {\abc} stands out as a cost-efficient, energy-efficient (green), and environmentally friendly technology for low-power communications, making it an ideal choice for applications like large-scale passive IoT \cite{Qualcomm2022, Huawei, SA1, Dong2020}.

Nevertheless, the success of \abc heavily depends on meticulous design and implementation, primarily due to the unpredictable and uncontrollable nature of ambient RF sources. The presence of such sources introduces interference to the AmBC systems, necessitating careful planning and mitigation strategies to ensure optimal performance and reliability. Despite these challenges, \abc's potential for eco-friendly and resourceful communication makes it a highly promising solution for sustainable connectivity \cite{Huawei_ambient}.

\subsection{Contribution and Organization}
 
 \abc presents an innovative solution for establishing passive IoT-based networks, which are expected to play a significant role in future 6G wireless networks. This technology seamlessly integrates into existing communication systems, leverages diverse technologies, and supports numerous applications without requiring new infrastructure. It thus facilitates various applications across industry sectors such as logistics and supply chain, manufacturing, and agriculture. However, the challenges stemming from low backscattered power, double path losses, and interference from RF sources drastically impact the reliability of this innovative technology,  restricting its effectiveness in practical applications. Addressing these issues through advanced signal processing techniques is essential to improve the overall performance and robustness of \abc systems. Despite its potential, this research area has yet to be fully explored and developed.  This paper aims to comprehensively survey existing detection methods and explore potential solutions to enhance the reliability of \abc systems.

While previous surveys and tutorials have explored  a wide range
of \bc topics including architectures, physical layer,
benefits, limitations, and applications \cite{Diluka2022, Rezaei2023, Nguyen_Van, Fatemeh_Rezaei, Song2022,Usman_Saleh_Toro, Chenren_Xu}, they do not consider the critical aspect of \abc signal detection and  reliability. In particular, \cite{Nguyen_Van} provides a comprehensive overview of  \abc systems, covering fundamentals, general architecture, advantages, challenges, applications, limitations, and research efforts. Reference \cite{Usman_Saleh_Toro} presents the potential of \bc for enabling green IoT through joint communication and sensing. The work \cite{Chenren_Xu}  covers the basics and modulation of \bc while summarizing recent research on \bbc systems. Reference \cite{Fatemeh_Rezaei} addresses the integral aspects of {\bc} wireless-powered networks and reviews their performance improvement techniques,  emphasizing large-scale networks. The work \cite{Diluka2022} presents a comprehensive link budget analysis for \bc, addressing the limitations and challenges. Reference \cite{Rezaei2023} specifically investigates coding techniques for \bc, surveying the existing coding schemes and highlighting the potential approaches to address the implementation complexity and reliability. Reference \cite{Song2022}  also specifically focuses on antenna design, RF system integration, and advanced packaging technologies for \bc systems.

In contrast, our manuscript exclusively focuses on detection methodologies, aiming to bridge this gap in the literature. Our paper provides a comprehensive analysis of various signal detection methods and their specific applications. We have thoroughly investigated and evaluated the effectiveness of different detection schemes, offering valuable insights into their strengths and limitations.
By filling this research void, our manuscript fosters the development of improved detection techniques for \abc systems.  To provide a holistic perspective, Table \ref{summary} compares our survey paper with other relevant literature reviews, highlighting the unique contributions and the specific areas of focus in each work.

We summarize the contributions of this paper as follows:

\begin{itemize}
     \item  Our survey begins with the basis of \abc.  We briefly discuss the primary tasks performed by passive tags, namely backscatter modulation and EH. Subsequently, we address the challenges associated with \abc signal detection. Furthermore, we discuss \bc channel characteristics and delve into different \abc channel estimation (CE) methods, including pilot-based techniques and blind estimation approaches.

    \item We address the general principle of signal detection and describe various detection schemes suited for \bc systems, including coherent, non-coherent, and semi-coherent techniques. We also describe learning-based signal detection schemes that leverage machine learning (MaL) algorithms to enhance detection performance. Their practical applications and the advantages and disadvantages associated with each approach are highlighted.

     \item We then conduct a comprehensive review and analysis of previous research findings on signal detection in \abc.  Additionally, we present simulation examples to provide practical insights into the effectiveness of these schemes.

     \item Furthermore, our study entails an in-depth review and investigation of signal detection methods in diverse system setups, including symbiotic radio (SR), orthogonal frequency division multiplexing (OFDM), and intelligent reflecting surface (IRS)-aided \abc systems. We also provide simulation results to facilitate a thorough evaluation of these methods. We thus present a comprehensive spectrum of practical scenarios, offering rich insights into the versatility of signal detection methodologies.

     \item Finally, we highlight open issues, challenges, and future research directions. This includes technical obstacles, theoretical conundrums, and practical implementation hurdles. We anticipate that this direction-setting discourse will inspire future scholarly explorations and contribute to the advancement of this vital field.
 \end{itemize}

\begin{figure*}[t]
\centering
\includegraphics[width=7in]{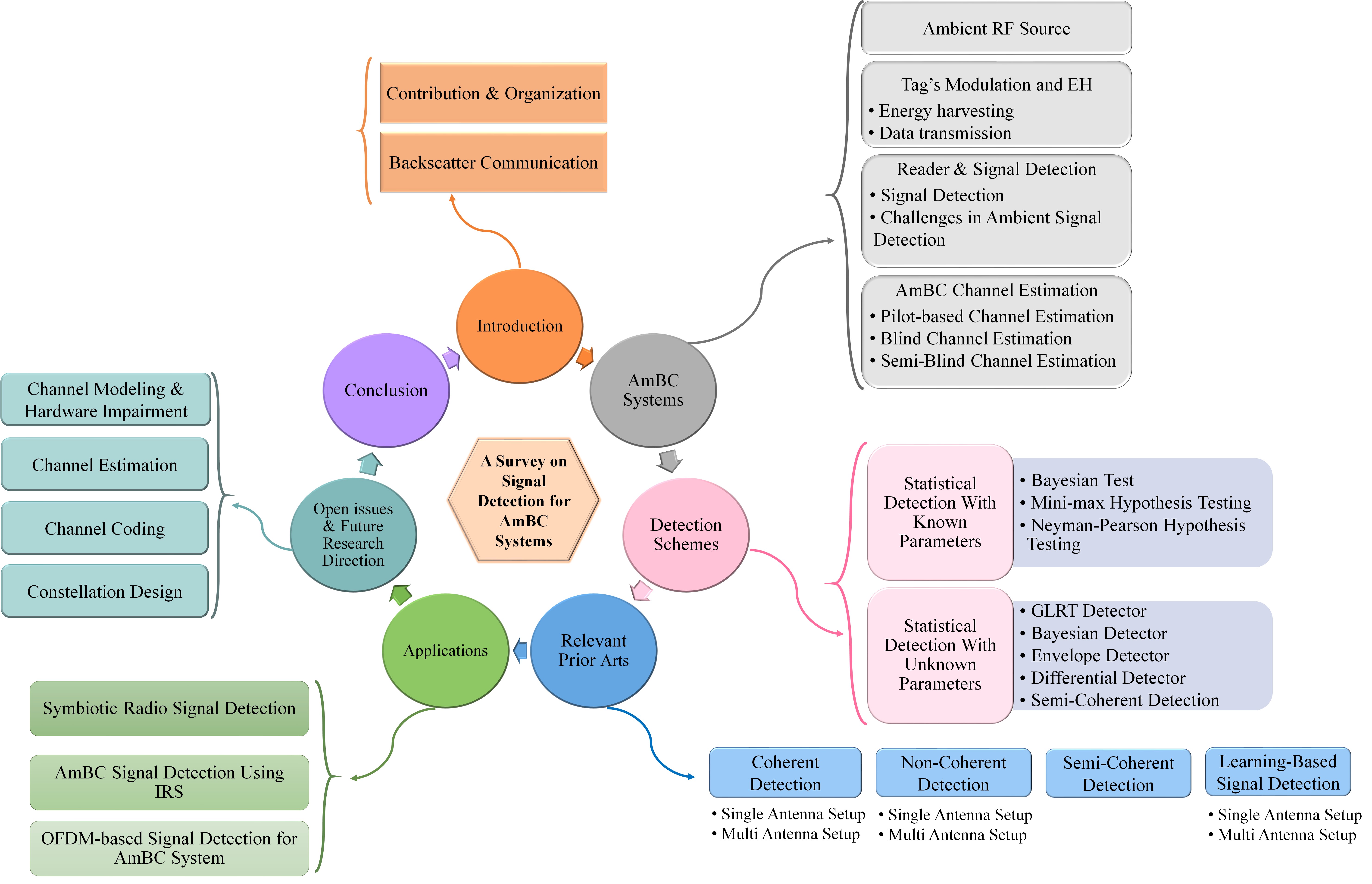}
\caption{Structure of this paper.}\label{outline}
\end{figure*}

The organization of this paper is presented in Fig. \ref{outline}. Section \ref{General_AmBC System_Model} provides an overview of \abc systems, covering crucial tasks performed by passive tags, such as backscatter modulation and EH. Additionally, this section discusses reader signal detection and various CE schemes. In Section \ref{Detection_Background}, we delve into the fundamental principles of signal detection, encompassing coherent, semi-coherent, and noncoherent detection techniques. We also explain deep learning-based signal detection methods. Section \ref{Sec:prior_art} offers a comprehensive summary of the detection schemes adopted for \abc, presenting an analysis of the existing literature in this domain. Moving forward, Section \ref{Sec:application} explores \abc signal detection within different system configurations, including scenarios involving SR, OFDM, and IRS-aided systems. Section \ref{Sec:open_issue} focuses on addressing open issues and highlighting potential future research directions in the context of \abc signal detection. Finally, we conclude the paper in Section \ref{Sec:conclude}, summarizing the key findings and contributions. For the reader's convenience, Table \ref{tab:acronyms} and Table \ref{Table:notation} provide succinct summaries of the abbreviations and notations used throughout this article.

\section{ \abc Systems }\label{General_AmBC System_Model}

As per Fig. \ref{configurations}, the tag harvests energy from the surrounding ambient FR signals to power its circuit operation. The tag also transmits its information to the reader by modulating and reflecting the incident RF signal. In this setup, the tag can transmit data without initiation from the reader when it harvests sufficient energy from the RF source.

\subsection{Ambient RF Source}\label{Section:ambient_source}
Ambient RF sources play a crucial role in the performance of communication systems. These sources are generally classified into static and dynamic types based on their operational characteristics. Static RF sources, such as TV towers and FM radio BSs, primarily emit high-powered signals in the VHF and UHF bands, which due to their inherent obstruction-penetrating capabilities, are widely used for large-area broadcasting \cite{Lu2015}. On the other hand, dynamic RF sources, like Wi-Fi access points and cellular networks, sporadically transmit in bands such as $2.4$ GHz, $5$ GHz, and millimeter-wave domains, usually with lesser power than static sources.

The effective communication range in RF systems depends on various attributes, both inherent to the RF sources and external factors like the environment. Among the significant determinants are:
\begin{itemize}
    \item \textbf{Transmit Power:} Owing to their high-power transmissions, static RF sources can facilitate communications spanning extensive distances, often ranging from several hundred meters to multiple kilometers \cite{Vincent_Liu, Anran2017}. Conversely, dynamic sources, with their attenuated power profiles, are more suited for short-range communications, typically within the ambit of $1$-$5$ meters \cite{Bharadia2015}.

    \item \textbf{Frequency Considerations:} Bands in the higher frequency domains, especially millimeter-wave frequencies, are inherently susceptible to increased absorption and scattering. This challenge is accentuated in urban terrains, characterized by dense infrastructural elements leading to frequent signal obstructions \cite{Diluka2022,Fatemeh_Rezaei}.

    \item \textbf{Antenna Design and Configuration:}  The efficacy of communication is also influenced by the employed antenna design. Directional antennas, by focusing the RF energy in a specific direction, can potentially extend communicative ranges. Additionally, the deployment of multiple-input multiple-output (MIMO) configurations, leveraging an array of antennas, can harness spatial diversity, thereby augmenting received power, enhancing symbol detection fidelity, and subsequently improving communication ranges.
\end{itemize}

\subsubsection{Implications on Symbol Detection:}
Ambient RF sources, based on their static or dynamic nature, exert specific effects on symbol detection within communication systems. The inherent properties of these sources bring forth challenges and considerations that communication engineers must address. Some effects and challenges are detailed below.
\begin{itemize}
    \item \textbf{Noise Floor and SNR:} High-powered transmissions from static sources lead to an elevation in the ambient noise floor \cite{Krishnan,Zexiong}. Consequently, while the received signal strength (RSS) may be notably high, the corresponding SNR may be degraded due to this elevated noise, challenging precise symbol detection. Contrarily, dynamic sources, owing to their intermittent transmission nature, can be sources of sporadic interference. This is particularly evident in dense deployment scenarios, leading to challenges such as the ``hidden node problem" \cite{Shengbo}.

    \item \textbf{Multipath Propagation:} The signals emitted by static sources have wide coverage and can interact with multiple obstacles like buildings or mountains. This leads to the signal reaching the receiver via different paths, creating multipath fading. A direct consequence of this multipath effect can be inter-symbol interference (ISI), where the symbols interfere with each other, corrupting the received signal. In contrast, dynamic sources, which typically have more localized and constrained transmissions, face multipath effects primarily from immediate surroundings, making their multipath profile distinct from that of static sources.

    \item \textbf{Interference from Co-located Systems:} In environments where multiple communication systems co-exist, interference from one system can adversely affect the symbol detection of another. Especially in shared spectrum scenarios, the management of interference becomes pivotal to maintaining system integrity.

    \item \textbf{Doppler Effect:} In scenarios where dynamic sources are in motion, such as vehicular networks, Doppler shifts can be introduced. This necessitates the deployment of compensation mechanisms to prevent potential symbol misinterpretation.
\end{itemize}

\begin{figure}
\centering
	\includegraphics[width=3.5in]{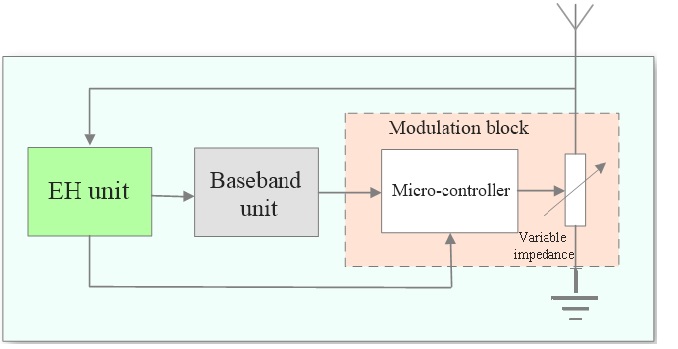}
 	\caption{Tag circuit diagram.} \label{Tag}
\end{figure}

\subsection{Tag's Backscatter Modulation \&  Energy Harvesting} \label{tag_reflection}
Despite the differences among the backscatter configurations, the tag has the same analog front-end in different setups which includes an EH unit and a modulation block (Fig. \ref{Tag}). The operation of the baseband processing unit depends on the \bc configuration. Specifically, in \mbc, the tag processes commands from the reader, controlling both data decoding and encoding. In contrast, in \abc and \bbc, the tag simply reflects the RF signal from the emitter. This makes the baseband processor design for such tags simpler.

\subsubsection{Energy harvesting}
At the analog front-end, the tag exploits power switching protocol to perform EH and data transmission, simultaneously \cite{Diluka2022}. In particular, the received RF signal power ($P_T$) is divided into two parts based on a power splitting ratio, which is the power reflection coefficient, $\xi \in (0,1)$. The tag thus reflects  $\xi P_T$ for data transmission and absorbs the remaining part, i.e., $P_b= (1-\xi) P_T$ for EH. The harvested power can be modeled as a linear or nonlinear function of $P_b$.  The widely-used linear model has  $P_{h} = \rho P_b$, where $\rho \in (0,1]$ is the power conversion efficiency. Despite its simplicity, the linear model does not account for saturation and other non-linear effects.  Thus,  the nonlinear models are developed to characterize the non-linear effects \cite{Wang2020, Wang2017, boshkovska2015practical, dong2016performance}.

\subsubsection{Passive modulation} To transmit data, the tag,  without generating an RF signal, tunes its load impedance to impinge its data into the incident RF signal. The tag thus performs what is known as load modulation. In this process, the tag switches its load impedance to modulate the incident RF signal.  For example, for binary phase shift keying (BPSK), the tag may switch between $Z_1 $ and $Z_2 $ to generate bit “$0$'' or “$1$'', leading to absorbing or reflecting (i.e., non-backscattering and backscattering states), respectively. The  reflection coefficient of the tag is expressed as follows \cite{Nguyen_Van}
\begin{align}\label{tag:ref}
\Gamma_{i} = \frac{Z_i-Z_{a}^\star}{Z_i+Z_{a}}, 
\end{align}
where $Z_{a}$ denotes the antenna impedance of the tag, which depends on the structure of the antenna, $Z_i$ is the load impedance of the state $i=\{1,2,\ldots, Q\}$, and  $\xi_i = \vert \Gamma_{i} \vert^2$ is the power reflection coefficient.

Although binary backscatter modulators have been widely investigated, they limit the data rate and are spectrally inefficient. Thus, it is desirable to use higher-order modulation $(Q > 2)$ by using more than two different impedance values. We  denote $ \Gamma_{i} = |\Gamma_i| e^{j \theta_i}$ where $ \theta_i \in [0, 2\pi]. $ Thus, the tag can use distinct  $\theta_i$ values to send its data, which amounts to a $Q$-ary phase shift keying (PSK) constellation and the reflection coefficients of all other impedance values have a constant magnitude, i.e., $|\Gamma_i|^2=\xi_i=\xi$. To design the constellation points, the load impedance $Z_i$ can then be computed via the Smith chart techniques \cite{ReedPH21}. Similarly, one can generate non-PSK constellations, e.g., amplitude shift keying (ASK) or quadrature amplitude modulation (QAM),  by varying $ \theta_i$, $|\Gamma_{i}|$, or both, respectively \cite{ReedPH21}.

\begin{table}[]
\caption{List of notations.}\label{Table:notation}
\centering
\begin{tabular}{@{}ll@{}}
\toprule
\textbf{Notation}  & \textbf{Definition}        \\ \midrule

$P_D(\cdot)$   & Detection probability   \\
$P_F(\cdot)$      &  False alarm probability   \\
$s(n)$ 	&RF source transmit signal at time instance $n$\\
$b(n)$    &Tag bit at time instance $n$\\
$P_T$     &RF source signal power\\
$N$       &RF source packet size\\
$M$ 	&Number of the antenna at the reader\\
$T_b$  &Tag symbol duration\\
$T_s$  &RF source symbol duration\\
$\xi$       & Power reflection coefficient of the tag \\
$\rho$      & Energy conversion efficiency \\
$\Gamma$   & Complex reflected coefficient of the tag       \\
$\mathbf{h}_{sr}$  &Direct RF source-to-reader channel\\
$\mathbf{h}_{tr}$  &Backscatter tag-to-reader channel\\
$h_{st}$          &Forward RF source-to-tag channel\\
$\sigma^2_{sr}$   &Direct channel power gain  \\
$\sigma^2_{tr}$   &Backscatter channel power gain \\
$\sigma^2_{st}$   &Forward channel power gain \\
$\sigma^2_{w}$    &Noise power \\
$F_c$            &Carrier frequency\\
$\Delta F$ 	&Carrier frequency offset between the RF source and the reader\\
$\Delta \phi$ 	&Phase offset between the RF source and the reader\\  
$\Delta \gamma$	&  SNR difference between backscatter and direct link. \\
$\tau$ 		&  Detection threshold\\
$N_{\text{cp}}$  & CP length\\
$N_s$   	 & Number of subcarriers	\\
Ntr      	 & Number of training sequence\\
$\eta_1$ 	 & Signal-to-interference ratio\\
$\eta_2$ 	 & Interference-to-noise ratio \\
$I$ & Number of IRS reflecting elements\\
$ \boldsymbol{\Theta}$     &IRS reflection-coefficients matrix\\
$\varrho_i$  			&Amplitude of the $i$-th reflecting element at the IRS \\
$\boldsymbol{h}$  &Channel link from the tag-to-IRS\\
$\boldsymbol{f}$  &Channel link from the IRS-to-reader\\
$\mathbf{A}^H$  & Hermitian conjugate transpose  \\
$\mathbf{A}^T$  & Transpose    \\
$\mathbf{A}^{\star}$     & Conjugate    \\
$\mathbf{I}_M$  & Identity matrix of size $M$. \\
$\|. \|$& Euclidean norm    \\
$|.|$           & Absolute value\\
$\mathbb{E}[\cdot]$   & Expectation   \\
$\otimes$  & Addition modulo $2$ \\
 diag(·)  & Diagonalization operation   \\
$\mathbb{C}^{M\times N}$   & $M\times N$ dimensional complex matrix.      \\
$ \mathcal{C}\mathcal{N}(\boldsymbol{\mu},\,\mathbf{C})$ & \begin{tabular}[c]{@{}l@{}}Circularly symmetric complex Gaussian (CSCG)\\ random vector with mean  $\boldsymbol{\mu}$ and covariance matrix $\mathbf{C} $.\end{tabular} \\ \bottomrule
\end{tabular}
\end{table}

\subsection{Reader \& Signal Detection}\label{reader_detection}

The reader performs complex RF operations to recover the tag's data, affected by the strong direct link interference from the ambient RF source. The efficacy of data detection at the reader, as well as the associated interference and distortion, are deeply influenced by the channel characteristics. Compared to the conventional one-way channels, \bc channel is more prone to deep fades. This means there is a heightened probability of severe attenuation (detailed in Section \ref{channel_estimation}), resulting in communication outages and increased error rates. Employing readers with multiple antennas can mitigate the impact of deep fades, enhancing link reliability. It will leverage power gain, and receive diversity gain to enhance signal detection, thereby improving the BER and outage performance (Section \ref{Sec:prior_art}).

\subsubsection{Signal Detection}

\begin{figure}[t]
\centering
    	\includegraphics[width=3.5in]{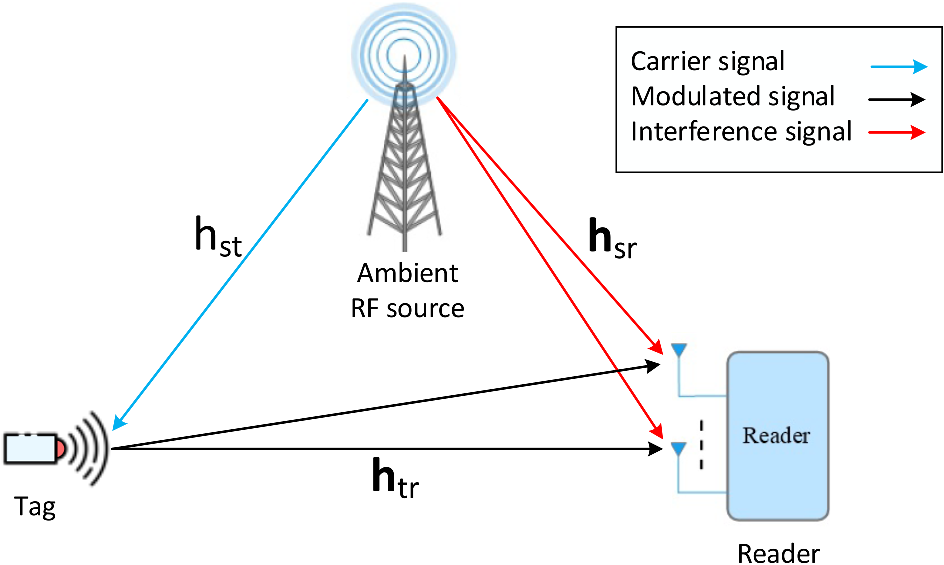}
 	\caption{AmBC system model with the multiple-antenna reader.} \label{system_model_multi}
\end{figure}

As before, we consider an ambient RF source, a backscatter tag, and a reader (Fig.~\ref{system_model_multi}). The reader is equipped with multiple antennas $M$, to enhance the signal detection. The RF source transmits the signal $s(n)$ at the $n$-th time slot, where $\mathbb{E}[|s(n)|^2 ]=1$, and both reader and tag receive it. The signal $s(n)$ could be either a complex Gaussian signal or a modulated one. 
\begin{enumerate}
    \item[(a)] Complex Gaussian ambient source: Complex Gaussian signals are random signals that follow a normal distribution and have both magnitude and phase components. They are often used as a source of interference in wireless communication systems, e.g., artificial noise \cite{Lu2022}. Thus, it is possible to represent the transmitted signal as a complex Gaussian random variable with zero mean and power $P_T$, denoted as $s(n)\sim \mathcal{CN}(0, P_T)$ \cite{Qian2017}.

    \item[(b)] Modulated ambient source: In contrast, modulated signals are signals that have been altered in a specific way to carry information, e.g., signals of TV towers, cellular BSs, and Wi-Fi APs. Accordingly, the symbol $s(n)$ is assumed to be modulated with $Q$-ary modulation, with a power of $P_T$, and is drawn from a constellation set $\mathcal{S} = \{S_1, S_2, \dots, S_Q\}$, where each symbol in the set has an equal probability of occurrence \cite{Qian2017, Youyou_Zhang}.  
\end{enumerate} 

Upon receiving the signal at the tag, it harvests energy from it and reflects the signal to the reader after modulating it with its own data $\Gamma_{i}$. Note that the ambient RF source typically transmits at much higher rates than the tags, i.e., $T_b = N T_s $, where $T_b$ and $T_s$ are respectively the symbol periods of the tag and RF source \cite{Fatemeh_Rezaei}. We assume that the direct RF source-to-reader channel, $\mathbf{h}_{sr} = [{h}_{sr,1}, \ldots {h}_{sr,M}]^{\rm{T}}$, the forward RF source-to-tag channel, $h_{st}$, and the backscatter tag-to-reader channel, $\mathbf{h}_{tr}= [{h}_{tr,1}, \ldots, {h}_{tr,M}]^{\rm{T}}$, are modeled as $v=\zeta_v^{1/2} \tilde{v}$, where $v=\{{h}_{sr,m},h_{st},{h}_{tr,m}\}, m\in\mathcal{M}=\{ 1, 2, \cdots, M\}$, $\tilde{v}\sim \mathcal{CN}(0,1)$ captures the quasi-static Rayleigh fading, and $\zeta_v$ accounts for pathloss, which is modeled via free-space pathloss model \cite{Hassan2014LTECN}. Moreover, we assume a constant tag reflection coefficient, i.e., $\vert \Gamma_i \vert^2 = \xi$ (Section \ref{tag_reflection}). Then, the signal received at the $m$-th antenna of the reader during the $n$-th time slot can be expressed as 
\begin{equation}\label{eqn_received}
y_m(n) = [h_{sr,m}  + \Gamma_{i} h_{st} h_{tr,m}]x(n)   + w_m(n), \quad m\in\mathcal{M},
\end{equation}
where $x(n) =\sqrt{P_T}s(n)e^{-j(2\pi \Delta F t+\Delta \phi)}$. Here, $\Delta F$ denotes the carrier frequency offset (CFO), and $\Delta \phi$ represents the phase offset between the RF source and the reader. Additionally, $w_m(n) \sim \mathcal{CN}(0,\sigma^2_{w})$ is the additive white Gaussian noise (AWGN) with mean $0$ and variance $\sigma^2_{w}$.

The accurate decoding of transmitted data is frequently hampered by the CFO. Techniques such as periodogram-based methods offer a solution by capturing the power distribution of the signal across its frequency components, aligning these estimates with maximum likelihood (ML) methods to fine-tune parameters based on observed data {\cite[p. 542]{kay1993fundamentals}}. In  AmBC systems, the unknown nature of the RF source poses a unique challenge. Blind CFO estimation methods, such as those leveraging the orthogonality among subcarriers or the second-order statistics of demodulated data, present viable solutions \cite{Besseghier,Harsha}. Differential estimation, which capitalizes on phase differences between consecutive symbols, offers a pilot-free approach, though it may be less effective in noisy environments \cite{Jing_Qian_3}. Additionally, machine learning techniques, harnessing pattern recognition in received signals, emerge as promising tools (Section \ref{mach_detection}). Overall, CFO compensation could be imperfect due to the unknown RF source parameters. Thus, much more research is needed on this topic.

Finally, the received signal can then be expressed as a function of the transmitted signal and the backscatter modulation symbol as follows:
\begin{equation}\label{received_signal_2}
y_m(n) =\left\{ \begin{array}{l}
h_{0,m} s(n) + w_m(n),~\quad \text{if decide on} ~H_0,\\
h_{1,m} s(n) + w_m(n), ~\quad \text{if decide on} ~H_1,\\
\vdots \\
h_{Q,m} s(n) + w_m(n), ~\quad \text{if decide on} ~H_Q,\\
\end{array} \right.
\end{equation}
where $\{H_0, H_1, \ldots, H_Q\}$ represent $Q+1$ hypotheses. Here, $H_0$ refers to the null hypothesis, and $H_i$, $i \in \{1,2,\ldots,Q\}$ corresponds to the hypothesis that the transmitted symbol is $\Gamma_i$. In addition, $h_{0,m} = h_{sr,m}$ denotes the direct channel link and $h_{i,m} = h_{sr,m} + \Gamma_{i} h_{st} h_{tr,m}$ indicates the composite channel link. Finally, the reader can decode the transmitted symbols of the tag based on the received signals by using detection techniques (Section \ref{Detection_Background}).

\subsubsection{Challenges in AmBC signal detection}
Signal detection in AmBC systems is severely difficult due to  several challenges:
\begin{enumerate}
    \item \textbf{Low backscatter signal strength}: The backscattered signals experience deeper fades due to the double path losses. Thus, the received signal power at the reader, i.e., signal-to-noise ratio (SNR), is usually very low. On the other hand, the direct-link signal is typically strong and causes direct interference to the reader. Additionally, the backscattered signals may be corrupted by other signals present in the environment, such as RF signals from other communication systems.

    \item \textbf{Unknown ambient RF source}: The knowledge of the ambient RF source parameters (such as bandwidth, transmit power, and waveform) is not typically available. However, to detect the tag's signal, the reader needs to first cancel out the direct-link interference from the RF source.

    \item \textbf{Channel state information at the reader}: The reader requires accurate channel state information (CSI) to detect the tag's data. Precise CSI estimation demands long enough training signals, which is time-consuming and energy-intensive \cite{Zhao2019}. However, passive tags inherently possess limited resources, have strict energy constraints, and cannot support such training or pilot signals \cite{Qian2017}.

    \item \textbf{Multipath propagation:} The backscattered signals may experience multipath propagation, which can cause signal fading and ISI.
    
    \item \textbf{Limited bandwidth:} Backscattered signals are typically transmitted over a limited bandwidth, which can limit the transmission rate and signal detection accuracy.
\end{enumerate}

We aim to address these challenges by presenting a comprehensive analysis of signal detection in \abc systems. Note that for coherent (or semi-coherent) detection of the tag's data, the reader requires perfect (or partial) CSI, which could be obtained through pilot-based or blind techniques. In the following, we briefly describe these methods, which ultimately affect the signal detection quality of the reader.

\subsection{\abc Channel Estimation}\label{channel_estimation}
CE is a vital aspect of \abc systems, as it provides the reader with the parameters necessary for tag signal detection.  The presence of the backscattered signals makes CE a fundamentally different problem as compared to that in traditional wireless communication \cite{Abdallah2021, Zhao2019, Liu2021, Abdallah2023, Wang2022, Ma2018, Zhao2018}. A typical \abc system has two sets of channels (Fig. \ref{system_model_multi}) to be estimated for interference suppression and successful tag data detection, (i) the direct channel from the RF source to the reader ($\mathbf{h}_{sr}$), (ii) the cascaded (or dyadic) channel,  $h_{st} \mathbf{h}_{tr}$, which describe the signal propagation from the RF source to the tag and the tag to the reader. The dyadic channel has significantly different fading characteristics from conventional one-way wireless links and results in deeper fades \cite{griffin2010fading, Griffin2007, Rezaei2023, Diluka2022}. It also suffers from double pathloss, which can significantly decrease the received signal power and degrade the reliability of the CE.
 
When estimating channels, the ambient RF source may transmit known pilot symbols or unknown information symbols. Accordingly, pilot-based and blind CE techniques have been developed \cite{Abdallah2021, Zhao2019, Liu2021, Abdallah2023, Wang2022, Ma2018, Zhao2018}. Another strategy is the semi-blind estimation,  which exploits both pilot and data transmissions in the estimation process \cite{Abdallah2021, Abdallah2023}.

\begin{table*}[ht]
\caption{Summary of channel estimation methods in \abc Systems}
\label{table_channel_estimation}
\centering
\begin{tabular}{|p{2cm}|p{3.2cm}|p{3.5cm}|p{3.8cm}|p{3cm}|}
\hline
\rowcolor[HTML]{ECF4FF}  \textbf{Estimation Type} & \textbf{Description} & \textbf{Techniques/Methods} & \textbf{Pros} & \textbf{Cons} \\
\hline
Pilot-based  & Uses known pilot sequences & LS, ML, MMSE, LMMSE, DFT, DNNs & Provides a better estimate of the channel response, and incorporates prior knowledge of channels & High computational complexity \\
\hline
Blind & Leverages statistical properties of the received signal & SOS, Subspace, EM, EVD, Compressed Sensing & Eliminates the need for pilot symbols, simple and computationally efficient & Lower accuracy levels due to absence of pilot symbols \\
\hline
Semi-blind & Utilizes both pilot symbols and data transmissions & EM and DD  estimation strategy & Higher accuracy than blind techniques, lower training overhead compared to pilot-based method & Involves complexity in treating unknown data as hidden variables \\
\hline
\end{tabular}
\end{table*}

\subsubsection{\textbf{Pilot-based channel estimation}}
In this scheme, the RF source transmits known symbols, i.e., pilots, and the tags backscatter known information symbols over the RF source signal, which are then captured and recognized by the reader. Exploiting classical and MaL techniques, the reader thus estimates the direct and cascaded channels based on the transmitted known pilot sequences of the RF source and the tags' data. The classical CSI estimation methods include least square (LS) estimation, ML estimation, minimum mean-squared error (MMSE) estimation, and linear MMSE (LMMSE) estimation \cite{kay1993fundamentals, rezaei2023timespread}. LS estimation minimizes the squared difference between estimated and actual channels. It does not require any prior knowledge of the channels and is straightforward and computationally efficient. ML estimation, which seeks to find the most likely estimate of the channel based on the received signal, is computationally more complex than LS estimation. However, it provides a better estimate of the channel response and achieves a lower mean squared error (MSE). On the other hand, the optimal MMSE estimation incorporates the prior knowledge of the channels by considering the statistical properties of the channel and noise to achieve the minimum MSE.  The suboptimal LMMSE estimation simplifies the computational complexity of MMSE CE and provides good accuracy. Other CSI estimation methods include discrete Fourier transformation (DFT) \cite{Zhao2019}, deep neural networks (DNNs) \cite{Liu2021}, and iterative estimation \cite{Zhu2018}. In particular, DFT is used to derive coarse estimates of the direction-of-arrivals and channel gains, which are then refined through angular rotation \cite{Zhao2019}. Modeling the CE as a denoising problem, DNNs are used to recover the channel coefficients from the received noisy pilot signals  \cite{Liu2021}. Moreover, having initial coarse channel estimates through LS, an iterative process is used to refine the estimator \cite{Zhu2018}. To use a  pilot-based scheme, some degree of cooperation between the ambient RF source and the reader is required, e.g., cooperative \abc \cite{Yang2018}. Otherwise, blind schemes can be applied for channel estimation purposes (Section II-D-2).

\subsubsection{\textbf{Blind channel estimation}}
Various blind CE techniques leverage the statistical properties of received signals to estimate channels without using pilot symbols. However, achieving high accuracy levels, as measured by the MSE, may be challenging without pilot symbols \cite{Ozdemir2007}. Blind CE methods include second-order statistics (SOS), subspace, expectation maximization (EM), eigenvalue decomposition (EVD), and compressed sensing approaches \cite{Yi_Zhang, Eric_Moulines, zeng1997blind}. While SOS and subspace techniques are simple and computationally efficient, they may not provide highly accurate channel estimates compared to EM and compressed sensing methods.

An EM-based estimator is proposed in \cite{Ma2018} to obtain absolute values of channel parameters in single-antenna \abc nodes. Moreover, in \cite{Zhao2018}, a blind channel estimator is developed for a multi-antenna reader, utilizing the EVD of the covariance matrix of received signals. This method requires no knowledge of RF signals, except for the transmit power.

\subsubsection{\textbf{Semi-blind channel estimation}} 

These methods combine both pilot symbols and data for wireless CE \cite{Abdallah2023}. Pilot symbols enable initial CE, while data symbols, forming a significant portion of the transmitted signal, refine the estimation.

Semi-blind estimation offers higher accuracy than blind techniques and reduces training overhead compared to purely pilot-based methods \cite{Carvalho2004}. Examples include the EM technique and decision-directed (DD) estimation. The EM technique treats unknown data as hidden variables, yielding estimates with monotonically non-decreasing likelihood \cite{Abdallah2013, Abdallah2023}. On the other hand, the DD method initially estimates the channel using pilot transmission, then uses the decoded data as additional pilots to re-estimate the channel \cite{Abdallah2023}. This technique achieves greater accuracy than pilot-based estimators and lower computational complexity than EM-based estimators, as it incorporates transmitted data into estimation without computing their posterior probabilities \cite{Abdallah2019}.

Table \ref{table_channel_estimation} briefly outlines the estimation methods along with their respective advantages and disadvantages.

\subsection{Synchronization}
Synchronization is vital, ensuring precise alignment of transmitter and receiver timing and frequency. Offset errors, even slight ones, can disrupt synchronization. Accurate synchronization is critical for efficient multiple-access operations, impacting functions like location determination, energy optimization, and mobility. Yet, synchronization grows intricate with numerous devices due to hardware variations and communication delays. Poor synchronization leads to reduced range, lower throughput, and heightened ISI. 

For  BackCom networks, tags must be able to synchronize using incident RF signals. However, this process is computationally expensive, and its power consumption increases exponentially at lower incoming signal power levels \cite{Manideep}. Tags may not synchronize well because of their limited hardware and the passive modulation \cite{Manideep}. Thus, it is important to ensure the highest sensitivity with minimal power consumption, ideally at the $\mu$W level \cite{Manideep, jia2021passive }. To this end, \cite{zhang2016hitchhike} employs a low-power energy detector synchronization method for Wi-Fi backscattering. However, this approach offers limited accuracy, with a deviation of up to 2 $\mu$s at an input power of {\qty{-20}{\dB m}}. On the other hand, an integrated circuit is presented in \cite{Manideep} to achieve synchronized backscatter communication with ambient Wi-Fi signals, supporting multiple tags. It can achieve desirable sensitivity while consuming microwatts of power. Note that the reader can send queries for the synchronization process. This is in accordance with RFID standards, which demonstrate how such communication can be achieved \cite{epcglobal2013epc}.

\section{Detection Schemes}\label{Detection_Background}

\begin{figure*}[t]
\centering
	\includegraphics[width=7in]{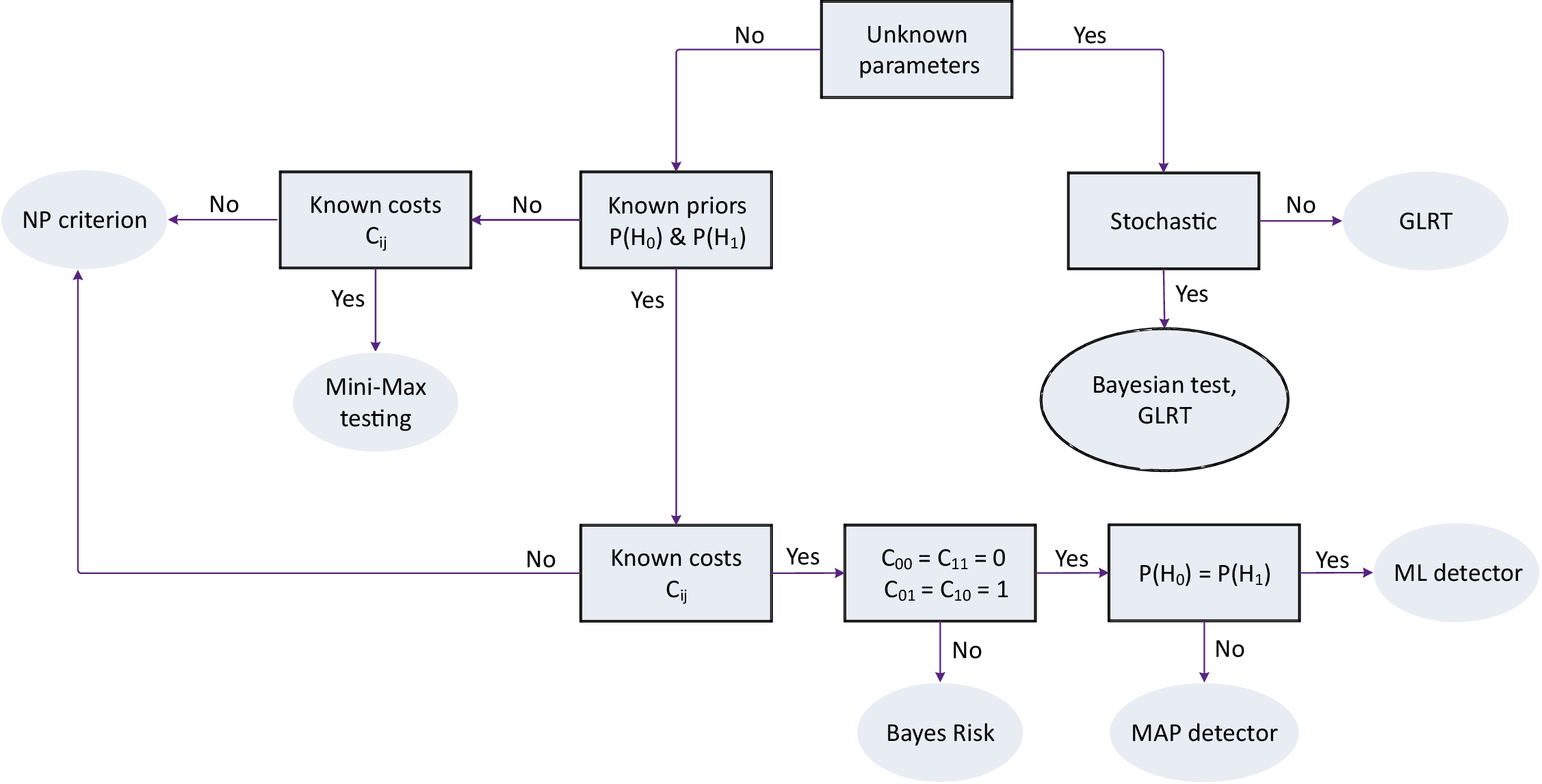}
	\caption{Summary of detection methods for binary hypothesis testing. This diagram can also be extended to multiple hypothesis testing by replacing $P(H_0)$ and  $P(H_1)$ with  $P(H_i)$, $i=\{0,1,\ldots, Q-1\}$ \cite{kay1993fundamentals}.  }\label{summary_detection_methods}
\end{figure*}

Signal detection is a fundamental technique in many fields such as digital communications, radar, and image processing. In general,  the goal is to determine whether a particular event is occurring in noisy data \cite{kay1993fundamentals,poor2013introduction}. This task is achieved through hypothesis testing and decision theory. In the context of a wireless communication system, the transmitter sends the modulated signal through a wireless channel where the signal can be distorted by noise, interference, and other factors. The receiver then demodulates the signal and performs detection over each signaling interval to eliminate the effects of noise, interference, channel fading, and other impairments \cite{kay1993fundamentals}.

For instance,  in BPSK, with  carrier frequency $f_0$ [Hz],  $s_0(t)$ and $s_1(t)$ represent bits ``$0$" and ``$1$", respectively. The receiver removes the carrier frequency and outputs the baseband signal $y(t)$, which consists of noise and other distortions. The detector must decide which of the two signals, or bits, were transmitted based on the continuous signal  $y(t)$. Due to limited bandwidth, additive noise, and wireless communication challenges, $y(t) $ can be distorted, making detection challenging.

As signals and noise are inherently random, statistical methods are necessary for detection. The probability density functions (PDFs) of the signal and noise characteristics determine the quality of the detection process. Hypotheses are associated with a-priori probabilities $\pi_i=\Pr(H = H_i)$, which represent our knowledge about the applicable hypothesis before any data is observed. Each hypothesis corresponds to a different model (PDF) for the observed data, which is used to distinguish among the hypotheses. An optimal detector can be designed when complete knowledge of the PDFs is available. However, the detection performance is significantly affected by the channel characteristics and the availability of CSI. Depending on the receiver's ability to exploit knowledge of the carrier's phase, detection can be classified as coherent, noncoherent, or semi-coherent, which are explained in the following sections.

\begin{enumerate}
 \item[$\bullet$]  {Coherent detection}: This technique requires the exact knowledge of the carrier phase (well-synchronized transceiver) as well as the CSI at the receiver. Coherent detectors are optimal in terms of error probability. However, the availability of CSI and the carrier phase knowledge is a challenging task in certain practical applications.
 
 \item[$\bullet$]  {Non-coherent detection}: The knowledge of the carrier phase and the CSI is not required in noncoherent detection, which ultimately reduces the corresponding receiver complexity at the expense of a decreased spectral efficiency or a performance penalty.
 
 \item[$\bullet$] {Semi-coherent detection}: It combines aspects of both coherent and non-coherent detection which relies on a limited number of training symbols to estimate the required detection parameters, rather than estimating the full CSI.

 \item[$\bullet$] {MaL-based detection}: MaL techniques are leveraged to enhance the signal detection process. These techniques can learn the patterns in the signal without the need for a complete carrier phase or CSI knowledge, thereby potentially improving the performance of the system over time as more data is processed. The MaL-based approach can be adaptive to changes in the signal environment and may overcome some limitations associated with traditional detection techniques.
\end{enumerate}

The upcoming subsections provide a concise overview of several popular detection methods used in \abc systems. A visual summary of these methods can be found in Fig. \ref{summary_detection_methods}.

\subsection{Statistical Detection with Known Parameters}
 
In this section, we briefly summarize some well-known detection methods, assuming complete knowledge of the PDFs under each of the possible scenarios. In the following, we restrict our attention to the binary ($Q = 2$)  case {\cite{levy2008principles}}. 

\subsubsection{\bf{Binary hypothesis testing}}
It decides between two hypotheses, i.e., $H_0$ and $H_1$, based on the observation data, $\mathbf{y}$. Two pieces of information are used, (i) a-priori probabilities: $\pi_0=\Pr(H = H_0)$ and $\pi_1=\Pr(H = H_1)$; $\pi_0+\pi_1=1$, (ii)  measurement model, corresponding to the probability density for $\mathbf{y}$, conditioned on each of the hypothesis, i.e.,
\begin{equation}\label{distribution}
\begin{array}{l}
H_0:  \quad  \mathbf{Y} \sim f_{\mathbf{Y}}(\mathbf{y}|H_0),\\
H_1:  \quad  \mathbf{Y} \sim f_{\mathbf{Y}}(\mathbf{y}|H_1),
\end{array}  
\end{equation}
which are referred to as likelihood functions. Here, $\mathbf{Y}$ is a random vector with sample space $\mathcal{Y}$.

A decision is made by partitioning the  range of $\mathbf{Y}$, i.e., $\mathcal{Y}$, into two disjoint decision regions, $\mathcal{Y}_0$ and $\mathcal{Y}_1$.  Then, if $\mathbf{y} \in \mathcal{Y}_i$, we decide that $H_i$
is the best match to the data.  Hence, the design of the decision region is a key factor. The decision function for this test is expressed as $\delta (\mathbf{y}) = i$ if hypothesis $H_i$ is decided.

There are four possible outcomes in a test of this form, depending on the decision we make ($H_0$ or $H_1$), and the true distribution of the data (also $H_0$ or $H_1$). We denote these as $(0, 0), (0, 1), (1, 0)$, and $(1, 1)$, where the first argument denotes the decision based on the regions $\mathcal{Y}_0$ and $\mathcal{Y}_1$ and the second denotes the true distribution that generated the data. It is obvious that, the outcomes $(0, 1)$ and $(1, 0)$ are mistakes or errors.

The detection performance of any decision rule can be measured by calculating the false alarms and detection probabilities, which are respectively given as
 \begin{align}
     &P_D(\delta)=\Pr (\text{decide } H_1|H_1 ~\text{is true})= \int_{\mathcal{Y}_1} f_{\mathbf{Y}}(\mathbf{y}|H_1) d\mathbf{y},\label{P_D}\\
     &P_F(\delta)=\Pr (\text{decide } H_1|H_0 ~\text{is true})= \int_{\mathcal{Y}_1} f_{\mathbf{Y}}(\mathbf{y}|H_0) d\mathbf{y}.\label{P_F}
 \end{align}

\subsubsection{\bf{Bayesian test}} \label{Bayesian}
We can improve the selection of decision regions by assigning a cost to both correct and incorrect decisions, reflecting their relative importance. The cost of each outcome $(i, j)$, where $i, j \in \{0, 1\}$ are binary decisions, can be represented by $C_{ij} \ge 0$. The overall cost associated with a test (i.e., with decision regions $\mathcal{Y}_0$ and $\mathcal{Y}_1$) is usually called the Bayes Cost, and it is defined as
\begin{equation}\label{label:cost}
R (\delta,H) = \sum_{i=0}^1\sum_{j=0}^1 \pi_j C_{ij} \Pr (\text{decide } H_i|H_j ~\text{is true}),
\end{equation}
where, the a-priori probabilities $(\pi_j)$, cost structure $(C_{ij})$, and distribution of $\mathbf{Y}$ \eqref{distribution} under each hypothesis are assumed to be known. 

The goal is then designing the decision regions to minimize the Bayes risk, i.e., $\delta_B=  {\text{argmin}_{\delta} } \: \mathbb{E} \{R(\delta, H)\}$. According to \eqref{label:cost}, the optimal decision is thus given by
\begin{equation}\label{Bayesian_decision}
L(\mathbf{y}) \overset{\Delta}{=} \frac{f(\mathbf{y}|H_1) }{f(\mathbf{y}|H_0)}  \underset{H_0}{\overset{H_1}{\gtrless}}  \frac{(C_{10}-C_{00})\pi_0}{(C_{01}-C_{11})\pi_1} \overset{\Delta}{=} \tau,
\end{equation}
where $ L(\mathbf{y})$ is the likelihood ratio and $ \tau \in [0, \infty]$ is the threshold, which is determined from the a-priori probabilities and costs. The overall decision rule is referred to as a likelihood ratio test (LRT).

Since this test aims to make a few mistakes, zero cost is often assigned to correct decisions, i.e., $C_{00} = C_{11} =0$. Moreover, when  $C_{01} = C_{10} = 1$, the Bayesian decision rule \eqref{Bayesian_decision} reduces to 
\begin{equation}\label{MAP_detector}
\pi_1  f(\mathbf{y}|H_1)  \underset{H_0}{\overset{H_1}{\gtrless}}  \pi_0 f(\mathbf{y}|H_0),
\end{equation}
where, the hypothesis corresponding to the largest a-posteriori probability, i.e., $\text{argmax}_{H_i}{\rm{Pr}}(H_i | \mathbf{y})$, is chosen to minimize the probability of a decision error. This test is referred to as the maximum a-posteriori probability (MAP) decision rule. 

When the hypotheses are equally likely, i.e.,  $\pi_0 = \pi_1$, MAP detector \eqref{MAP_detector} is further simplified to ML detector, and the optimum decision rule chooses the hypothesis for which the corresponding likelihood function is largest, i.e.,
 \begin{equation}
  f(\mathbf{y}|H_1)   \underset{H_0}{\overset{H_1}{\gtrless}}  f(\mathbf{y}|H_0).
\end{equation}
ML detection is widely used in the design of receivers for digital communication systems.

Bayesian approach requires meaningful assignment of not only costs ${C_{ij}} $, but a priori probabilities $\pi_i$. However, in some applications, e.g.,  failure testing, it is difficult to determine appropriate a priori probabilities {\cite{levy2008principles}}.

\subsubsection{\bf{Mini-Max hypothesis testing}}  
To make Bayesian hypothesis testing robust with respect to uncertainty in the a priori probabilities, the Mini-Max decision rule is developed. This test uses a LRT based on an assumed prior $\pi$ chosen so that the worst-case performance is as good as possible. Thus, the problem to find the test ($\delta_Q$) and a-priori value ($\pi_{0Q}$) is formulated as {\cite{levy2008principles}}
\begin{equation}
(\delta_Q, \pi_{0Q}) =  \underset{\delta}{\text{argmin} }  \: \underset{\pi_0 \in \{0,1\}}{\text{max} }  \: R(\delta, \pi_0),
\end{equation}
which can be solved by applying the saddle point method {\cite{levy2008principles}}.

\subsubsection{\bf{Neyman-Pearson hypothesis testing}}\label{NP}
When there is no obvious set of cost assignments ($C_{ij}$), the optimization criterion is to choose the decision rule to maximize $P_D(\delta)$  \eqref{P_D} subject to a constraint on the maximum allowable $P_F(\delta)$ \eqref{P_F}, i.e., 
\begin{align}\label{eqn:NP_test}
\delta_{\text{NP}}=  \:\: &\underset{\delta}{\text{argmax} }   \: P_D(\delta) \nonumber \\
&\text{s.t.}~P_F(\delta)\leq \Xi. 
\end{align}
The Neyman-Pearson (NP) method uses the receiver operating characteristic (ROC) curve to investigate the performance of arbitrary decision rules. ROC is a graphical plot that describes the detection probability $(P_D)$ versus the false alarm probability $(P_F)$ of a detector, where the upper and lower boundaries represent achievable and unachievable regions in the $(P_F, P_D)$-square. The ROC curve has the following properties. First, it is concave. Second, all points on the ROC curve meet $P_D \geq P_F$. Third, the feasible test region is symmetric around the point $(0.5, 0.5)$. Fourth, the slope of the ROC at point $(P_D(\tau), P_F(\tau))$ indicates the threshold $\tau$ of the corresponding LRT. 
 Typical applications of using NP criterion are sonar and radar communication systems \cite{kay1993fundamentals}. 

\begin{rem} 
For \textit{$Q$-ary Hypothesis Testing} with the set of $Q$ hypotheses: $\{H_0, H_1, \ldots, H_{Q-1}\}$,  and associated a-priori probabilities $\{\pi_0, \pi_1, \ldots, \pi_{Q-1}\}, \sum_{i=1}^Q{} \pi_i = 1$, the decision function and decision region are respectively expressed as $\delta (\mathbf{y}) = i$ if $H_i$ is decided, and  $\mathcal{Y}_i \overset{\Delta}{=}\{\mathbf{y}| \delta(\mathbf{y})=i \}$, where the intersection of $ \mathcal{Y}_i$, $\forall i$, is null and the union of these regions constructs the sample space   $\mathcal{Y}$. In this case, the likelihood function for the observed data under each hypothesis takes the form $f_{\mathbf{Y}}(\mathbf{y}|H_i), i =\{1, \ldots, Q\}$. Following the same principle as Section \ref{Bayesian}, the MAP and ML decision rules can be respectively obtained as
\begin{align}
 & \delta_{\text{MAP}}(\mathbf{y}) = \underset{i}{\text{argmax} }   \:    f(\mathbf{y}|H_i)\pi_i, \nonumber\\
& \delta_{\text{ML}}(\mathbf{y})= \underset{i}{\text{argmax} }   \:    f(\mathbf{y}|H_i),
\end{align}
where $\pi_0, \pi_1, \ldots , \pi_{Q-1}=\frac{1}{Q},$ for ML detection.
\end{rem}

\subsection{Statistical Detection with Unknown Parameters}\label{Statistical_Detection}

For all hypothesis testing problems considered in the previous section, it has been assumed that a complete description of the probability distribution of observations is available under each hypothesis. However, for practical signal detection problems, even though an exact model of the received signal may be available, it often includes some unknown parameters, such as amplitude, phase or time delay, frequency, channel statistics,  or an additive noise variance \cite{kay1993fundamentals, van2004detection}. It is therefore challenging to detect signals with unknown parameters, so designing a detector with an acceptable BER is of great importance.

In the following, we describe various well-known detection methods that consider the PDFs with unknown parameters for the binary case, i.e., $H_i: \mathbf{Y} \sim f_{\mathbf{Y}}(\mathbf{y}|\mathbf{x}, H_i), \mathbf{x} \in \mathcal{X}_i $ for $i =0,1$, which are parametrized by an unknown parameter vector $\mathbf{x}$. These methods are the generalized likelihood ratio test (GLRT), and Bayesian theory (for statistically unknown parameters). Additionally, two widely used non-coherent suboptimal receivers are presented: (a) Envelope or ED; and (b) Differential detector. These approaches do not require knowledge of channel phases or polarities. However, they are highly susceptible to interference, ISI, and noise.

\subsubsection{\bf{GLRT}}

It combines estimation and detection by replacing the unknown parameters with their ML estimate under each hypothesis and then forming a likelihood ratio as if the estimates were the true parameter values. The GLRT test can thus be stated as \cite{levy2008principles}
\begin{equation}
L_G(\mathbf{y}) \overset{\Delta}{=} \frac{f(\mathbf{y}|\hat{\mathbf{x}}_1,H_1) }{f(\mathbf{y}| \hat{\mathbf{x}}_0,H_0)}  \underset{H_0}{\overset{H_1}{\gtrless}} \tau ,
\end{equation}
where  $\hat{\mathbf{x}}_i = \text{arg max}_{\mathbf{x}\in \mathcal{X}_i} f(\mathbf{y}|\mathbf{x},H_i)$ is the  ML estimate of  unknown parameter vector $\mathbf{x}$ under $H_i$ for $i=0,1$. 

The GLRT is asymptotically optimal in the sense that it maximizes the rate of decay of the probability of a miss or of a false alarm. Thus, both for ease of implementation and performance reasons, the GLRT is an attractive default test in situations where no uniformly most powerful test (UMPT) test exists \cite{levy2008principles}.

\subsubsection{\bf{Bayesian}}
The Bayesian approach is based on Bayes' theorem, where the unknown parameter vector $\mathbf{x}$ is assigned a probability density $f(\mathbf{x}|H_i)$ under hypothesis $H_i$ for $i=0,1$, which can then be used to evaluate the marginal density. The Bayesian test is thus expressed as \cite{levy2008principles}
\begin{equation}
L_G(\mathbf{y}) \overset{\Delta}{=} \frac{\int_{\mathcal{X}_1} f(\mathbf{y}|{\mathbf{x}}_1,H_1)f(\mathbf{x}_1) d\mathbf{x}_1}{\int_{\mathcal{X}_0} f(\mathbf{y}| {\mathbf{x}}_0,H_0)f(\mathbf{x}_0)d\mathbf{x}_0}  \underset{H_0}{\overset{H_1}{\gtrless}} \tau.
\end{equation}
where $f(\mathbf{x}_i) = f(\mathbf{x}|H_i)$.

The Bayesian approach has advantages such as the ability to incorporate prior knowledge and handle complex signal models. It can also update estimates with new observations. However, it can be computationally intensive, especially when the signal model is complex and the number of unknown parameters is large. In addition, it can be sensitive to the choice of prior distributions, and the choice of threshold for the decision rule can have a significant impact on the performance of the signal detector.

Following the same principle as Section \ref{Bayesian}, the MAP and ML decision rules can be respectively obtained as (for a general $Q$-ary hypothesis) 
\begin{align}
\delta_{\text{MAP}}(\mathbf{y})
& = \underset{i}{\text{argmax} }   \:     \pi_i \int_{\mathcal{X}} f(\mathbf{y}|{\mathbf{x}},H_i)f(\mathbf{x}) d\mathbf{x}.\nonumber\\ 
\delta_{\text{ML}}(\mathbf{y})
& = \underset{i}{\text{argmax} }   \:    \int_{\mathcal{X}} f(\mathbf{y}|{\mathbf{x}},H_i)f(\mathbf{x}) d\mathbf{x}. 
\end{align}
where $\pi_0 = \ldots =\pi_{Q-1}=\frac{1}{Q}$ for ML detection.

\subsubsection{\bf{Envelope detector}}

This method considers the test statistic based on the average energy of the received signal samples. Let $y(n)$, $n=\{1,2,\ldots N\}$, be the received signal at the reader, and then the reader takes the average of the received signal energy over $N$ samples to detect the data given as follows:

\begin{equation}
 E =\frac{1}{N}\sum^N_{n=1} |y(n)|^2.
\end{equation}

\begin{figure}
\centering
	\includegraphics[width=3.5in]{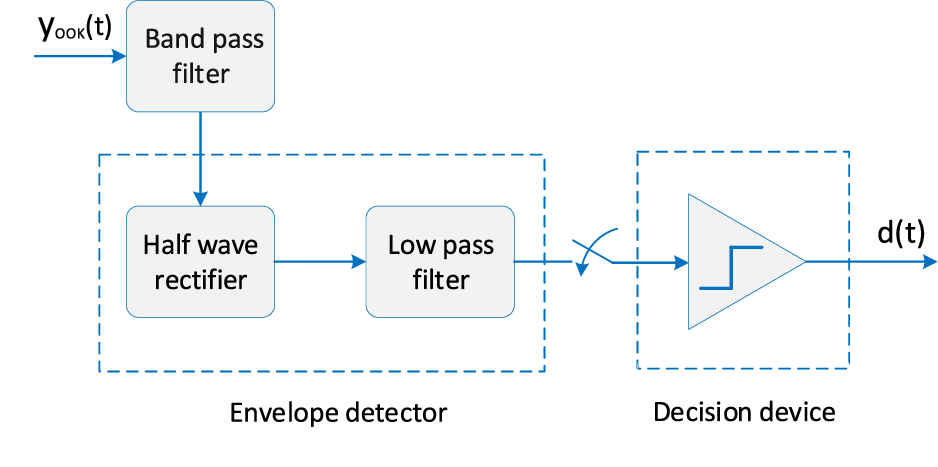}
 	\caption{Non-coherent detection of OOK modulation \cite{proakis2001digital}.} \label{Envelope_detector}
\end{figure}

\begin{figure}
\centering
	\includegraphics[width=3.5in]{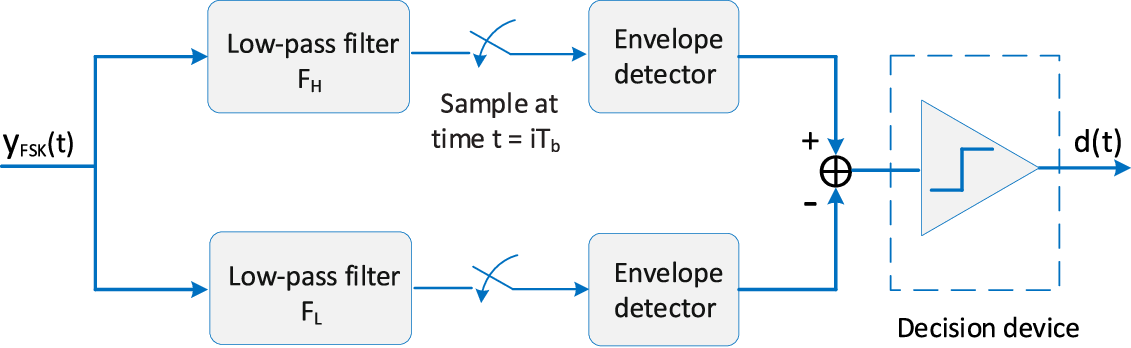}
 	\caption{Non-coherent detection of BFSK modulation \cite{proakis2001digital}.} \label{Envelope_detector_fsk}
\end{figure}
An envelope detector is a device or circuit used to extract the envelope of a signal. The envelope of the signal contains the original baseband signal, and it can be used to recover the information that was modulated onto the carrier. An envelope detector typically works by rectifying the input signal to remove the negative-going portions of the waveform and then filtering the resulting signal to remove the carrier frequency. The filtered signal is then a replica of the original baseband signal, with the same amplitude and frequency content but with a DC offset.

As a result, the envelope detector simplifies frequency and phase synchronization at the cost of performance compared to coherent detection. Fig. \ref{Envelope_detector} and Fig. \ref{Envelope_detector_fsk} respectively show the non-coherent detectors for on-off keying (OOK) and binary frequency shift keying (BFSK), the two well-known modulations employed at the backscatter tags. The BFSK signal is processed through two band-pass filters and envelope detectors, with the outputs being inverted to each other. The decision is made based on a simple test of $P_H > P_L$. However, envelope detectors cannot be used for phase modulation schemes like PSK, as they remove phase information. A differential detector will be discussed as an alternative.

\begin{figure}[t]
\centering
\begin{minipage}[b]{.5\textwidth}
\centering
\includegraphics[width=2.4in]{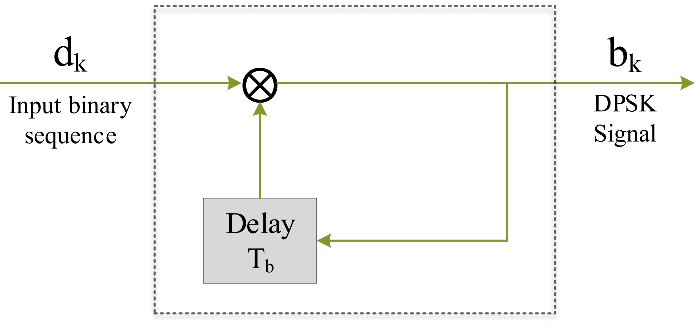}
 \subfigure{Differential encoding.} \vspace{5mm}
\end{minipage} 
\begin{minipage}[b]{.5\textwidth} 
\centering\includegraphics[width=3.5in]{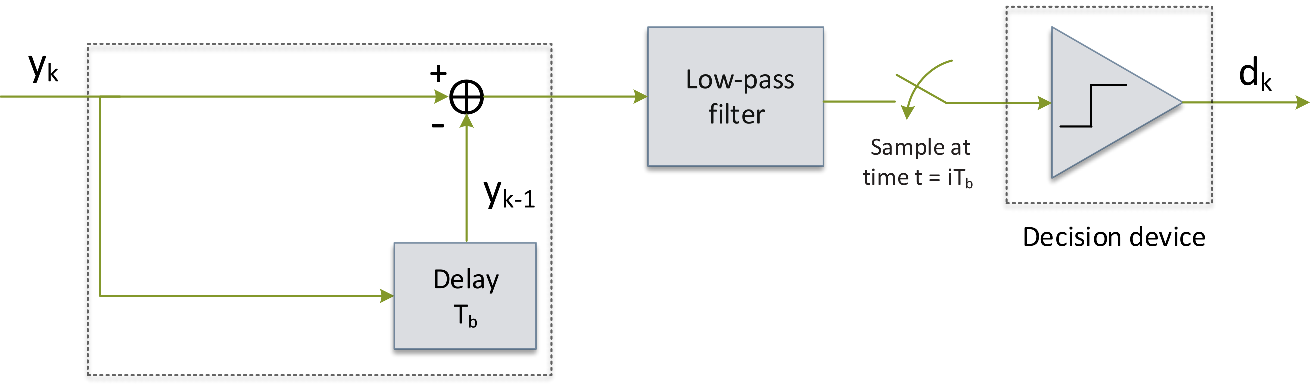}
\subfigure{Differential decoding.} 
\end{minipage}
\caption{DPSK encoding and decoding diagram \cite{proakis2001digital}.} 
\label{differential_modulation}
\end{figure}

\subsubsection{\bf{Differential detector}}
It is a suboptimal non-coherent receiver that does not require knowledge of channel phases or polarities. That will reduce the complexity of the receiver and avoid power-consuming
training sequences needed for CE \cite{Wang2016, Devineni2019, Qian2017, Qin_Tao}. This scheme works when the channel is unknown but is unchanged over a coherence interval that spans several (two) symbol durations. The differential detector compares the current received symbol with the previously received symbol and makes a decision based on the difference between them, which is known as differential phase shift keying (DPSK) modulation (Fig. \ref{differential_modulation}). Differential detection is commonly applied in systems with PSK-signalling, where the information bits are encoded as the phase difference between current and previous symbols.

For the differential scheme, the transmitted symbols $d_k$ are first differentially encoded to produce the modulated symbols $b_k$, selected from a $Q$-ary modulated symbols. Specifically, for binary modulation, the differential encoder generates the modulated symbol $b_k$ by taking the XOR (denoted by $\otimes$) of the previous modulated symbol $b_{k-1}$ and the current transmitted symbol $d_k$, i.e., $b_k=b_{k-1} \otimes d_k$ (Fig. \ref{differential_modulation}a), where, $b_0$ is chosen as a reference symbol with a value of \num{1} and it is assumed that the transmitted symbols $d_k$ have an equal probability of being either 0 or 1.  This can thus resolve $180^{o}$ phase ambiguity, which is enough for the binary case.
For a general $Q$-array case, however, a $Q$-fold phase ambiguity occurs. The differential encoder  is thus formulated as \cite[Section 3.5.2]{vitetta2013wireless}
\begin{eqnarray}
    b_k = b_{k-1} \exp (j\frac{2 \pi}{Q} d_k),
\end{eqnarray}
which is equivalent to
\begin{eqnarray}
    \varphi_{k} - \varphi_{k-1} = \frac{2 \pi}{Q} d_k,
\end{eqnarray}
{where $b_k = e^{j \varphi_{k}}$, and $\varphi_{k} = 2\pi(k-1)/Q, k = 1,\ldots,Q $. Therefore, differential encoding maps the $k$-th information symbol $d_k$ into the phase change from the last transmitted symbol $b_{k-1}$ to the following one $b_k$.

At the receiver, the differential detector computes the difference between the current received symbol $y_k$ and the previously received symbol $y_{k-1}$ (Fig. \ref{differential_modulation}b) :
\begin{equation}
    \Delta y_k = y_k - y_{k-1}.
\end{equation}

The decision for the transmitted symbol $\hat{d}_k$ is then made by choosing the symbol from the constellation set $\mathcal{S}$ that minimizes the distance between $\Delta y_k$ and the difference between each possible symbol and the previously transmitted symbol $\hat{d}_{k-1}$ \cite{proakis2001digital}:
\begin{equation}
    \hat{d}_k = \underset{d_i \in \mathcal{S}}{\operatorname{argmin}}\:\: |\Delta y_k -  (d_i - \hat{d}_{k-1})|^2
\end{equation}
The symbol $\hat{d}_k$ is then used to recover the transmitted message.

Differential detection is simple and improves noise immunity. However, accurate synchronization between the receiver and transmitter is required, as the reference signal must be correctly derived from the transmitted symbol. The receiver must also ensure that the received signals are correctly mapped to the transmitted symbols, especially when using $Q$-ary modulation.

\begin{table*}[]
	\caption{Summary of detectors in \abc systems}\label{table_detector}
	\centering
\begin{tabular}{|p{1.8cm}|p{6cm}|p{3.8cm}|p{3.8cm}|}
\hline
\rowcolor[HTML]{ECF4FF}  Detectors & Description & Pros & Cons \\
\hline
Bayesian  & Uses Bayes' theorem to update the probability of a hypothesis based on new data & Incorporates prior information, flexible, updates as new data arrives & Computationally expensive, requires prior knowledge \\
\hline
Mini-Max  & Minimizes the maximum possible loss that could occur when a decision is made based on the test result & Robust to model failure, minimizes maximum loss & Assumes a fixed loss function, may be conservative \\
\hline
NP& Uses a fixed level of significance and a LRT statistic to decide whether to accept or reject a null hypothesis & Provides a clear decision rule, widely used, well-established & Assumes a fixed level of significance, may lead to errors \\
\hline
GLRT & Uses the LRT statistic and does not assume a specific distribution for the data & Does not require distributional assumptions, flexible & Can be computationally expensive, may not exist in all cases \\
\hline
Differential  & Identifies a signal presence by comparing the difference between the received signal and a reference signal to a threshold value & Can detect weak signals in noisy environments, versatile & Requires prior knowledge, may be sensitive to noise \\\hline
Energy & Detects the presence of a signal based on its energy in a particular frequency band & Simple to implement, effective for detecting signals with high energy & May be susceptible to noise \\
\hline
Semi-Coherent & A combination of both coherent and non-coherent detection & Improves signal detection performance, versatile & Requires coherent signals, can be computationally expensive \\
\hline
 MaL & A detector that maximizes the likelihood function of the received signal & Asymptotically optimal, efficient for large sample sizes & Requires knowledge of signal statistics, computationally expensive  \\
\hline
\end{tabular}
  \end{table*}

The detection methods and their respective advantages and disadvantages are briefly described in Table \ref{table_detector}.

\subsection{Learning-Based Signal Detection}
MaL was introduced in the late 1950s \cite{Mohammad_Abu_Alsheikh} and is now gaining attention in both the academic and industrial communities for next-generation networks \cite{Mihaela_van,Chunxiao_Jiang}. In the last decade, MaL techniques have become increasingly popular for various tasks, including image/audio processing, spam detection, computer vision, fraud detection, etc \cite{Mohammad_Abu_Alsheikh}. Over time, the research focus has changed and become centered on reliable and computationally feasible algorithms. These come from statistics, mathematics, neurology,  computer science, and other disciplines.  

 \begin{figure}[t]
 \centering
 	\includegraphics[width=3.5in]{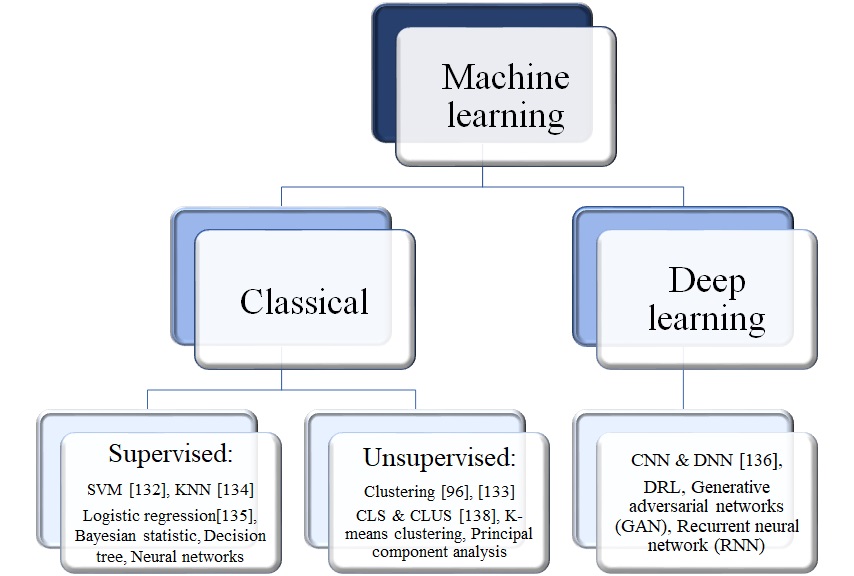}
 	 \caption[]{MaL architecture\footnotemark .}
  \label{summary_ML}
 \end{figure}

One of the advantages of the MaL-based signal detection approach is that it can be used in a variety of AmBC systems, including those with multiple antennas, varying channel conditions, and different modulation schemes. Additionally, the MaL-based approach can be adapted to changing conditions by updating the training data and retraining the model as needed. In conclusion, the MaL-based signal detection approach is a powerful and flexible method for signal detection in AmBC systems. It provides high accuracy, robustness, and adaptability, making it a valuable tool for a wide range of applications.

Several types of MaL algorithms can be used for signal detection, including decision trees, random forests, support vector machines (SVMs), artificial neural networks, and others \cite{Mohammad_Abu_Alsheikh}. The choice of algorithm will depend on the specific requirements of the system, such as the dimensionality of the input signals, the number of classes, and the desired accuracy of the model. The accuracy of the predictions will depend on the quality of the training data, the choice of features, and the specific MaL algorithm used. Supervised learning, unsupervised learning, and deep learning are all subfields within the larger field of MaL, but they differ in terms of their approach and applications (Fig. \ref{summary_ML}).
\footnotetext{The deep learning method can also be classified into two main categories: supervised and unsupervised techniques.}
\begin{enumerate}
    \item[$\bullet$] \textbf{Supervised learning}: Supervised learning algorithms are trained on labeled data and make predictions based on the relationship between the input and output variables. The goal is to learn a mapping from inputs to outputs so that the algorithm can make predictions for new, unseen data.  There are two types of supervised learning, linear and discrete (logistic) regression. Linear regression uses a linear regression function, whereas logistic regression uses a logistic function based on a sigmoid curve. 

    \item[$\bullet$] \textbf{Unsupervised learning}: Unsupervised learning algorithms are trained on unlabeled data and aim to find patterns or structures in the data without the use of a label. The goal is to uncover hidden relationships and dependencies in the data, rather than making predictions. A particularly popular unsupervised learning technique is $k$-means clustering. Based on Euclidean distance, the clustering method assigns each item to the cluster whose centroid is closest to it, then iteratively updates the cluster-centroid to reduce differences between clusters until convergence has been achieved

    \item[$\bullet$] \textbf{Deep learning}: Deep learning algorithms can be trained in a supervised or unsupervised manner, depending on the problem at hand and the availability of labeled data. Fig. \ref{deep_learning_arch} illustrates the structure of a simple DNN, which is composed of multiple layers of interconnected nodes, known as artificial neurons. The input layer takes in data and the output layer provides the final prediction. Between the input and output layers are multiple hidden layers that perform transformations on the data to extract features and make predictions. 
\end{enumerate}

\begin{figure}[t]
\centering
		\includegraphics[width=3.3in]{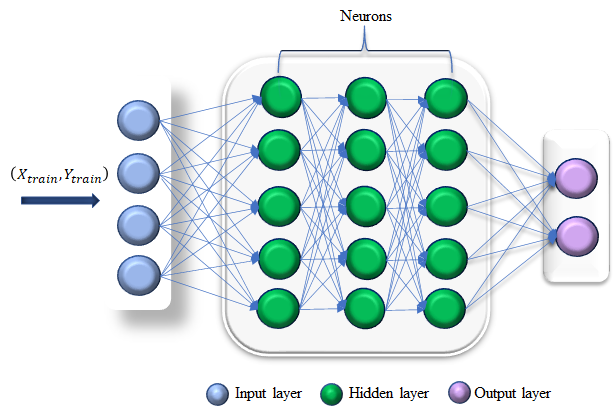}
	\caption{Architecture of a classical deep neural network.}\label{deep_learning_arch}
\end{figure}

\section{Relevant Prior Art}\label{Sec:prior_art}
This section provides a comprehensive overview of detection methods for \abc, covering coherent, non-coherent, and semi-coherent techniques. Simulations are included to gain deeper insights into the performance of various schemes. By thoroughly examining state-of-the-art techniques and algorithms, we aim to identify current limitations of detection strategies and propose future research directions.

\begin{table*}[]
\caption{Summary of coherent detection in \abc systems. }\label{table_choheent}
 \centering
\begin{tabular}{|c|c|l|l|c|}
\hline
\rowcolor[HTML]{FFFFFF} 
Works& Tag modulation type & Detection design & Setup parameters  & Achieved BER 
\\ \hline
\rowcolor[HTML]{ECF4FF} 
\multicolumn{5}{|c|}{\cellcolor[HTML]{ECF4FF}Single-antenna } \\ \hline
\rowcolor[HTML]{E0FFFF} 
\cite{Chen_Chen_2}& OOK    & \begin{tabular}[c]{@{}l@{}}ML detector\\ Practical ML detector\end{tabular}  &  \begin{tabular}[c]{@{}l@{}}$N = 100$ \\ Ntr = $10$\\ SNR = $20$ dB\end{tabular}    & $\sim \hspace{0.3em}> 10^{-2} $  \\

\cite{Chen_Chen}   & OOK    & ML detector  & \begin{tabular}[c]{@{}l@{}} \begin{tabular}[c]{@{}l@{}}  $f_c= 900$ MHz \\ $P_s = 30$ dB 
\end{tabular} \end{tabular} &  \cellcolor[HTML]{FFFFFF}$\sim \hspace{0.3em}< 10^{-2} $  \\
 
\rowcolor[HTML]{E0FFFF} 
\cite{William_Barott}  & OOK    & ML detector  & \xmark  &   \cellcolor[HTML]{E0FFFF} \xmark  \\
 
\cite{zargari2023improved}&  OOK   & JCED, IED, ED,  ML detector  & \begin{tabular}[c]{@{}l@{}}$N = 512$\\ $P_F = 0.05$ \\ SNR = $20$ dB\end{tabular} & \cellcolor[HTML]{FFFFFF} $\sim \hspace{0.3em}> 10^{-2} $  \\

\rowcolor[HTML]{E0FFFF} 
\cite{Youyou_Zhang}& \begin{tabular}[c]{@{}l@{}}$M$-PSK at RF source\\ OOK at tag\end{tabular}& ML detector  & \begin{tabular}[c]{@{}l@{}}$N = 50$\\ $\Delta \gamma = -20$ dB\\ SNR = $20$ dB\end{tabular} & \cellcolor[HTML]{E0FFFF} $\sim \hspace{0.3em}> 10^{-2} $  \\

\cite{Yunkai_Hu} & \begin{tabular}[c]{@{}l@{}}LDPC-coded RF source\\ BPSK at tag\end{tabular} & ML detector  & \begin{tabular}[c]{@{}l@{}}$N = 101$\\ 
$\Delta \gamma = -10$ dB\\ SNR = $10$ dB\end{tabular}  &\cellcolor[HTML]{FFFFFF}  $\sim \hspace{0.3em}> 10^{-4}$\\

\rowcolor[HTML]{E0FFFF} 
\cite{Jing_Qian_6}   & \begin{tabular}[c]{@{}l@{}} $4$-PSK \\ OOK \end{tabular}   & ML detector  & \begin{tabular}[c]{@{}l@{}} $N = 50$ \\ $\Delta \gamma$ = $0.2$\\ Ntr $ = 5$\\ SNR = $20$ dB \end{tabular}   &\cellcolor[HTML]{E0FFFF}  $\sim 10^{-2}-10^{-3} $  \\

\cite{Georgios_Vougioukas_1,Georgios_Vougioukas} & \begin{tabular}[c]{@{}l@{}}Frequency-shifted  BPSK\\ Pseudo-FSK \\ FM\end{tabular} & ML detector  & \begin{tabular}[c]{@{}l@{}}$F_s = 2$ MHz\\ $N = 2000$\\ Ntr $= 4$\\ SNR = $20$ dB\end{tabular}  &\cellcolor[HTML]{FFFFFF}  $\sim \hspace{0.3em}> 10^{-2} $  \\

\rowcolor[HTML]{E0FFFF} 
\cite{Jixiang} &\begin{tabular}[c]{@{}l@{}}$8$-PSK RF source\\ TSK at tag\end{tabular}  & ML and energy detectors  & \begin{tabular}[c]{@{}l@{}}  $N = 40$\\  Ntr $= 14$\\  SNR = $20$ dB\end{tabular}  &  $\sim \hspace{0.3em}> 10^{-2} $  \\
\hline

\rowcolor[HTML]{ECF4FF} 
\multicolumn{5}{|c|}{\cellcolor[HTML]{ECF4FF} Multi-antenna} \\ \hline
\rowcolor[HTML]{E0FFFF} 
\multicolumn{1}{|c|}{\cellcolor[HTML]{E0FFFF}\cite{zhang2020}} &
\multicolumn{1}{c|}{\cellcolor[HTML]{E0FFFF}OOK} &
\multicolumn{1}{l|}{\cellcolor[HTML]{E0FFFF}\begin{tabular}[c]{@{}l@{}}Direct link averaging detector\\ Distance detector\\ Modified distance detector\end{tabular}} &
\multicolumn{1}{l|}{\cellcolor[HTML]{E0FFFF}\begin{tabular}[c]{@{}l@{}}$N =  50 $ \\ $\Delta \gamma = -30$dB\\SNR = 20 dB\end{tabular} } & $\sim \hspace{0.3em}< 10^{-1}$\\
\rowcolor[HTML]{FFFFFF} 
\multicolumn{1}{|c|}{\cellcolor[HTML]{FFFFFF}\cite{YangZ2018}} &
\multicolumn{1}{c|}{\cellcolor[HTML]{FFFFFF}BPSK} &
\multicolumn{1}{l|}{\cellcolor[HTML]{FFFFFF}\begin{tabular}[c]{@{}l@{}}ML detector\\ Linear detectors\\ SIC-based detectors\end{tabular}} &
\multicolumn{1}{l|}{\cellcolor[HTML]{FFFFFF}\begin{tabular}[c]{@{}l@{}} $\Delta \gamma$ = -$10$ dB \\SNR = $20$ dB\end{tabular}}&\cellcolor[HTML]{FFFFFF}
$\sim \hspace{0.3em} 10^{-2}$  \\
\rowcolor[HTML]{E0FFFF} 
\multicolumn{1}{|c|}{\cellcolor[HTML]{E0FFFF}\cite{Ma2019}} &
\multicolumn{1}{c|}{\cellcolor[HTML]{E0FFFF}OOK} &
\multicolumn{1}{l|}{\cellcolor[HTML]{E0FFFF}\begin{tabular}[c]{@{}l@{}}Linear detector\\ Bayesian learning detector\end{tabular}} &
\multicolumn{1}{l|}{\cellcolor[HTML]{E0FFFF}\begin{tabular}[c]{@{}l@{}} $32\times 12$ S-R CH$^{*}$ \\$N =  5$ \\SNR = $20$ dB\end{tabular}}&
$\sim \hspace{0.3em} 10^{-3}$ \\
\rowcolor[HTML]{FFFFFF} 
\multicolumn{1}{|c|}{\cellcolor[HTML]{FFFFFF}\cite{GuoZXL19}} &
\multicolumn{1}{c|}{\cellcolor[HTML]{FFFFFF}OOK} &
\multicolumn{1}{l|}{\cellcolor[HTML]{FFFFFF}\begin{tabular}[c]{@{}l@{}}Beamforming ED\\Liklihood ratio \end{tabular}} &\multicolumn{1}{l|}{\cellcolor[HTML]{FFFFFF}\begin{tabular}[c]{@{}l@{}}$N =  50 $\\$\Delta \gamma = -25$ dB \\ SNR = $20$ dB\end{tabular}}&\cellcolor[HTML]{FFFFFF} $10^{-2}-10^{-3}$ \\

\rowcolor[HTML]{E0FFFF} 
\multicolumn{1}{|c|}{\cellcolor[HTML]{E0FFFF}\cite{ma2018signal}} &
\multicolumn{1}{c|}{\cellcolor[HTML]{E0FFFF}OOK} &
\multicolumn{1}{l|}{\cellcolor[HTML]{E0FFFF}ML detector } &
\multicolumn{1}{l|}{\cellcolor[HTML]{E0FFFF}\begin{tabular}[c]{@{}l@{}} SNR = $10$ dB \end{tabular}}&
$\cellcolor[HTML]{E0FFFF}\sim \hspace{0.3em} 10^{-1}$  \\ 

\rowcolor[HTML]{FFFFFF} 
\multicolumn{1}{|c|}{\cellcolor[HTML]{FFFFFF}\cite{Liu2015}} & \multicolumn{1}{c|}{\cellcolor[HTML]{FFFFFF}OOK} &\multicolumn{1}{l|}{\cellcolor[HTML]{FFFFFF}ML detector with differential encoder} &\multicolumn{1}{l|}{\cellcolor[HTML]{FFFFFF}\begin{tabular}[c]{@{}l@{}} $N = 1000$ \\ SNR = $20$ dB  \end{tabular}}&
\cellcolor[HTML]{FFFFFF}$\sim \hspace{0.3em} 0.005$  \\ 

\rowcolor[HTML]{E0FFFF} 
\multicolumn{1}{|c|}{\cellcolor[HTML]{E0FFFF}\cite{LiuSTM21}} &
\multicolumn{1}{c|}{\cellcolor[HTML]{E0FFFF}BPSK} &\multicolumn{1}{l|}{\cellcolor[HTML]{E0FFFF}\begin{tabular}[c]{@{}l@{}}ML detector\\ Linear detector \\ Differential OSTBC detector\end{tabular}} &\multicolumn{1}{l|}{\cellcolor[HTML]{E0FFFF}\begin{tabular}[c]{@{}l@{}}$2\times 2$ T-R CH$^{*}$\\ $\Delta \gamma = 40$ dB \\ SNR = $20$ dB\end{tabular}} & $\sim \hspace{0.3em} 10^{-3}$  \\ 

\rowcolor[HTML]{FFFFFF} 
\multicolumn{1}{|c|}{\cellcolor[HTML]{FFFFFF}\cite{Tao2019}} &
\multicolumn{1}{c|}{\cellcolor[HTML]{FFFFFF}OOK} &\multicolumn{1}{l|}{\cellcolor[HTML]{FFFFFF}Maximum eigenvalue detector} & \multicolumn{1}{l|}{\cellcolor[HTML]{FFFFFF}\begin{tabular}[c]{@{}l@{}}$N =  100$ \\ SNR = $20$ dB\end{tabular}}
& \cellcolor[HTML]{FFFFFF} $\sim 10^{-1}-10^{-2}$ \\ \hline

\rowcolor[HTML]{ECF4FF} 
\multicolumn{5}{|c|}{\cellcolor[HTML]{ECF4FF}Multi-antenna Tag} \\ \hline

\rowcolor[HTML]{E0FFFF} 
\multicolumn{1}{|c|}{\cellcolor[HTML]{E0FFFF}\cite{HuZZJJN19}} & 
\multicolumn{1}{c|}{\cellcolor[HTML]{E0FFFF}OOK} &
\multicolumn{1}{l|}{\cellcolor[HTML]{E0FFFF}NP, GLRT} &
\multicolumn{1}{l|}{\cellcolor[HTML]{E0FFFF}\begin{tabular}[c]{@{}l@{}} Tag antenna: $10$ \\$N =  1000$ \\SNR = $20$ dB\end{tabular}}&
\cellcolor[HTML]{E0FFFF} $\sim \hspace{0.3em} 0.3$  \\ 

\rowcolor[HTML]{FFFFFF} 
\multicolumn{1}{|c|}{\cellcolor[HTML]{FFFFFF}\cite{Niu0J19}} &
\multicolumn{1}{c|}{\cellcolor[HTML]{FFFFFF}QPSK} &
\multicolumn{1}{l|}{\cellcolor[HTML]{FFFFFF}ML detector
} &\multicolumn{1}{l|}{\cellcolor[HTML]{FFFFFF}\begin{tabular}[c]{@{}l@{}} Tag antenna: $4$ \\$N =  1000$ \\$\Delta \gamma$ = $0$ dB\end{tabular}}&\cellcolor[HTML]{FFFFFF}
\cellcolor[HTML]{FFFFFF}  $\sim \hspace{0.3em} 10^{-3}$ \\ \hline

\hline
\rowcolor[HTML]{ECF4FF} 
\multicolumn{5}{|c|}{\cellcolor[HTML]{ECF4FF}Multi-antenna RF source} \\ \hline

\rowcolor[HTML]{E0FFFF} 
\multicolumn{1}{|c|}{\cellcolor[HTML]{E0FFFF}\cite{hu2020ambient}} &
\multicolumn{1}{c|}{\cellcolor[HTML]{E0FFFF}OOK} &
\multicolumn{1}{l|}{\cellcolor[HTML]{E0FFFF}ML detector with differential encoder} &
\multicolumn{1}{l|}{\cellcolor[HTML]{E0FFFF}\begin{tabular}[c]{@{}l@{}} 64 antenna-FDA RF source \\$N =  1000$ \\SNR = $20$  dB \\ CFO = $3$MHz\end{tabular}}&\cellcolor[HTML]{E0FFFF}
\cellcolor[HTML]{E0FFFF}  $\sim \hspace{0.3em} 10^{-1}$ \\ \hline

\end{tabular}

\vspace{5pt} 
\noindent $^{*}$ S-R CH and T-R CH denote the source-to-reader and tag-to-reader channels, respectively.

\end{table*}

\subsection{Coherent Detection}
In the following, we consider two system setups: (i) a single-antenna setup where each node (RF source, tag, and reader) has a single antenna, and (ii) a multi-antenna setup where any node can have multiple antennas (Table \ref{table_choheent}). Multi-antenna systems offer additional degrees of freedom, leading to diversity and multiplexing gains, thereby enhancing detection performance.

\subsubsection{\bf{Single-antenna setup}} 
In \cite{Chen_Chen_2}, the authors examine the detection performance of an AmBC reader equipped with full-duplex (FD) technology, using a two-antenna Wi-Fi AP. They recommend training signal interleaving and LMMSE-based CE to mitigate self-interference and estimate the channel. They propose two ML detectors founded on theoretical and practical thresholds and optimize the pilot symbols number to maximize system throughput. Another practical implementation discussed in \cite{William_Barott} involves correlation processing as the ML detector in AmBC systems. While this method augments data rates and ranges, it demands more intricate readers.

In \cite{Yiwen_Tao}, a solution to the low SNR challenge in AmBC systems, stemming from unpredictable and variable RF signal characteristics, is proposed. The authors advocate the adoption of nonlinear filtering methods for the reader, which amplify the output SNR and necessitate simpler analog circuits and slower ADC devices compared to linear filters. They introduce a novel analog circuit grounded in stochastic resonance, a nonlinear filtering technique that boosts SNR while diminishing potent noise. Moreover, they present a new sampling method for sub-symbol level samples. The reader subsequently leverages the sampled signal and an automatically computed decision threshold to retrieve transmitted symbols, leading to a notable decline in BER due to the enhanced output SNR. In \cite{zargari2023improved}, the authors propose two innovative detectors to address challenges in AmBC systems arising from unknown channel states, RF source parameters, and interference. These detectors are the joint correlation-energy detector (JCED) and the improved energy detector (IED), which are benchmarked against the conventionally used ED. Notably, the JCED incorporates a test statistic melding two factors: the energy of received signal samples and their correlation. This design facilitates a broader signal evaluation. Conversely, the IED, also termed the $p$-norm detector, integrates an arbitrary positive power function represented by $p$. The research reveals that both the JCED and IED surpass the TED.

In \cite{Wenjing_Zhao}, backscatter technology is proposed as a potential solution for high-speed rail (HSR) wireless communications. HSR environments introduce challenges including high penetration losses, significant Doppler shifts, fast time-varying channels, and swift handovers owing to transceiver mobility. Although techniques like beamforming, massive MIMO, and FD can curtail signal penetration loss, the authors propose a cost-effective backscatter method as an alternate solution. Their proposed strategy encompasses outfitting the train with two antennas: an exterior one for radio signal reception and an interior one for backscatter signal acquisition. The paper delves into channel capacity by considering design facets like CFO, channel estimators, and signal detectors, and compares three types of detectors: ML, successive interference cancellation (SIC), and zero-forcing (ZF). The ZF detector outperforms the SIC detector, whereas the ML detector, although optimal, is computation-intensive

In \cite{Yunkai_Hu}, the BER performance of the AmBC reader is assessed using low-density parity-check code (LDPC)-coded RF source signals, marking a unique contribution to existing literature. The research introduces a joint detection-decoding algorithm based on cascade channel coefficients, aiming to detect both the tag and RF source signals. CE unfolds in two stages, employing either the LS algorithm or the MMSE estimator. In the first stage, the tag sends a ``$-1$" symbol sequence, while in the second, it transmits a ``$1$" sequence. LDPC-coded RF source signals are decoded through a belief propagation decoding algorithm, necessitating only raw and encrypted bits for tag signal detection. Subsequently, MMSE estimation is invoked to detect the tag signal based on RF source signal knowledge. Analytical upper and lower BER performance bounds are established for both tag and RF source signals under ML decoding. The findings indicate that AmBC systems deploying LDPC-coded RF source signals showcase superior BER performance. This underscores the pivotal role of LDPC coding in amplifying the reliability and resilience of AmBC systems.

In \cite{Jing_Qian_6}, multi-level energy detection is explored for AmBC systems using $M$-PSK modulation schemes. This differentiates the study from earlier ones emphasizing OOK or PSK modulations exclusively for the tag. The objective is to enhance data rate and robustness without increasing power consumption. By integrating $M$-PSK modulation at the tag level, data transmission accelerates without necessitating a power-hungry high-frequency oscillator. The reader utilizes ML detection and selects the higher conditional PDF  based on multiple hypothesis tests to detect the transmitted data. Concurrently, an ED approach is evaluated, slightly surpassing the ML detector in terms of symbol error rate (SER). Detection thresholds for both systems are deduced by approximating specific unknown parameters, notably the variance of the received signal by leaning on pilot-based channels instead of the actual channels. Moreover, a hardware prototype is developed based on a phasor diagram illustration to select the load impedance at the tag, demonstrating the feasibility of the $4$-PSK AmBC protocol. Experimental outcomes reveal that a data rate reaching $20$ kbits/s is attainable spanning a $2.5$-foot distance.

\begin{figure}
\centering
	\includegraphics[width=3.5in]{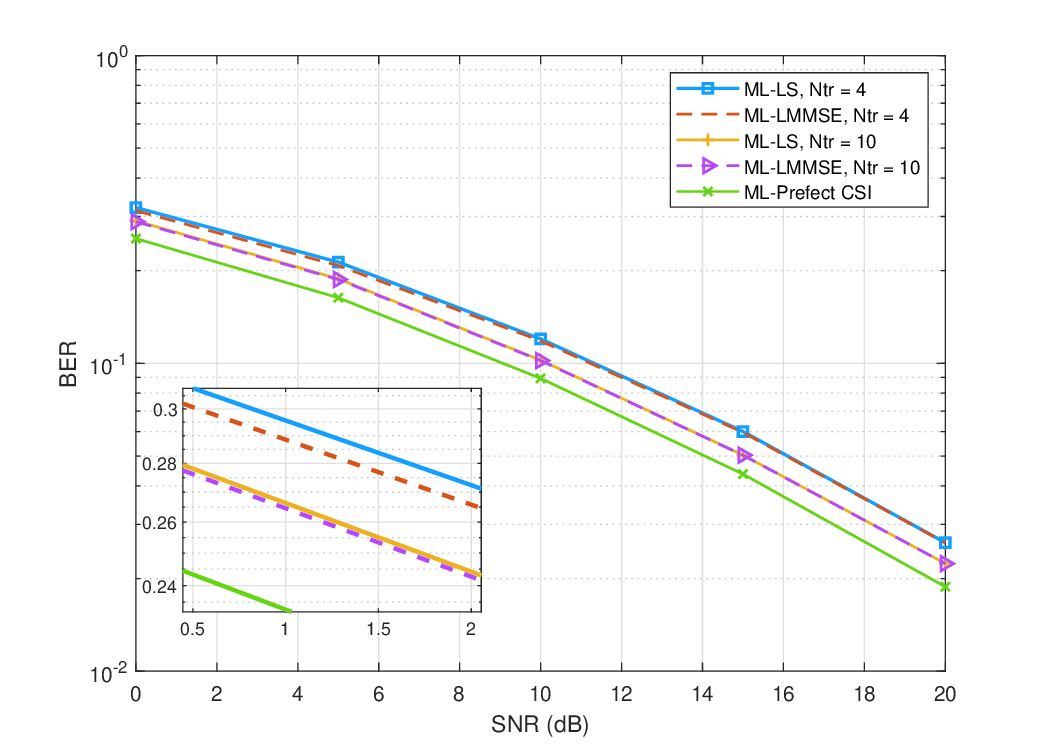}
 	\caption{BER versus SNR for different channel estimation techniques and training sequence (Ntr) under Rayleigh fading. \cite{Georgios_Vougioukas_1}.} \label{pseudo_fsk_fig}
\end{figure}
 
AmBC systems typically use OOK modulation due to their low cost and simplicity in circuit design. However, more advanced modulation schemes, like binary frequency-shift keying (BFSK), have also been introduced as shown in \cite{Anran_Wang, Pengyu_Zhang, Georgios_Vougioukas}. Practical applications employ various techniques such as IC \cite{Anran_Wang}, ring oscillators \cite{Pengyu_Zhang}, or the Fourier decomposition of periodic rectangular pulses \cite{Georgios_Vougioukas} to generate the needed frequency shift. The tag circuit, which includes an inductor, capacitor tank oscillator, and negative metal oxide semiconductor (NMOS) transistor switch, emits square waves at varied frequencies that correspond to tag information. Using an NMOS transistor driven by a square wave, the tag antenna then alternates between open and short impedance \cite{Anran_Wang}. One significant advantage of BFSK modulation is its ability to negate interference from direct ambient RF signals, thus preventing error floors in high SNR scenarios. Reference \cite{Jixiang} proposes high-order time shift keying (TSK) for enhanced time domain modulation. The study assesses its performance in AmBC with Gaussian and MPSK sources, introducing ML and ED detectors. Remarkably, error rates improve with specific training symbols, but even without them, symbol detection remains accurate. Simulations show larger modulation orders work for MPSK sources, while complex Gaussian sources suit the low SNR region. 

Moreover, BFSK modulation has several drawbacks, including high energy consumption due to large tag frequency shifts and system complexity when the CFO is unknown. It struggles to support high tag data rates, and AmBC's frequency shift keying  (FSK) decoding differs from traditional FSK, making signal detection algorithm design challenging.

In \cite{Georgios_Vougioukas_1}, the BER performance of the tag is scrutinized using pseudo-FSK modulation with an ML detector and LS/LMMSE CE techniques. Fig. \ref{pseudo_fsk_fig} depicts the BER performance versus SNR for different CE techniques. The data reveals that the ML detector, when paired with perfect CSI, surpasses the LMMSE and LS detectors. When utilizing a training sequence length of $4$, the ML detector attains a performance edge of roughly $2$ dB over the LS and LMMSE. The LMMSE estimator exhibits a $0.4\%$ uplift in BER performance compared to the LS estimator. Boosting the number of bits for CE diminishes the disparity between LS and LMMSE estimations and the ML detector, but this adjustment introduces a trade-off between data rate and BER performance.

In \cite{Georgios_Vougioukas}, a reader is designed using both digital and analog modulation schemes, complemented by a fully coherent detector. Digital modulation employs pseudo-FSK, a frequency-altered version of BFSK, while FM serves the analog modulation's purpose. The research assesses the BER performance and introduces short packet error correction coding for the frequency-shifted form of BFSK. Notably, the introduced frequency-shifted BFSK type necessitates a considerable frequency shift at the tag, resulting in significant power consumption.

\begin{figure}
\centering
	\includegraphics[width=3.5in]{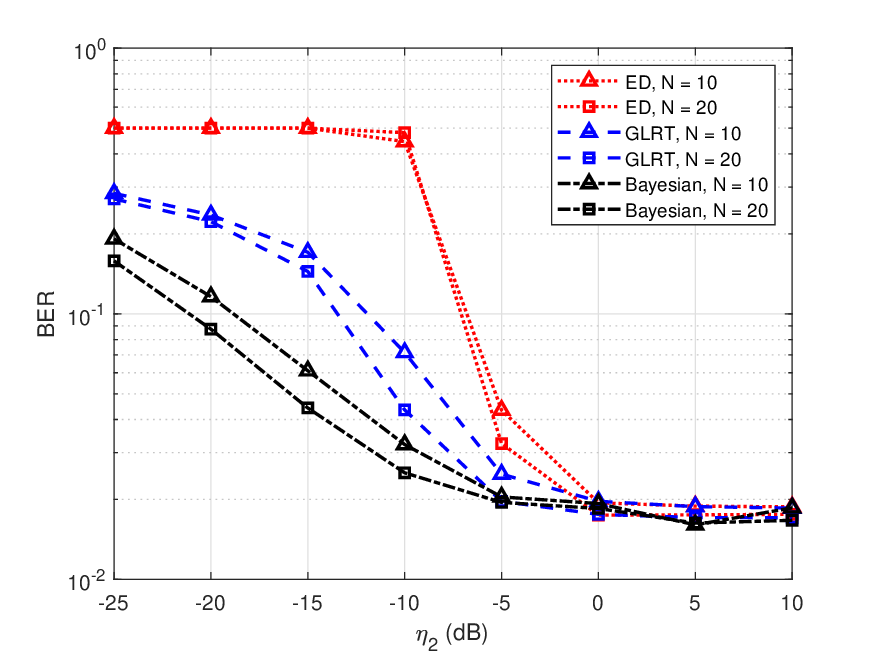}
 	\caption{BER versus INR ($\eta_2$) with given $\eta_1 = 30$ dB,  $\sigma^2_{st}=\sigma^2_{sr}$,  $\eta_1=\xi^2\sigma^2_{tr}$, $\xi=1$, and $\sigma^2_{w}=1$ \cite{Sudarshan_Guruacharya}.} \label{simluation_1}
\end{figure}
\begin{figure}
\centering
	\includegraphics[width=3.5in]{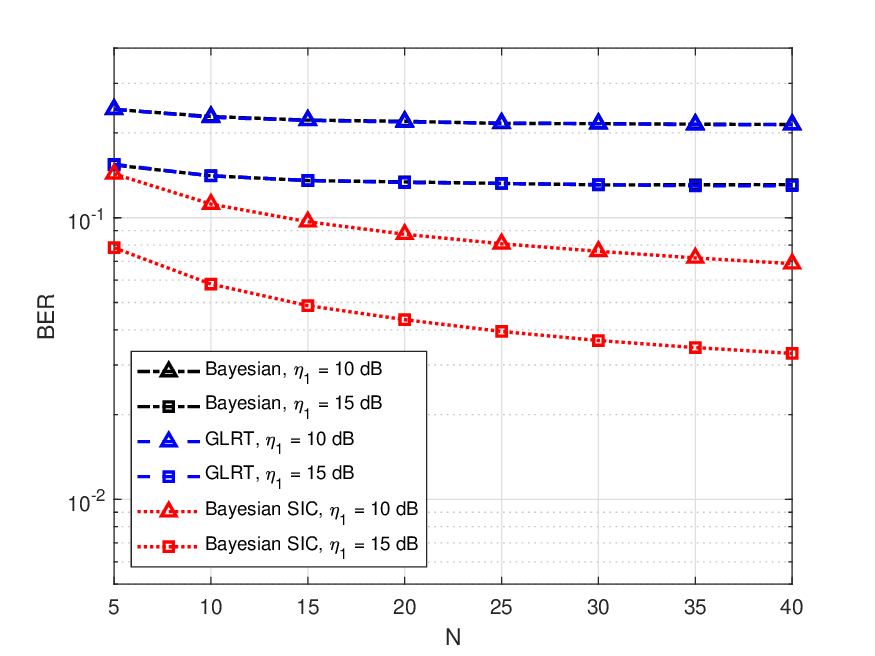}
 	\caption{BER versus N, with given $\eta_2=0$,  $\sigma^2_{st}=\sigma^2_{sr}$,  $\eta_1=\xi^2\sigma^2_{tr}$, $\xi=1$, and $\sigma^2_{w}=1$ \cite{Sudarshan_Guruacharya}.} \label{simluation_2}
\end{figure}

The legacy receiver grapples with interference stemming from the tag's backscattered data. This raises concerns about the range, intensity, and influence of this interference on the legacy receiver's BER. \cite{Chen_Chen} addresses these issues by formulating a BER expression for a legacy receiver equipped with an ML detector, taking into account the backscatter interference. The study establishes that the BER performance is affected by the distances and angles among the tag, RF source, and legacy receiver.  

\textbf{Takeaways}: In summary, several techniques have been introduced to boost the signal detection efficacy of single-antenna \abc readers. Nonetheless, every method possesses its unique constraints, underscoring the need for further scrutiny in subsequent studies. Topics like semi-blind detection, nonlinear filtering approaches, LDPC-coded RF source signals, multi-tiered energy detection, and interference mitigation require additional investigation. The development of more robust estimation methods that are capable of handling imperfect channel knowledge can also enhance the performance of  \abc systems. Recognizing the interference's effective range and magnitude induced by legacy receiver tags can facilitate the design of mitigation techniques.

\subsubsection{\bf{Multi-antenna setup}}\label{Sec:multi_antenna_coherent}
In \cite{zhang2020}, CE and data detection are investigated assuming $M$-PSK RF symbols transmitted by the RF source. The tag transmits two pilot symbols for CE, allowing the reader to remove source interference and estimate the channel by averaging observations. An algorithm utilizing Viterbi-Viterbi and the sample mean is proposed for estimating channel amplitude and phase based on data symbols in the RF packet. Additionally, a distance detector is proposed, minimizing reader complexity by relying on well-calibrated channel estimates. Cognitive \abc, where an \abc network coexists with a legacy system and utilizes the legacy spectrum, is introduced in \cite{GuoZXL19}. Direct link interference from the legacy RF source transmitter poses a challenge, and conventional EDs may reach an error floor. Two error floor-free detection schemes are proposed based on the spatial diversity of a multi-antenna reader in \cite{GuoZXL19}. The first method employs receive beamforming to remove direct link interference at the reader, while the ED recovers the backscatter signal. The second approach utilizes the optimal ML test, both considering the CSI. A clustering-based detection algorithm is also explored, using an iterative algorithm with a short pilot sequence for joint CSI features and backscatter symbol detection. To address noise power density uncertainty, \cite{Tao2019} introduces a maximum-eigenvalue detector derived from the GLRT, with approximate BER performance characterization and simulation results showing superior performance over the ED. The ratio detector proposed in \cite{ma2018signal} leverages the received signal strength ratio at the reader to detect tag backscattered bits. An antenna selection scheme is also presented to improve reader BER.

\begin{figure}
\centering
	\includegraphics[width=3.5in]{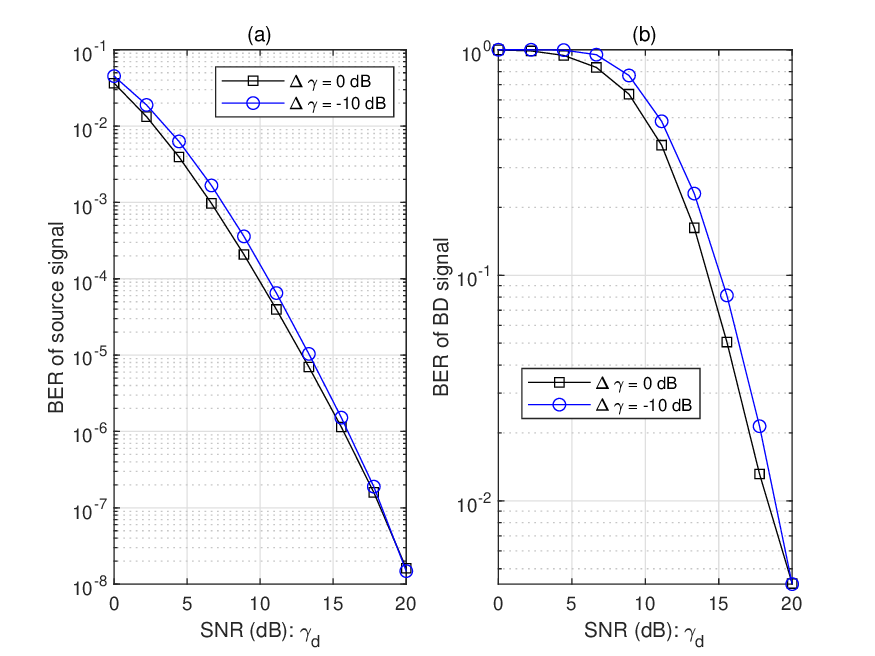}
 	\caption{BER of the ML detector versus SNR for $M = 4$ (number of the reader antennas) \cite{YangZ2018}.} \label{BER_CRx}
\end{figure}

References \cite{YangZ2018, DattaC13} discuss cooperative multi-antenna receiver designs that simultaneously detect information from both the ambient RF source and the tag. \cite{YangZ2018} proposes a detector design for channels with flat and frequency-selective fading. The work introduces optimal ML, linear, and SIC-based detectors for flat fading channels. The SIC-based detector estimates the RF source signal and subtracts direct-link interference to detect the tag signal. The detected tag signal is then utilized to refine the detection of the RF source symbol. Simulation results demonstrate that the SIC-based detector achieves near-optimal ML detection performance when the RF source and tag signal periods are equal. Furthermore, the ML detector is derived for the cooperative receiver, considering ambient OFDM carrier signals under frequency-selective fading channels. Fig. \ref{BER_CRx} (a) and Fig. \ref{BER_CRx} (b) illustrate the achieved BER of the ML detector for the RF source signal and the backscatter signal, respectively, at varying relative SNR between the backscatter link and the direct link ($\Delta \gamma$).

The RF chain circuits, which enable baseband processing of passband communication signals, are crucial parts of digital receivers. However, RF chains can consume substantial power and increase the costs of the radio system. In general, for a low-power RF chain, the power consumption for the entire RF chain could range from a few milliwatts to tens of milliwatts. Spatial modulation (SM) is a novel modulation technique that leverages multi-antenna wireless systems \cite{DattaC13} to reduce the number of RF chains at the transmitter and receiver. It employs a single transmit RF chain and selects a single-antenna element for information transmission. Generalized SM (GESM) is proposed to enhance the spectral efficiency of SM  by enabling the activation of multiple antennas. In the context of AmBC, a cooperative detection method is discussed in \cite{Ma2019}, involving a multi-antenna generalized spatial modulation (GESM) source, a multi-antenna reader, and a single-antenna tag. The authors propose linear and sparse Bayesian detectors based on Bayesian detection to detect both tags and GESM sources. In \cite{hu2020ambient}, the AmBC detection problem is explored when the RF source is a multi-antenna frequency diverse array (FDA). The FDA elements introduce a small frequency increment to the carrier frequency compared to the phase array (PA). The ML detector, utilizing a differential encoder at the tag, is employed to detect the tag's symbol. Simulation results indicate that AmBC FDA outperforms AmBC PA in terms of BER. Assuming full CSI at the reader and perfect synchronization, \cite{Niu0J19} proposes an ML-based detector for jointly estimating the backscatter signal and RF source signal in a multi-antenna AmBC network. The authors consider spatial modulation at the tag, where only one antenna is active, and the index of the active antenna along with the information is transmitted.

\textbf{Takeaways:} 
In conclusion, multi-antenna configurations in coherent detection greatly enhance the signal detection capability of AmBC readers, offering spatial degrees of freedom and advancements in diversity and multiplexing. Various detection methods, such as Viterbi-Viterbi and sample means, EDs, maximum-eigenvalue detectors, ratio detectors, linear and sparse Bayesian detectors, and ML detectors, have been proposed. Cooperative multi-antenna receiver designs have been introduced to simultaneously detect data from the tag and the ambient RF source. Moreover, the use of multi-antenna frequency diverse arrays and GESM can reduce the number of RF chains at the transmitter and receiver, respectively. While multi-antenna AmBC systems have limitations, they also offer advantages that warrant further research and improvement.

\subsection{Non-Coherent Detection for AmBc}
Similarly, we examine two types of system configurations:  (i) a single-antenna setup, and (ii) a multi-antenna setup  (Table \ref{table1_non_semi})

\begin{table*}[]
\caption{Summary of non-coherent and semi-coherent detection in AmBC systems.}\label{table1_non_semi}
\centering
\begin{tabular}{|c|c|l|l|c|}
\hline

\rowcolor[HTML]{FFFFFF} 
Works& Tag modulation type & Detection design & Setup parameters  & Achieved BER  \\ \hline
\rowcolor[HTML]{ECF4FF} 
\multicolumn{5}{|c|}{\cellcolor[HTML]{ECF4FF}Single-antenna } \\ \hline
\rowcolor[HTML]{E0FFFF}  \cite{Wenjing_Zhao}& OOK    & \begin{tabular}[c]{@{}l@{}}ML detector\\ SIC detector\\ ZF detector\end{tabular}          & \begin{tabular}[c]{@{}l@{}}$N = 10$\\ 
SNR = $20$ dB\end{tabular} & $\sim \hspace{0.3em}< 10^{-2}  $    \\

\cite{Jing_Qian_3}& OOK    & \begin{tabular}[c]{@{}l@{}}ML detector \\ Energy detector\end{tabular}& \begin{tabular}[c]{@{}l@{}}$N = 100$\\ $\Delta \gamma$ = $0.5$\\ SNR = 20 dB\end{tabular}  & \cellcolor[HTML]{FFFFFF}  $\sim \hspace{0.3em}> 10^{-4} $  \\

\rowcolor[HTML]{E0FFFF } 
\cite{Wang2016} & \begin{tabular}[c]{@{}l@{}}Differential encoder\\ OOK\end{tabular}    & Energy detector & \begin{tabular}[c]{@{}l@{}}$N = 50$\\ SNR = 20 dB\end{tabular}     &   $\sim \hspace{0.3em}> 10^{-2} $  \\

\rowcolor[HTML]{FFFFFF} 
\cite{Kartheek_Devineni} & OOK    & \begin{tabular}[c]{@{}l@{}}Energy detector\\ Mean threshold detector\end{tabular}   & \begin{tabular}[c]{@{}l@{}}$N = 50$\\ SNR = $20$ dB\end{tabular}  & \cellcolor[HTML]{FFFFFF}$\sim \hspace{0.3em}> 10^{-2} $  \\

\rowcolor[HTML]{E0FFFF} 
\cite{Zeng_Tengchan}     & OOK    & Statistical covariance detector   & \begin{tabular}[c]{@{}l@{}}$P_{F} = 0.1$\\$ N = 5$\end{tabular}  & \cellcolor[HTML]{E0FFFF}  $\sim \hspace{0.3em}> 10^{-2} $  \\

\rowcolor[HTML]{FFFFFF} 
\cite{Daskalakis_1}& 4-PAM  & Energy detector& \begin{tabular}[c]{@{}l@{}}$R_b = 345 \text{b/s}$ \\ 
$P_s = -20$  dBm \end{tabular}& \cellcolor[HTML]{FFFFFF}$\sim \hspace{0.3em}> 10^{-2} $  \\

\rowcolor[HTML]{E0FFFF}
\cite{Yiwen_Tao}   & OOK    & Automatic thresholding detector & \begin{tabular}[c]{@{}l@{}}$N = 50$\\ SNR = $2$ dB\end{tabular}  & \cellcolor[HTML]{E0FFFF}  $\sim \hspace{0.3em}> 10^{-4}  $ \\

\rowcolor[HTML]{FFFFFF} 
\cite{Sudarshan_Guruacharya} & OOK    & \begin{tabular}[c]{@{}l@{}}Energy detector\\ GLRT\\ Bayesian\end{tabular} & \begin{tabular}[c]{@{}l@{}} $N = 20$\\$\eta_1 = 30$ dB\\ $\eta_2 = 20$ dB\end{tabular}& \cellcolor[HTML]{FFFFFF}$\sim 10^{-2}  $    \\

\rowcolor[HTML]{E0FFFF} 
\cite{Xiaoyu_Zhou} & OOK  & Energy detector  & \begin{tabular}[c]{@{}l@{}}$N = 5$\\ $K = 2$\\ SNR = $20$ dB\end{tabular}     & \cellcolor[HTML]{E0FFFF}  $\sim 10^{-2}  $   \\

\rowcolor[HTML]{FFFFFF} 
\cite{Devineni_Kartheek_2} & \begin{tabular}[c]{@{}l@{}}Differential encoder\\ OOK\end{tabular}  & \begin{tabular}[c]{@{}l@{}}Average signal energy \\ Mean threshold detector \\Approximate ML detector \\ ML detector\end{tabular} & \begin{tabular}[c]{@{}l@{}}$N = 500$\\ SNR = $20$ dB\end{tabular}  & \cellcolor[HTML]{FFFFFF}$\sim 10^{-2}$ \\

\rowcolor[HTML]{E0FFFF} 
\cite{shayan_Qin}  & $M$-FSK& \begin{tabular}[c]{@{}l@{}}ML detector\\ Energy detector\end{tabular}& \begin{tabular}[c]{@{}l@{}}$N = 10$\\  BFSK\\ SNR = $15$ dB\end{tabular}   & \cellcolor[HTML]{E0FFFF}  $\sim 10^{-3}   $   \\

\rowcolor[HTML]{FFFFFF} 
\cite{Wanchun_Liu} & \begin{tabular}[c]{@{}l@{}}QPSK at RF source\\ OOK at tag\end{tabular} & \begin{tabular}[c]{@{}l@{}}ML detector\\ Energy detector\end{tabular} & \begin{tabular}[c]{@{}l@{}}$N = 4$\\   SNR = $20$ dB\end{tabular} & \cellcolor[HTML]{FFFFFF}$\sim 10^{-5}   $   \\

\rowcolor[HTML]{E0FFFF} \cite{Jing_Qian}$^*$   & \begin{tabular}[c]{@{}l@{}}$4$-PSK  at RF source\\ OOK at tag\end{tabular}    & \begin{tabular}[c]{@{}l@{}}Semi-coherent detector \\ ML detector\\ Energy detector\end{tabular}  & \begin{tabular}[c]{@{}l@{}}$N = 40$\\  Ntr $ = 4$\\ $\Delta \gamma$ = $0.5$\\ SNR = $20$ dB\end{tabular}                     &\cellcolor[HTML]{E0FFFF}   $\sim 10^{-3}-10^{-4} $      \\

\rowcolor[HTML]{FFFFFF} 
\cite{Qin_Tao}$^*$ & \begin{tabular}[c]{@{}l@{}}$8$-PSK  at RF source\\ OOK at tag\end{tabular}    & \begin{tabular}[c]{@{}l@{}}Semi-coherent Manchester detector\\ Non-coherent Manchester detector\\ Energy detector\\ ML detector\end{tabular} & \begin{tabular}[c]{@{}l@{}}$N = 20$\\ $\Delta \gamma$ = $0.5$\\ SNR = $20$ dB\end{tabular}& \cellcolor[HTML]{FFFFFF}$\sim 10^{-2}-10^{-4} $      \\  

\rowcolor[HTML]{E0FFFF} \cite{Jing_Qian_2}$^*$& \begin{tabular}[c]{@{}l@{}}$8$-PSK at RF source\\ OOK at tag\end{tabular} & \begin{tabular}[c]{@{}l@{}}ML detector \\ Energy detector\end{tabular} & \begin{tabular}[c]{@{}l@{}}$N = 30$\\ $\Delta \gamma$ =$ 0.5$\\ SNR = $20$ dB\end{tabular}& $\sim 10^{-4} $   \\

\rowcolor[HTML]{FFFFFF} \cite{YeZCSL23}& \begin{tabular}[c]{@{}l@{}}Differential encoder\\ OOK\end{tabular}  & \begin{tabular}[c]{@{}l@{}}Energy detector  \end{tabular} & \begin{tabular}[c]{@{}l@{}}$N = 100$\\  SNR = $20$ dB\end{tabular}& $\sim 10^{-3} $   \\ \hline

\rowcolor[HTML]{ECF4FF} 
\multicolumn{5}{|c|}{\cellcolor[HTML]{ECF4FF}Multi-antenna reader} \\ \hline
\rowcolor[HTML]{E0FFFF} 
\multicolumn{1}{|c|}{\cellcolor[HTML]{E0FFFF}\cite{LiuSTM21}} &
\multicolumn{1}{c|}{\cellcolor[HTML]{E0FFFF}BPSK} &
\multicolumn{1}{l|}{\cellcolor[HTML]{E0FFFF}\begin{tabular}[c]{@{}l@{}}Differential OSTBC detector\end{tabular}} &
\multicolumn{1}{l|}{\cellcolor[HTML]{E0FFFF}\begin{tabular}[c]{@{}l@{}}$2\times 2$ T-R CH$^{*}$\\ $\Delta \gamma = 40$ dB \\ SNR = $20$ dB\end{tabular}} &\cellcolor[HTML]{E0FFFF} $\sim \hspace{0.3em} 10^{-2}-10^{-3}$  \\

\rowcolor[HTML]{FFFFFF} 
\multicolumn{1}{|c|}{\cellcolor[HTML]{FFFFFF}\cite{Devineni2021}} &
\multicolumn{1}{c|}{\cellcolor[HTML]{FFFFFF}OOK} &
\multicolumn{1}{l|}{\cellcolor[HTML]{FFFFFF}
Direct averaging }
& \multicolumn{1}{l|}{\cellcolor[HTML]{FFFFFF}\begin{tabular}[c]{@{}l@{}}$N =  2000$ \\ SNR = $20$ dB\end{tabular}}&
$\sim \hspace{0.3em} 10^{-1}$ \\

\rowcolor[HTML]{E0FFFF} 
\multicolumn{1}{|c|}{\cellcolor[HTML]{E0FFFF}\cite{DevineniD20}} &
\multicolumn{1}{c|}{\cellcolor[HTML]{E0FFFF}OOK} &
\multicolumn{1}{l|}{\cellcolor[HTML]{E0FFFF}Sample mean of received signal} &
\multicolumn{1}{l|}{\cellcolor[HTML]{E0FFFF}\begin{tabular}[c]{@{}l@{}}$N =  2000$ \\ SNR = $20$ dB\end{tabular}} &\cellcolor[HTML]{E0FFFF} $\sim 10^{-1}-10^{-2}$ \\

\rowcolor[HTML]{FFFFFF} 
\multicolumn{1}{|c|}{\cellcolor[HTML]{FFFFFF}\cite{Mohamed_ElMossallamy_3}} &
\multicolumn{1}{c|}{\cellcolor[HTML]{FFFFFF}OOK} &
\multicolumn{1}{l|}{\cellcolor[HTML]{FFFFFF}\begin{tabular}[c]{@{}l@{}}Non-coherent post-detection\\ EGC for ED OFDM symbol\end{tabular}} & \multicolumn{1}{l|}{\cellcolor[HTML]{FFFFFF}\begin{tabular}[c]{@{}l@{}}$N =  1024$\\ SNR = $20$ dB\end{tabular}}&
 $\sim \hspace{0.3em} 10^{-4}$ \\

\rowcolor[HTML]{E0FFFF} 
\multicolumn{1}{|c|}{\cellcolor[HTML]{E0FFFF}\cite{Ma2015}} &
\multicolumn{1}{c|}{\cellcolor[HTML]{E0FFFF}OOK} &
\multicolumn{1}{l|}{\cellcolor[HTML]{E0FFFF}Energy detector} &
\multicolumn{1}{l|}{\cellcolor[HTML]{E0FFFF}\begin{tabular}[c]{@{}l@{}}$N =  2$\\ SNR = $20$ dB\end{tabular}}&\cellcolor[HTML]{E0FFFF}
$\sim \hspace{0.3em} 0.3$ \\

\rowcolor[HTML]{FFFFFF} 
\multicolumn{1}{|c|}{\cellcolor[HTML]{FFFFFF}\cite{Tao2020}} &
\multicolumn{1}{c|}{\cellcolor[HTML]{FFFFFF}OOK} &
\multicolumn{1}{l|}{\cellcolor[HTML]{FFFFFF}MAP detector} &
\multicolumn{1}{l|}{\cellcolor[HTML]{FFFFFF}\begin{tabular}[c]{@{}l@{}}$N =  20$\\ SNR = 10 dB\end{tabular}}& $\sim \hspace{0.3em} 10^{-2}$ \\

\hline

\rowcolor[HTML]{ECF4FF} 
\multicolumn{5}{|c|}{\cellcolor[HTML]{ECF4FF}Multi-antenna Tag} \\ \hline
\rowcolor[HTML]{E0FFFF} 
\multicolumn{1}{|c|}{\cellcolor[HTML]{E0FFFF}\cite{Chen2018}} &
\multicolumn{1}{c|}{\cellcolor[HTML]{E0FFFF}OOK} &
\multicolumn{1}{l|}{\cellcolor[HTML]{E0FFFF}F-test detector} &
\multicolumn{1}{l|}{\cellcolor[HTML]{E0FFFF} \hspace{-0.3em} \xmark} &\cellcolor[HTML]{E0FFFF}\xmark \\

\rowcolor[HTML]{FFFFFF} 
\multicolumn{1}{|c|}{\cellcolor[HTML]{FFFFFF}\cite{chen2018detection}} &
\multicolumn{1}{c|}{\cellcolor[HTML]{FFFFFF}OOK} &
\multicolumn{1}{l|}{\cellcolor[HTML]{FFFFFF}Bartlett detector} & 
\multicolumn{1}{l|}{\cellcolor[HTML]{FFFFFF} \xmark} & \xmark\\

\rowcolor[HTML]{E0FFFF} 
\multicolumn{1}{|c|}{\cellcolor[HTML]{E0FFFF}\cite{ChenWGLT20}} &
\multicolumn{1}{c|}{\cellcolor[HTML]{E0FFFF}OOK} &
\multicolumn{1}{l|}{\cellcolor[HTML]{E0FFFF}\begin{tabular}[c]{@{}l@{}}Chi-squared\\ F-test\\ Bartlett test\end{tabular}} &
\multicolumn{1}{l|}{\cellcolor[HTML]{E0FFFF} \xmark } & \cellcolor[HTML]{E0FFFF}\xmark \\

\rowcolor[HTML]{FFFFFF} 
\multicolumn{1}{|c|}{\cellcolor[HTML]{FFFFFF}\cite{ChenWDHKT20}} &
\multicolumn{1}{c|}{\cellcolor[HTML]{FFFFFF}OOK} &
\multicolumn{1}{l|}{\cellcolor[HTML]{FFFFFF}GLRT} &
\multicolumn{1}{l|}{\cellcolor[HTML]{FFFFFF} \xmark }&\cellcolor[HTML]{FFFFFF} \xmark \\

\hline
\end{tabular}

\vspace{5pt} 
\noindent $^{*}$ T-R CH denotes the  tag-to-reader channel.

\end{table*}

\subsubsection{\bf{Single-antenna setup}} 
In \cite{Sudarshan_Guruacharya}, an optimal non-coherent detection detector is derived based on the joint PDF of the received signal. Two types of non-coherent detectors, Bayesian and GLRT, are designed. Optimal decision thresholds and test statistics are derived for both detectors, with the Bayesian detector exhibiting superior performance at the expense of increased computational complexity. The Signal-to-interference-to-noise ratio (SINR) is defined as $\eta = \frac{\xi^2 \sigma^2_{st}\sigma^2_{tr}P_T}{\sigma^2_{sr}P_T+\sigma^2_{w}}=\frac{\eta_1}{1+\frac{1}{\eta_2}}$, where $\eta_1$ represents the signal-to-interference ratio (SIR) and $\eta_2$ represents the interference-to-noise ratio (INR). Fig.  \ref{simluation_1} depicts the BER versus INR, illustrating that increasing INR initially decreases BER until it saturates. The Bayesian approach outperforms the other proposed schemes at lower INR values. For large INR values, all three detectors exhibit similar performance due to direct-link interference, indicating that increasing the number of samples or transmit power alone does not enhance non-coherent detection performance. Improving system performance can be achieved by reducing the distance between the tag and the reader, thereby increasing the SIR. Fig. \ref{simluation_2} demonstrates the impact of the number of samples, $N$, on BER performance, revealing that direct-link interference causes BER to saturate without improvement as $N$ increases. However, the SIC technique can eliminate the error floor by eliminating direct-link interference.

In \cite{Jing_Qian_3}, the performance of the AmBC reader and a tag that uses differential modulation is evaluated in terms of BER. The authors propose a suboptimal detector to reduce computational complexity and derive upper and lower bounds for BER to provide more accurate insights. They also present a practical approach for estimating the required parameters without assuming the availability of CSI. In \cite{Jing_Qian_2}, the joint PDF of the received signal is utilized to derive the ML  detector using a blind estimation method. To mitigate computational complexity, the authors propose differential encoding. The performance of complex Gaussian RF signals and PSK-based signals is compared, and the complexity of the detectors is discussed. Furthermore, \cite{Wang2016} introduces an optimal detector for signal detection and BER analysis. Closed-form expressions for the detection thresholds are derived, resulting in minimum BERs and balanced error probabilities for bits ``$0$" and ``$1$". An equiprobable error threshold is proposed to achieve the same error probability in detecting transmit bits. The authors also suggest the use of differential encoding to enhance CE in the absence of training symbols. Fig. \ref{DPSK_single_antenna} presents the BER versus SNR with three thresholds, highlighting the good performance of $\text{T}_{\text{h}}^{\text{apx}}$ and $|\delta|/2$ in both low and high SNR regions. It is observed that increasing the number of averaging samples leads to lower BER, as expected.

\begin{figure}
\centering
	\includegraphics[width=3.5in]{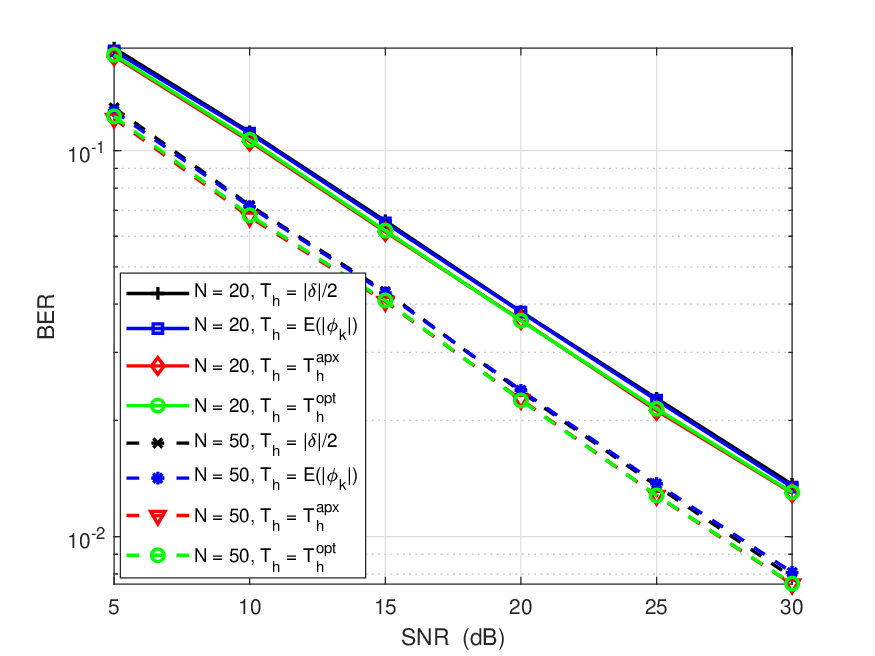}
 	\caption{ BER versus transmit SNR for four different thresholds and number of averaging samples under Rayleigh fading  \cite{Wang2016}.} \label{DPSK_single_antenna}
\end{figure}

Unlike previous works that primarily consider ML and suboptimal detectors, \cite{Kang_Lu, Gongpu_Wang_3} introduce an ED  suitable for high SNRs. However, \cite{Zeng_Tengchan} presents a statistical covariance method that outperforms the ED at low SNRs. This method involves computing auto-correlations of the received signal and constructing an approximation of the statistical covariance matrix. The study also derives the probability of detection and BER  and proposes two statistics for detecting reflected signals.
 
In \cite{Daskalakis_1}, the $4$-pulse amplitude modulation (PAM) technique is introduced for AmBC systems. A proof-of-concept prototype is presented, achieving a low bit rate of $345$ bit/s with a power consumption of $27$ µW. The communication range is reported as $1$ m between the tag and reader and $34.5$ m between the tag and RF source. The $4$-PAM modulation is implemented at the tag using a complex load with four distinct impedance values, generating four symbols through four different reflection coefficients. The optimal performance of $M$-PAM modulation is obtained when the reflection coefficients are equidistantly aligned on a straight line on the Smith chart. The packet structure consists of a preamble, tag number, sensor number, and tag data. The preamble marks the start of a frame, and four bits are assigned to both the tag and sensor numbers to support multiple tags. To address CFO, an ED is utilized at the reader, followed by synchronization and DC removal procedures to determine the frame's starting point.

Reference \cite{Wanchun_Liu} proposes a new channel model for AmBC systems, referred to as the backscatter multiplicative multiple access channel (MAC). This MAC enables tags to use both direct interference and indirect links to recover information simultaneously, thereby improving throughput. In high SNR regimes, where the direct channel is much stronger than the backscatter channel, this MAC achieves a larger achievable rate region compared to conventional time division multiple access (TDMA). A joint coherent and non-coherent detection technique is employed by the reader to decode information from the RF source and the tag. Both synchronous and non-synchronous scenarios are considered for detection, with the latter introducing significant ISI and increased BER.  More specifically, \cite{YeZCSL23} takes I/Q imbalance into account, whereas the AmBC transmitter uses differential encoding and the AmBC receiver uses the energy difference as its detection mechanism. This work derives a closed-form BER expression as a function of the detection threshold and I/Q imbalance factors. The results show that I/Q imbalance reduces the BER, and that phase imbalance degrades the BER of the examined detector more than amplitude imbalance.

In \cite{Devineni_Kartheek_2}, the exact BER analysis presented in \cite{Kartheek_Devineni} is extended by considering three different detectors under assumptions of perfect and imperfect CSI. The mean threshold, ML, and approximate ML detectors are analyzed under perfect CSI, while the mean threshold and differential coding are utilized for signal detection under imperfect CSI. Considering a multi-tag scenario, \cite{Xiaoyu_Zhou} investigates the effectiveness of tag selection and the corresponding detection method. A tag selection protocol is designed, and the ED  is employed as the detection method.

In \cite{Kartheek_Devineni}, it is demonstrated that the exact conditional density functions of average signal energy follow a noncentral chi-squared distribution. The ergodic BER is derived under various hypotheses and with different detection thresholds.

In \cite{shayan_Qin}, two optimal AmBC detectors are designed, and the analytical performance characterization is conducted using $M$-FSK  modulation. The first detector utilizes the ML  approach, despite direct interference from the RF source, which results in an error floor at high SNRs. The second detector is a simple ED  that eliminates the need for direct interference and CSI requirements due to the frequency shift characteristic of $M$-FSK modulation. The study indicates that $M$-FSK modulation outperforms OOK modulation, and the choice of modulation order significantly impacts the BER performance depending on the SNR.

\textbf{Takeaways}: 
In summary, using multiple antennas for signal detection in AmBC systems can significantly improve performance compared to single-antenna detection. The spatial diversity provided by multiple antennas enables diversity gain, leading to enhanced detection capabilities. However, several limitations need to be addressed. The use of multiple antennas requires additional hardware, which can be costly in practical implementations. Moreover, the performance improvement achieved with multiple antennas is dependent on the specific environment and signal characteristics, necessitating the optimization of antenna number and placement for each application. Potential solutions to these limitations include exploring cost-effective hardware solutions for multiple antennas and developing adaptive algorithms that can accommodate different environments and signal characteristics.

\subsubsection{{\bf{Multi-antenna setup}}}
In \cite{Ma2015}, an expression for the BER is derived when the reader employs an ED. However, even with a large number of receiving antennas, the BER remains unsatisfactory due to an error floor. In \cite{Tao2020}, an optimal detection method based on the MAP principle is proposed, considering complex Gaussian ambient RF signals. The performance of the optimal detector is compared with that of the ED, and the results confirm the superiority of the former.

In the work \cite{Liu2015}, the assumption of employing differential encoding at the tag is established, and the employment of the ML detector is advocated to circumvent the issue of CE. Additionally, the optimal threshold for the ML detector is derived.

The impact of Manchester encoding at the tag on the BER of the reader is investigated in \cite{Devineni2021}. Specifically, considering time-selective fading channels, the authors derive the BER expressions for both single and multi-antenna reader cases. With a multi-antenna reader and full direct link interference cancellation, Manchester encoding achieves an SNR gain over conventional uncoded OOK  modulation.

Similarly, \cite{DevineniD20} explores a time-selective fading channel where both the RF source and the tag are mobile simultaneously. The BER analysis is developed by tracking the angle of arrival (AoA) of the direct and backscatter links while using a uniform linear array antenna at the reader.

In \cite{Mohamed_ElMossallamy_3}, AmBC over orthogonal frequency-division multiplexing (OFDM) signals is investigated using a multi-antenna reader to improve error performance. The tags exploit the spectrum structure of ambient OFDM symbols to transmit their data bits. Non-coherent post-detection equal gain combining (EGC) is proposed for data detection at the reader side, enabling the combination of signals from multiple diversity branches with equal weights. An $M$-ary modulation scheme is considered to increase the communication rate, transmitting multiple bits per OFDM symbol. Unlike existing works that employ Gaussian approximations, this study derives exact and accurate approximate expressions for binary and $M$-ary error probabilities using the Meijer G-function.

Since the growth of IoT devices is exponential,  some of these devices may be deployed in high-mobility scenarios. This leads to lower channel coherence times and greater Doppler spread than static scenarios. The low coherence time makes coherent detection in such cases of high mobility a challenging task. Nevertheless, the formulation of the optimal test statistic for a time-selective channel remains an unsolved problem. In \cite{Kartheek_Devineni_2}, the non-coherent detection of ambient signals in a time-selective fading channel as a ﬁrst-order autoregressive process is studied by proposing a new architecture for the reader. The new architecture is based on the direct averaging of the received signal samples, which differs from the well-known adapted ED. A BER is derived for two scenarios, i.e., single- and multi-antenna readers, and the effect of timing errors is examined. Since the new architecture uses linear operations, synchronization, and timing errors are minimal. A unique result is that multi-antenna readers can overcome the error floor phenomenon by estimating the AoA of the direct link.

Multiple-antenna tags can overcome the low communication range between the tag and the reader. Multiple antennas also increase diversity gains and enhance the system's reliability to get a low BER \cite{ChenWGLT20}. They also increase the amount of power harvested by the tag. References \cite{Chen2018,chen2018detection,LiuSTM21,HuZZJJN19,ChenWDHKT20} discusses the detection problem in a multi-antenna tag scenario. For instance,  \cite{Chen2018} derives the detection probability of an AmBC network with a multi-antenna tag. The tag has  $M$ antennas out of $K$ for backscatter and $K-M $ for EH. This method uses the blind F-test detector without knowledge about the signal power, noise variance, and the CSI at the reader. This detector computes the sample variances of the received signal,  and comparing their ratios determines whether the tag symbol is ``$0$" or ``$1$". The final part of the paper proposes an algorithm for selecting the antenna.

Similarly,  \cite{chen2018detection, HuZZJJN19} study the same configuration. For instance, \cite{chen2018detection} uses Bartlett’s test for signal detection. Indeed, in the null hypothesis of Bartlett's test, all sample variances are equal against the other hypothesis, where at least two are different. However, \cite{HuZZJJN19} studies the NP and GLRT test to show the inefficiency of such a traditional method for signal detection in a multi-antenna tag scenario. Similar to \cite{chen2018detection, Chen2018}, \cite{ChenWGLT20} considers a multi-antenna tag for which one set of antennas is assigned to backscatter and another set is assigned to EH. In the backscattering antenna set, the system selects only one antenna for backscattering. As a result, the other silent backscatter antenna can be used for  EH. Since the null and alternative hypotheses have different covariance matrices, they proposed a Chi-squared-based detector. This requires the reader to know  CSI,  RF transmits, and noise power. Thus, optimal antenna selection is proposed to maximize variance difference. Additionally, two blind detections are investigated, the F-test and the Bartlett test, when the reader is unaware of the channel parameters (CSI, transmit power, and noise variance). 

In addition, the GLRT-based detection of the mentioned system is studied in \cite{ChenWDHKT20}. Specifically, two backscatter schemes are proposed here, a general backscatter scheme and a specially designed backscatter scheme. The arbitrary and quantized number of signals per antenna is used in the latter and former, respectively. The authors in  \cite{LiuSTM21} discuss orthogonal space-time block code (OSTBC) incorporating multi-antenna tag and reader.  They investigate both non-coherent and coherent OSTBC with and without CSI knowledge. Specifically, they study optimal ML and low-complex linear detectors using the accurate AmBC and linearized MIMO channel model. They also investigate differential non-coherent OSTBC to avoid the limitations of coherent detection. Simulated results show that the MIMO AmBC system based on OSTBC coherent and non-coherent detectors offers lower BER than a conventional single-input single-output (SISO) AmBC system.

\textbf{Takeaways:} Researchers have made noteworthy progress in optimizing communication by focusing on single-antenna systems for AmBC. Key findings center on innovative detection techniques and modulation strategies aiming to improve system reliability and lower the BER. Through these studies, it is evident that while single antennas offer a more streamlined setup, they come with their own set of challenges. Techniques like Bayesian and GLRT detectors, differential modulation, and the use of various thresholding methods have shown promise in tackling some of these issues. However, challenges like direct-link interference, the complexities of perfect vs. imperfect channel information, and the choice of optimal modulation methods remain. Continuous research is, therefore, essential to refine and enhance the capabilities of single-antenna AmBC systems.

\begin{table*}[]
\caption{Summary of MaL-based signal detection for AmBC system.}\label{table_learning}
\centering
\begin{tabular}{|lllll|}
\hline

\multicolumn{1}{|l|}{Works} & \multicolumn{1}{l|}{Tag modulation type} & \multicolumn{1}{l|}{Detection design} & \multicolumn{1}{l|}{Setup parameters} & Achieved BER \\ \hline
\rowcolor[HTML]{ECF4FF} 
\multicolumn{5}{|c|}{\cellcolor[HTML]{ECF4FF}Single-antenna reader } \\ \hline
\rowcolor[HTML]{E0FFFF} 
\multicolumn{1}{|l|}{\cellcolor[HTML]{E0FFFF}\cite{Yunkai_Hu_2} } & \multicolumn{1}{l|}{\cellcolor[HTML]{E0FFFF}BPSK} & \multicolumn{1}{l|}{\cellcolor[HTML]{E0FFFF}\begin{tabular}[c]{@{}l@{}}MMSE\\ SVM\\ Random forest\end{tabular}} & \multicolumn{1}{l|}{\cellcolor[HTML]{E0FFFF}\begin{tabular}[c]{@{}l@{}}$N = 256$\\ SNR $= 15$ dB\end{tabular}} & $\sim \hspace{0.3em}<  10^{-2}$ \\
\multicolumn{1}{|l|}{\cite{Qianqian_Zhang_1}} & \multicolumn{1}{l|}{OOK} & \multicolumn{1}{l|}{Clustering} & \multicolumn{1}{l|}{\begin{tabular}[c]{@{}l@{}}$N$ = 50\\ $\Delta \gamma = -10$ dB\\ SNR $= 15$ dB\end{tabular}} & \cellcolor[HTML]{FFFFFF}$\sim \hspace{0.3em}>  10^{-2}$ \\ \hline
\rowcolor[HTML]{ECF4FF} 
\multicolumn{5}{|c|}{\cellcolor[HTML]{ECF4FF}Multi-antenna reader } \\ \hline
\rowcolor[HTML]{E0FFFF} 
\multicolumn{1}{|l|}{\cellcolor[HTML]{E0FFFF}\cite{Xiyu_Wang}} & \multicolumn{1}{l|}{\cellcolor[HTML]{E0FFFF}BPSK} & \multicolumn{1}{l|}{\cellcolor[HTML]{E0FFFF}KNN} & \multicolumn{1}{l|}{\cellcolor[HTML]{E0FFFF}\begin{tabular}[c]{@{}l@{}}$M = 8$\\ SNR $= 20$ dB\end{tabular}} & \cellcolor[HTML]{E0FFFF}$\sim \hspace{0.3em}<  10^{-1}$ \\
\rowcolor[HTML]{FFFFFF} 
\multicolumn{1}{|l|}{\cellcolor[HTML]{FFFFFF}\cite{WangYDMJ22}} & \multicolumn{1}{l|}{\cellcolor[HTML]{FFFFFF}BPSK} & \multicolumn{1}{l|}{\cellcolor[HTML]{FFFFFF}\begin{tabular}[c]{@{}l@{}}KNN\\ SVM\\ LS-based classifier\\ Linear discriminant analysis\end{tabular}} & \multicolumn{1}{l|}{\cellcolor[HTML]{FFFFFF}\begin{tabular}[c]{@{}l@{}}$M = 4$\\ $\Delta \gamma = -29.8$ dB\\ SNR $= 15$ dB\end{tabular}} & \cellcolor[HTML]{FFFFFF}$\sim \hspace{0.3em}<  10^{-1}$ \\
\rowcolor[HTML]{E0FFFF} 
\multicolumn{1}{|l|}{\cellcolor[HTML]{E0FFFF}\cite{Chang_Liu}} & \multicolumn{1}{l|}{\cellcolor[HTML]{E0FFFF}OOK} & \multicolumn{1}{l|}{\cellcolor[HTML]{E0FFFF}\begin{tabular}[c]{@{}l@{}}Deep transfer learning\\ CNN\end{tabular}} & \multicolumn{1}{l|}{\cellcolor[HTML]{E0FFFF}\begin{tabular}[c]{@{}l@{}}$M = 8$\\ Ntr$ = 10$ \\ SNR $= 14$ dB\end{tabular}} & \cellcolor[HTML]{E0FFFF}$\sim 10^{-2}-10^{-4}$ \\
\rowcolor[HTML]{ECF4FF} 

\multicolumn{1}{|l|}{\cellcolor[HTML]{FFFFFF}\cite{Qianqian_Zhang,Qianqian_Zhang_3}} & \multicolumn{1}{l|}{\cellcolor[HTML]{FFFFFF}OOK} & \multicolumn{1}{l|}{\cellcolor[HTML]{FFFFFF}\begin{tabular}[c]{@{}l@{}}\cellcolor[HTML]{FFFFFF}GMM\\ \cellcolor[HTML]{FFFFFF}CLS, CLUS\\ \cellcolor[HTML]{FFFFFF}ED, OD\end{tabular}} & \multicolumn{1}{l|}{\cellcolor[HTML]{FFFFFF}\begin{tabular}[c]{@{}l@{}}\cellcolor[HTML]{FFFFFF}$N = 200$\\\cellcolor[HTML]{FFFFFF} $M = 2$\\ \cellcolor[HTML]{FFFFFF}$\Delta \gamma = - 30$ dB\end{tabular}} & \cellcolor[HTML]{FFFFFF}$\sim \hspace{0.3em}<  10^{-2}$ \\ \hline
\end{tabular}
\end{table*}

\subsection{Semi-Coherent Detection}

In AmBC systems, the acquisition of CSI is complicated by direct link interference, making detection at the reader a challenging task. To address this issue, \cite{Youyou_Zhang} proposes a semi-blind reader with a semi-blind channel estimator and a direct-link-averaging detector. Specifically, this reader only requires two pilot symbols from the tag to acquire CSI, and it can effectively remove direct-link interference while estimating the channel. The authors suggest several methods for estimating the phase and amplitude of the channel, including Viterbi-Viterbi and $M$-th power carrier phase estimators for phase estimation and sample mean or raw second-moment approaches for amplitude estimation. Although these approaches are suboptimal, they are easy to implement. The semi-blind reader marginalizes information over all possible RF source symbols using the direct-link averaging detector based on the estimated channel. Additionally, an iterative version of the semi-blind reader is proposed to improve its performance. The results indicate that the semi-blind reader achieves similar performance to a constellation learning algorithm proposed in \cite{Qianqian_Zhang} while having lower complexity.

In \cite{Jing_Qian}, a semi-coherent detection approach is proposed, which estimates the detection-required parameters from unknown received signals using a few pilot symbols when CSI is unknown. The authors investigate an ML detector based on a complex Gaussian RF source as a benchmark for comparison and examine its corresponding analytical BER and BER-based outage probability. The ML detector can be considered semi-coherent if the RF is assumed to be complex Gaussian, whereas it requires full CSI knowledge otherwise. Since the ML detector computes the energy of the received signal, the LRT corresponds to a modified ED. However, the detection threshold in this method depends on the variance of the received signals under different hypotheses. Therefore, a blind estimation technique is used to estimate the variance of the received signals.

Previous works, such as \cite{Jing_Qian,Jing_Qian_2}, have detection thresholds that depend on unknown variables, which increases power consumption and causes delay due to the estimation process preceding the detection. Additionally, these works assume equal prior probabilities, which is not valid for real-world applications. To address these issues, \cite{Qin_Tao} encodes information bits at the tag using Manchester and differential Manchester codes, allowing the reader to detect symbols individually. Manchester coding maps the original information bits ``$0$" and ``$1$" into ``$01$" or ``$10$", while differential Manchester coding expresses each bit at the tag by the presence or absence of a transition compared to the previous bit. The closed-form BER expressions are obtained for both detectors, considering complex Gaussian signals and unknown deterministic signals. The BER performance of Manchester coding is better than that of the semi-coherent detections \cite{Jing_Qian,Jing_Qian_2}, which require estimation and suffer from performance degradation due to estimation errors. The BER performance of the deterministic ambient signal is better than that of the complex Gaussian ambient signal.

\textbf{Takeaways:}  In AmBC systems, acquiring CSI is challenged by direct link interference, making detection complex. A semi-blind reader using a semi-blind channel estimator and direct-link-averaging detector is proposed to address this, demanding a few pilot symbols from the tag to obtain CSI. Various methodologies for channel phase and amplitude estimation are recommended, and the use of Manchester and differential Manchester codes for encoding at the tag improves BER performance. However, open challenges remain, such as the efficiency of CSI acquisition, the practical implementation of the semi-blind reader, and the effectiveness of the proposed encoding methods under various conditions. Future work could focus on refining CSI acquisition algorithms, enhancing the semi-blind reader's robustness, and improving encoding method performance.

 \subsection{Machine Learning-Based Detection}\label{mach_detection}
There has been an increased focus on the development of MaL-based signal \abc  detection algorithms  (Table \ref{table_learning}). The first step in MaL-based signal detection is to extract features from the received signals. These features may include time-domain and frequency-domain representations of the signals, as well as higher-level features such as wavelet coefficients, cyclostationary features, and other representations that capture the characteristics of the signals. Once the features have been extracted, the next step is to train a MaL model using labeled training data. The training data consists of input signals and the corresponding output decisions, and the MaL model is trained to learn the relationships between the input signals and the output decisions. The use of MaL can therefore solve interference and CE problems, as well as other significant problems in \abc networks. 

\subsubsection{\bf{Single-antenna setup}} 
As discussed in \cite{Yunkai_Hu_2}, a MaL-based detection technique for an AmBC system is proposed by converting the detection problem into a classification problem. When using BPSK modulation, the reader receives two energy levels and interprets these levels as ``$1$" or ``$-1$", respectively. The tag symbols can be viewed as training labels, while the received signals can be viewed as training data. The model is then trained using MaL algorithms based on SVM and random forest. After training the model, the reader classifies the input attributes into two categories, corresponding to symbols ``$1$" and ``$-1$". In low SNR regimes, MaL-based signal detection can achieve significantly higher data rates and lower BER than traditional MMSE detectors. Fig. \ref{fig_ml} illustrates the BER versus SNR for different detectors with the spreading gain $N = 256$ and the tag reflection coefficient $\xi = 0.8$. For a given tag symbol, the MMSE detector estimates the ambient RF signal first. Based on the estimated signal, the tag symbol is detected by using the constellation of signals with the least Euclidean distance to either symbol ``$1$" or symbol ``$-1$". Accordingly, the proposed MaL-based detectors outperform the MMSE detector in terms of the BER. Additionally, SVM produces slightly better results than random forests. 

An unsupervised learning approach is applied for AmBC systems to detect the tag bits using the energy features of the received signal in \cite{Qianqian_Zhang_1}. To further aid with the detection of signals, cluster-bit mapping is also performed utilizing the labeled bits from the tag. However, it should be noted that the energy feature results in performance loss as a result of direct link interference.

\textbf{Takeaways:}   Recent studies on BackComm systems have shifted focus to developing MaL-based signal detection algorithms. This involves feature extraction from signals, used to train a MaL model. Employing a single-antenna setup, MaL-based detection classifies the problem into a binary form, using tag symbols and received signals as training labels and data. SVM and random forest algorithms then enhance data rates and BER compared to MMSE detectors, especially in low SNR scenarios, with SVM slightly outperforming random forests. Unsupervised learning methods also contribute to AmBC systems but encounter performance loss due to direct link interference. 
Despite these strides, challenges persist in alleviating interference, perfecting feature extraction, and boosting the dependability of MaL algorithms. Looking ahead, future endeavors may concentrate on advancing feature extraction techniques, refining MaL algorithms, and probing alternative learning approaches. Moreover, an in-depth evaluation comparing various learning methods could provide valuable insights for continued innovation in this domain.

\subsubsection{\bf{Multi-antenna setup}} 

In \cite{WangYDMJ22}, a coherent reader is investigated in an AmBC system with multiple antennas. The study shows that prior knowledge of the ambient signal is unnecessary for the reader to decode the tag's modulated signal. By employing the preamble signal and MaL algorithms, the proposed design efficiently eliminates direct path interference. The tag's signal can be successfully decoded after demodulation using the logistic regression algorithm. The proposed detector achieves one dB higher SINR than the optimal coherent reader.

A statistical clustering framework for suppressing direct link signal is proposed in \cite{GuoZXL19} by using multiple receive antennas and the beamforming method. Motivated by \cite{GuoZXL19}, \cite{Qianqian_Zhang_3} proposes a method in which the received signals can be clustered directly with no need for pre-processing to retrieve the tag symbols, which leads to reducing the adverse effects of direct link signals. As a result, the received signals can be clustered directly without any pre-processing and then mapped to two sets of label symbols. Therefore, two clustering algorithms are proposed, namely clustering with labeled signals (CLS) and clustering with labeled and unlabeled signals (CLUS). According to the first clustering algorithm, the modulation-constrained Gaussian mixture model (GMM) clustering algorithm is used to cluster the labeled signals, and then the learned parameters are used to detect the unlabeled signals. By contrast, the second clustering algorithm uses a modulation-constrained GMM algorithm to cluster both labeled and unlabeled signals jointly. As compared to CLS, CLUS has a higher computational complexity but better performance.

Fig. \ref{fig_ml_2} illustrates that increasing the SNR of the direct link results in a significant improvement in BER performance. The optimal detector (OD), which is LRT with perfect CSI, performs better than other schemes. In particular, CLUS is more efficient than CLS, and both are more efficient than ED. It is also observed that the GMM clustering algorithm has a lower  BER than ED with perfect CSI. The reason for this is that the standard GMM algorithm estimates the cluster centroids without considering the information regarding the constellation of RF sources. Taking backscattered paths into account as interference, \cite{Xiyu_Wang} proposes a MaL-based AmBC detector that removes learnable features from the received signal by projecting it onto the orthogonal space of the direct path signal. An estimation of the ambient signal is made by augmenting the residual signal in the signal space and correlating it with the coarse estimate of the direct path interference. A k-nearest neighbors (KNN) classification algorithm is then applied to detect the tag bits. 

\begin{figure}[t]
\centering
\includegraphics[width=3.5in]{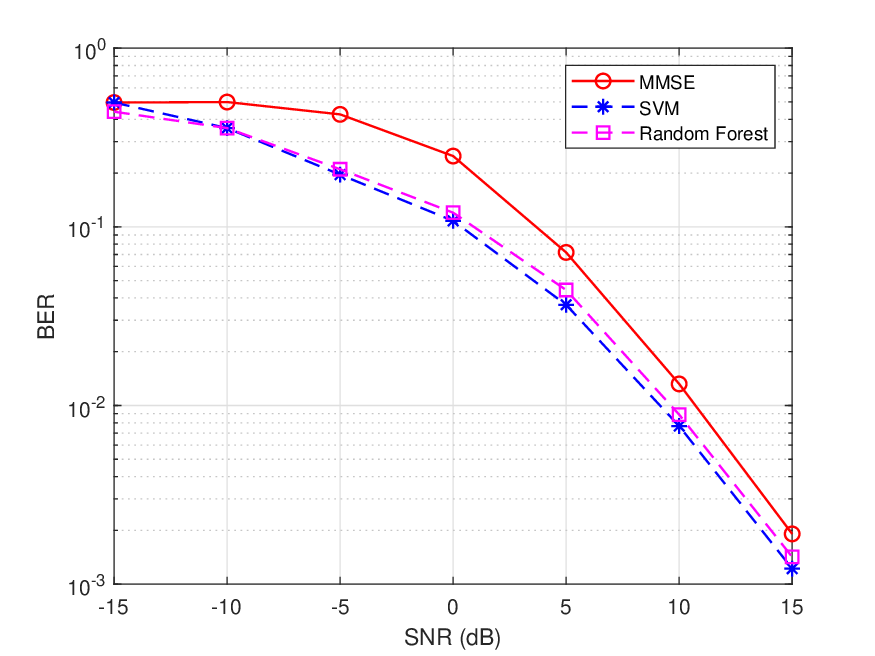}
\caption{  BER versus SNR  with supervised learning algorithms \cite{Yunkai_Hu_2}.}
\label{fig_ml}
\end{figure}

With a few online training data, the authors in \cite{Chang_Liu} propose a deep transfer convolutional neural network (CNN) method for extracting time-varying features consisting of offline learning, transfer learning, and online detection. Indeed, the information acquired from one tag detection task under the offline channel coefficients is transferred to another separate but related tag detection task in real-time using a DNN. As a further enhancement to the detection performance, the proposed framework allows the detector to be dynamically adjusted for various channel conditions. According to \cite{Qianqian_Zhang}, the authors have generalized the single-antenna reader described in \cite{Qianqian_Zhang_1}. Specifically, constellation learning tasks with labeled signals and constellation learning tasks with both labeled and unlabelled signals are proposed for tag symbol detection without requiring CSI knowledge.

\textbf{Takeaways:} Even with the significant strides that have been made, there are still several challenges that need to be tackled. Enhancing the performance of clustering algorithms remains a critical area of focus. It involves increasing the speed of these algorithms while maintaining or improving their accuracy, enabling them to handle larger datasets, and making them adaptable to diverse types of data structures. Furthermore, optimizing deep transfer learning methods is also essential. These methods need to be refined to better identify, extract, and utilize time-varying features in the data, which will improve their predictive performance across a wide range of scenarios. Boosting detector efficiency is another pivotal challenge. Detectors need to maintain high performance even under fluctuating channel conditions, and they should do so while minimizing power consumption, thereby extending battery life in wireless devices. In terms of future research, numerous opportunities exist. More advanced clustering algorithms could be developed that are not only efficient but also robust and scalable, capable of handling large-scale, high-dimensional, and noisy datasets. Refining deep learning methodologies could involve exploring novel architectures and training techniques that allow these models to learn more quickly, generalize more effectively, and provide more interpretable results. Finally, there is a great deal of potential in designing detectors that are more resilient and adaptable, able to handle varied channel conditions. This could involve exploring more sophisticated signal processing techniques or MaL approaches.

\begin{figure}[t]
\centering
\includegraphics[width=3.5in]{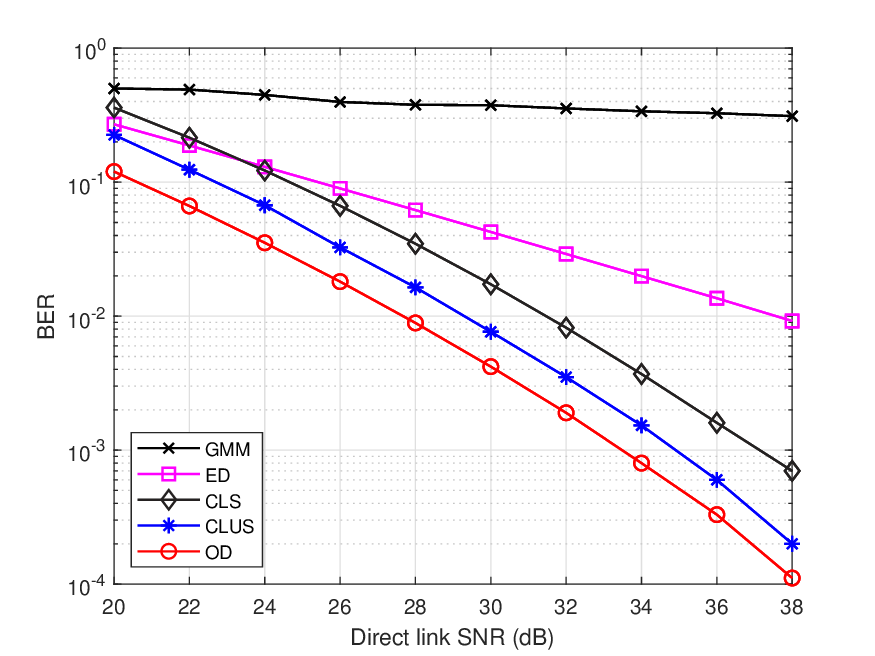}
\caption{ BER versus direct link SNR under the assumption of QPSK modulated ambient RF source with unsupervised learning algorithms  \cite{Qianqian_Zhang_3}.}
\label{fig_ml_2}
\end{figure}

\section{Applications}\label{Sec:application}
AmBC technology allows low-cost passive tags to harvest energy from their surroundings and connect to other networks. One promising application of AmBC is in the development of SR systems. AmBC can also be integrated with other emerging technologies such as OFDM, which can improve the performance of AmBC systems, and IRS, which can create more efficient and reliable wireless communication systems.  These issues are discussed next.

\subsection{Symbiotic Radio Signal Detection}
In wireless communication, the radio spectrum serves as a critical conduit for the transmission and reception of data. The allocation and utilization of this spectrum are subject to stringent regulatory frameworks established by regulatory bodies. Central to this framework is the concept of primary users (PUs), entities vested with exclusive privileges to operate within designated spectrum bands without encumbrance. In contrast, secondary users (SUs) lack this exclusive mandate but possess the liberty to opportunistically access unoccupied spectrum bands \cite{Haykin}.

The essence of this paradigm lies in its ability to seamlessly harmonize structured and adaptable utilization of the electromagnetic spectrum. While primary users uphold the organized and uninterrupted use of their allocated bands, secondary users dynamically navigate the spectrum landscape, availing themselves of unutilized frequency bands. Importantly, secondary users are bound by a critical stipulation: a requirement to relinquish occupancy if a primary user demands access to the spectrum. This orchestrated coexistence engenders an environment where both deliberate spectrum usage and resourceful flexibility converge, contributing to the optimization of airspace utilization \cite{Cabric,Liang4489760}.

\begin{table*}[]

\caption{Summary of  AmBC signal detection using IRS.} \label{table_IRS_Backcomm}
\centering 
\begin{tabular}{|l|l|l|l|l|}
\hline

\rowcolor[HTML]{FFFFFF} 
Works   & Tag modulation type & Detection design & Setup parameters  & Achieved BER      \\ \hline
\rowcolor[HTML]{ECF4FF} 
\multicolumn{5}{|c|}{\cellcolor[HTML]{ECF4FF}Single-antenna Reader} \\ \hline

\rowcolor[HTML]{E0FFFF} 
\cite{Wenjing_Yan}  & RPM  & \begin{tabular}[c]{@{}l@{}}Compressed sensing\\ Matrix factorization\end{tabular}  & \begin{tabular}[c]{@{}l@{}} $I = 8$  \\ $\beta = -20$ dB\\ SNR$ = 20$ dB\end{tabular} &  $\sim \hspace{0.3em}> 10^{-3}$ \\

\cite{Mahyar_Nemati,Yunfei_Chen} & OOK  & ML detector & \begin{tabular}[c]{@{}l@{}}$N_s = 64$\\ $N_{\text{cp}} = 16$\\  SNR = $15$ dB\end{tabular} & \cellcolor[HTML]{FFFFFF} $ \sim 10^{-4}-10^{-5} $         \\

\rowcolor[HTML]{E0FFFF}  
\cite{Yanan_Ma}   & RPM  & \begin{tabular}[c]{@{}l@{}}SVD detector\\ OMP detector\\ GAMP detector\end{tabular}  & 
\begin{tabular}[c]{@{}l@{}}\\
$ I = 64$\\ SNR $= 8$ dB\end{tabular} &  \cellcolor[HTML]{E0FFFF} $\sim 10^{-4}$    \\

\hline
\rowcolor[HTML]{ECF4FF} 
\multicolumn{5}{|c|}{\cellcolor[HTML]{ECF4FF}Multi-antenna Reader} \\ \hline

\cite{Ertugrul_Basar}   & SM   & \begin{tabular}[c]{@{}l@{}}ML detector\\ Energy detector\end{tabular}   & \begin{tabular}[c]{@{}l@{}}$I = 128$\\ $M = 8$ \\ SNR$ = -30$ dB\end{tabular}  & \cellcolor[HTML]{FFFFFF} $\sim \hspace{0.3em}> 10^{-3}$\\

\rowcolor[HTML]{E0FFFF} 
\cite{Xianfu_Lei}  & SM   & ML detector  & \xmark       & \xmark   \\
 
\cite{Mingjiang_Wu}   &   SM &  \begin{tabular}[c]{@{}l@{}}ML detector\\ Successive greedy detector\\ Energy detector\end{tabular} & \begin{tabular}[c]{@{}l@{}}$I = 128$\\ $M = 8$ \\ SNR $= -30$ dB\end{tabular}  & \cellcolor[HTML]{FFFFFF}$\sim \hspace{0.3em}< 10^{-3}$    \\ \hline
\end{tabular}
\end{table*}

Cognitive radio (CR) is a dynamic radio system that adjusts its transmission parameters in response to environmental feedback. Its core function is to identify unused spectrum portions and facilitate their access by secondary entities without causing disruptions. As the available wireless spectrum becomes increasingly limited due to rising wireless devices, applications, and stringent global spectrum allocation policies, the focus on spectrum sharing intensifies. One approach is to permit SUs to access the primary spectrum opportunistically or through other mechanisms. Previous research has looked into both CR and AmBC systems under spectrum reuse networks, sometimes termed cognitive backscattering communication \cite{Ying_Chang}.

The SR stands out due to the simultaneous presence of primary and secondary systems, reminiscent of CR. Unlike CR networks, which require active RF chains for both primary and secondary transmitters, SR networks incorporate both secondary and primary receivers, resulting in a cooperative receiver which is a recipient of both SU and PU.

\textbf{Key observations in SR networks:}
\begin{enumerate}
\item A cooperative operation exists between the primary and secondary systems. Consequently, the cooperative receiver can decode both primary and secondary information without causing interference, leading to dependable backscattering communications.
\item The secondary system offers multipath diversity to the primary system.
\item SR promotes both spectrum and energy efficiency by endorsing symbiotic spectrum sharing and ensuring reliable backscattering communication.
\end{enumerate}}

Reference  \cite{Ying_Chang} studies an SR network with multi-antenna primary and secondary receivers.  First, the channels are estimated using the pilot approach. Afterward, three coherent detection methods are presented to jointly recover the primary and secondary signals, namely, the ML detector, the linear detector, and the SIC-based detector. The linear detector is studied in more detail through the examination of maximum ratio combining (MRC), ZF, and MMSE. Compared to the ML detector, the linear detector has lower complexity at the cost of reduced performance. A SIC detector estimates the primary symbol at the secondary receiver by using a linear detector, then subtracts the direct link signal and estimates the secondary symbol by using a MRC estimator. There is a significant improvement in BER performance with SIC-based detectors compared to linear detectors at the expense of slightly higher computational complexity.

\subsection{AmBC Signal Detection Using IRS}
 
Elevating the efficiency of wireless networks necessitates the dynamic optimization of the propagation channel. In this context, the emergence of IRSs represents a recent and noteworthy advancement, heralding an efficient and cost-effective solution that has garnered significant attention across both academic and industrial spheres \cite{Qingqing_Wu_1,Shayan_Zargari_1,Shayan_Zargari_6}.

At its core, an IRS entails a metasurface adorned with an array of economically viable reflectors. These elements can be dynamically programmed to manipulate the wireless propagation environment, thereby enhancing both energy and spectral efficiency \cite{Shayan_Zargari_6,zhang2021,Shayan_Zargari_2}.

Unlike traditional relaying systems, which depend on active circuits and power sources,  the reconfigurable elements of an IRS  produce concentrated beams via passive reflections. These beams are directed to the receiver, ensuring precise signal guidance \cite{Hongliang_Zhang_2020}. This results in expanded coverage and a more dependable wireless communication network \cite{Emil_Bjornson_1}.

  \begin{figure*}
\centering
	\includegraphics[width=6.2in]{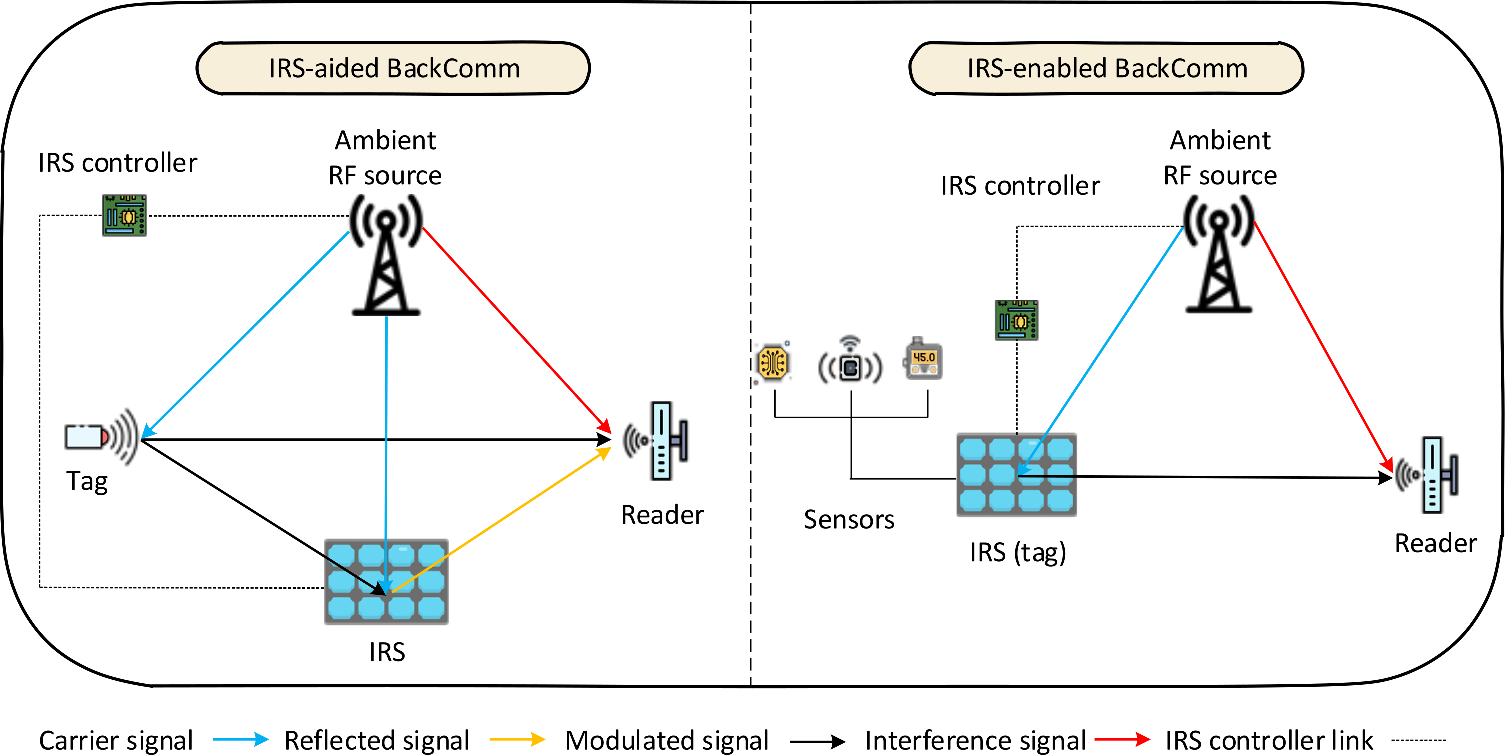}
 	\caption{BackComm using IRS system models.} \label{IRS_systm}
\end{figure*}

\textbf{Technical overview of the IRS System:} The channel extending from the tag, via the IRS, to the reader is represented as $g = \boldsymbol{h}^T\boldsymbol{\Theta}\boldsymbol{f}$, where $\boldsymbol{h}$ and $\boldsymbol{f}$ are the channel links from the tag-to-IRS and the IRS-to-reader, respectively. In addition, ${ {\boldsymbol{\Theta}}=\text{diag} (\varrho_1 e^{j\theta_{1}},\ldots,\varrho_i e^{j\theta_{i}},\ldots ,\varrho_I e^{j\theta_{I}})}$ is the IRS reflection-coefficients matrix where $\varrho_i\in [0,1]$ and $\theta_{i}\in (0,2\pi]$ are the amplitude and phase shifts of the $i$-th reflecting element at the IRS \cite{Shayan_Zargari_3,Qingqing_Wu_1,Syed_Junaid_Nawaz}
 
\textbf{Role of IRS in BackCom}: Given the impressive capabilities of IRS, it is predicted that IRS BackCom might be an effective solution for extensive connectivity \cite{Syed_Junaid_Nawaz}. This can significantly benefit the growth of wireless applications, such as smart cities and homes \cite{Xianfu_Lei,Park2020,GuoLL19a}, aiming to integrate various sensors, network technologies, and communication protocols for improved energy and spectral efficiency. In this scenario, merging AmBC systems with IRS offers a promising approach to achieve these goals and greatly enhance coverage, as illustrated in Table \ref{table_IRS_Backcomm}. As shown in Fig. \ref{IRS_systm}, there are two types of AmBC-IRS, either as a reflector or transmitter.

{\bf{1)  IRS-aided \bc}}: By reflecting incident signals smartly, the IRS can help \bc improve wireless propagation channels and achieve more reliable communications \cite{Jia2020, WZhao2020,Shayan_Zargari_4,Shayan_Zargari_5}. Therefore, the IRS serves as an intelligent reflector to facilitate wireless communication between the tag and the reader. Due to this, the IRS reflects the signals from the tag by altering their amplitude and phase. At the reader, the SINR or SNR can be increased by combining the reflected signals coherently. The BER performance of an IRS-enabled AmBC system is analyzed in \cite{Yunfei_Chen}, where IRS is used as a reflector to enhance wireless communication on several different links, including RF source-to-tag and RF source-to-reader. As part of this study, a ML detector is used as the reader, as well. An AmBC-based OFDM system is described in \cite{Mahyar_Nemati} where the tag transmits one bit per each OFDM subcarrier transmitted by a Wi-Fi AP. The reader utilizes the special structure of the OFDM signal to eliminate direct link interference and employs a ML detector to determine the information bits.

{\bf{2) IRS-enabled \bc:}}  The IRS plays the role of a transmitter in this concept. It also passively reflects signals for ongoing links as well as transmits its own data \cite{Hakimi_9877898}. Embedded sensors can provide this type of data, \cite{Park2020, Zhao2020}. As the IRS transmits its own data without generating the RF signal itself, it reduces both the cost and complexity of its design. Applications such as smart homes, smart cities, and IoT networks can benefit from this capability. To accomplish this, the AP transmits a superimposed signal of modulated and unmodulated components. Between these two parts, the total transmission power of the AP is divided. The IRS serves dual functions. First, it relays the AP signal to the users. Secondly, it uses the unmodulated part for backscattering its data to users. Each user must decode its own data as well as the data from the IRS. The transmission scheme works as follows \cite{Park2020}.

First, the AP  transmits $x(n)$ with the symbol period $T_s$. Second, the IRS uses the unmodulated part of $x(n)$  to modulate its data, $b(n)$ by applying BPSK modulation with symbol period $T_b\gg T_s$. Accordingly, for $b(n) = 1$ or $0$, the following phases: $0$ or $\pi$ are added to the optimal phase shifts at the IRS, respectively. As a result, the primary received signal at each user is the original version of the signal when $b(n) = 1$, and it gets a negative sign when $b(n) = 0$. Since the transmission rate of the primary system is much higher than the secondary system ($T_b = LT_s$, where the ratio $L$ should meet $L \gg 1$), the IRS data can be decoded by using a band-pass filter at each user since it remains constant over $LT_s$ symbol period. Thus, each user firstly decodes and subtracts IRS data, and then decodes its primary data based on the decoding order policy. The modulation schemes used by the IRS for \bc mainly fall into the following categories: 

\begin{enumerate}

\item[$\bullet$]  \textbf{Traditional modulation schemes}: \cite{Qianqian_Zhang_2} and \cite{Wankai_Tang} employ PSK and high-order QAM schemes at the IRS for information transmission in the \bc system, where \cite{Qianqian_Zhang_2} analyze the system performance in terms of resource allocation.

\item[$\bullet$] \textbf{Reflection pattern modulation (RPM)}: This is an index modulation technique that is being developed by IRS to transmit data \cite{Wenjing_Yan,Shuaishuai_Guo,Yanan_Ma}. In this way, the data is mapped into different phase shift matrices. To recover the information at the receiver, it is necessary to identify which phase shift matrix is being used. In this regard, \cite{Wenjing_Yan} sends information based on different ON/OFF states of the IRS phase shifts as RPM, which is then detected by the reader through compressed sensing and matrix factorization. In this system setup, the phase shifts in the OFF state contribute little to improving the system's performance. Therefore, \cite{Yanan_Ma} proposes an all-elements-active RPM in which finite phase shift matrices are optimally selected based on CSI and BER performance. As a means of detecting information, singular value decomposition (SVD), orthogonal matching pursuit (OMP), and generalized approximate message passing (GAMP) approaches are discussed.

\item[$\bullet$] \textbf{Spatial modulation}: By using this type of modulation, information is mapped to a traditional PSK/QAM symbol and an index of the transmit and/or receive antennas \cite{Ertugrul_Basar}. Through cooperation between the IRS and a primary transmitter, IRS-based SR allows antenna index-based SM. By using a PSK/QAM scheme, the primary transmitter facilitates passive beamforming towards the desired receiver antenna. A simple ED is then used by the reader to determine the antenna index. \cite{Xianfu_Lei} proposes a star-QAM-aided joint SM for IRS-based SR systems, in which the IRS information frame is divided into two blocks, with the first block using PSK modulation for information bits, and the second block indicating the receive antenna index. The reader employs a ML detector for detection purposes. There is a more comprehensive analysis of the detection performance of IRS-enabled \bc in \cite{Mingjiang_Wu}. A coherent SM scheme with a star-QAM constellation is first proposed based on a ML detector and a suboptimal successive greedy detector. Then, two non-coherent SM schemes are proposed and their performances are evaluated analytically.

\end{enumerate}

\subsection{OFDM-Based Signal Detection for AmBC systems}
A significant challenge with traditional AmBC systems is the interference from the external RF  (or ``direct-link") signals, which can degrade the quality of backscattered communication. One technique that has gained attention to address this challenge is the use of OFDM.  OFDM uses multiple orthogonal RF carriers to transport digital data. The OFDM signal has several characteristics that can be exploited to boost AmBC's performance. 

\begin{figure}
\centering
	\includegraphics[width=3.5in]{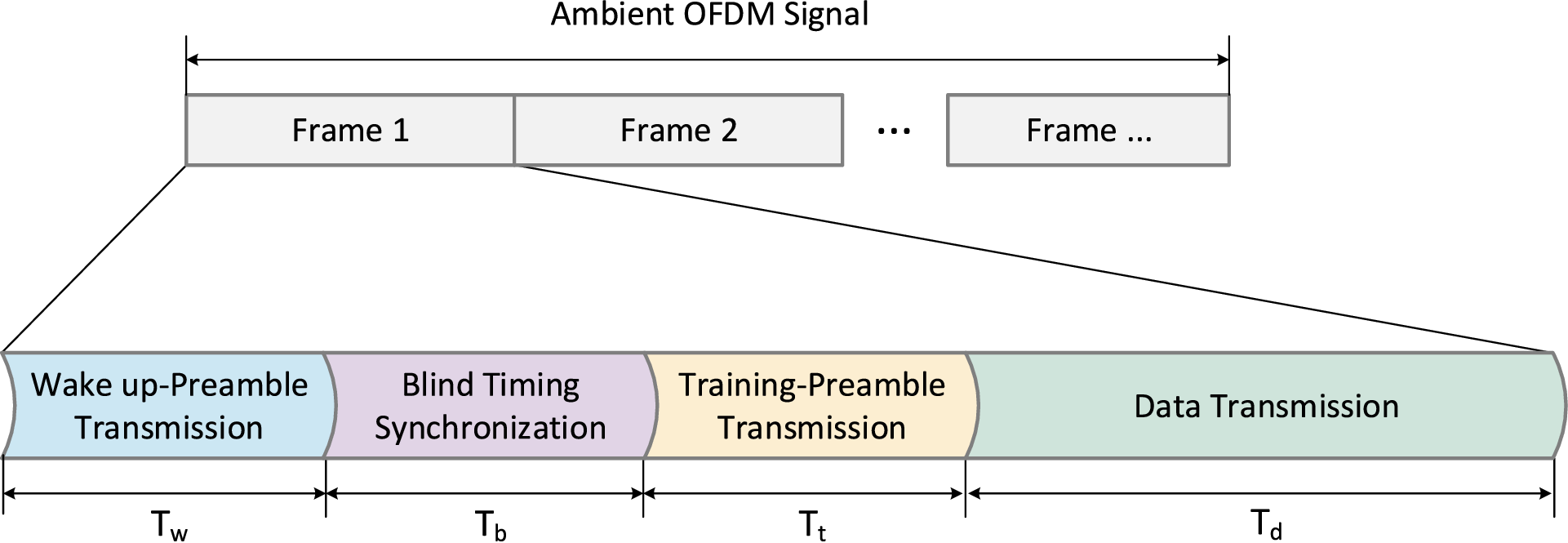}
 	\caption{OFDM-based AmBC system protocol \cite{Ying-Chang}.} \label{ofdm_fram}
\end{figure}

Study \cite{Gang_Yang} proposes a new AmBC system over OFDM signals to overcome this problem. Specifically, the RF source transmits OFDM signals to legacy receivers. The tag modulates its information over the OFDM carrier received from the RF source by adjusting its antenna impedance. Thanks to the repeating structure of the ambient OFDM signals coming from the cyclic prefix (CP), the strong direct-link interference can be canceled out at the reader, resulting in a much lower BER and higher data rate (see Table \ref{table_ofdm}). The proposed protocol for the OFDM-based AmBC system \cite{Ying-Chang} is represented in Fig. \ref{ofdm_fram} which consists of four phases. In the first phase, the tag is switched into the backscatter mode using a short sequence of alternating ``$0$" and ``$1$" \cite{Vincent_Liu}. The second phase of the process involves the tag switching its antenna to send information to the reader. Afterward, it performs blind timing estimation by estimating the channel propagation delay from the RF source based on the CP structure of the transmitted OFDM signal. During the third phase, the tag switches its antenna to backscatter mode. As a result, it transmits a training preamble that is known by the reader to estimate the average signal power, minimum propagation delay, and maximum channel spread. In the final phase, the tag transmits data bits to the reader.

\begin{figure}[t]
\centering
\includegraphics[width=3.5in]{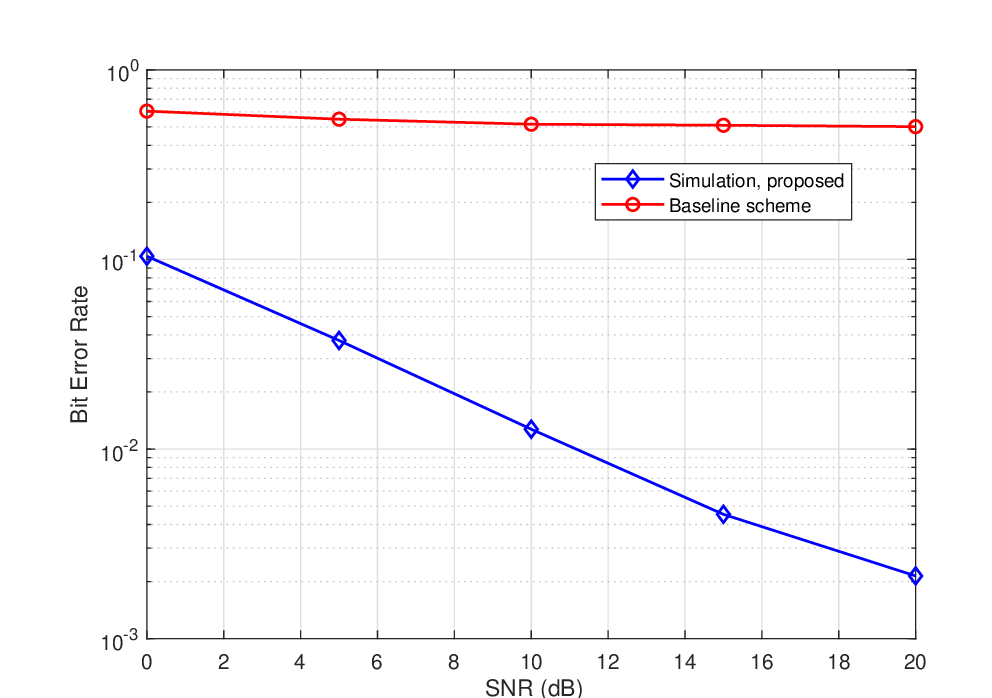}
\caption{BER comparison for the CP length $N_{\mathrm{c}\mathrm{p}}=64$ and the number of subcarriers $N=512$ \cite{Gang_Yang}.}
\label{shayan_fig_10}
\end{figure}

\begin{figure}[t]
\centering
\includegraphics[width=3.5in]{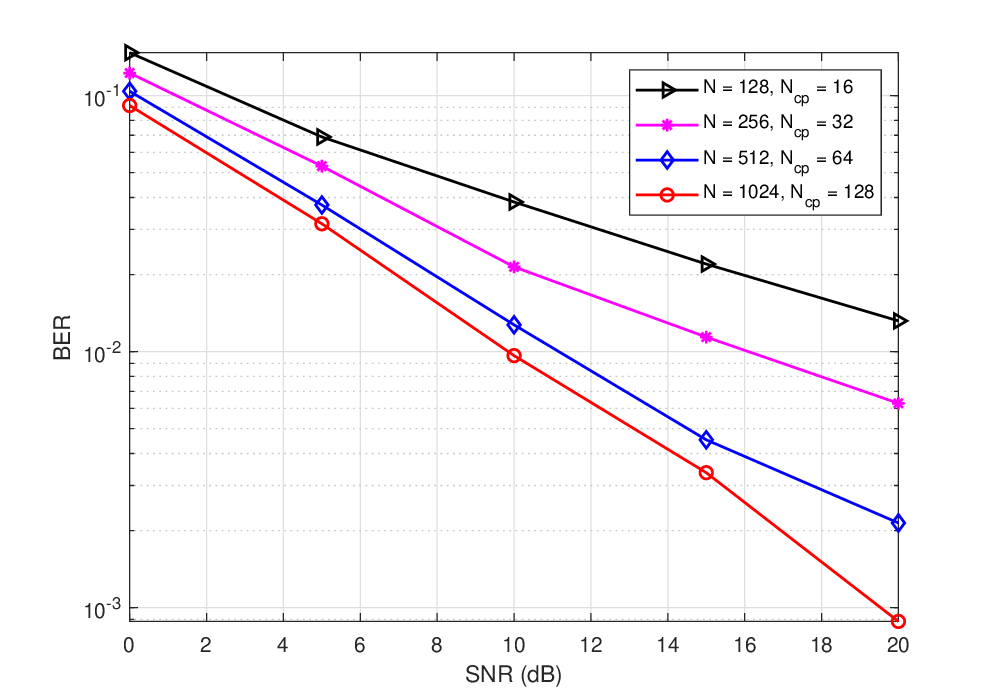}
\caption{BER comparison for different $N_{\text{cp}}$'s with $N=8N_{\text{cp}}$ \cite{Gang_Yang}.}
\label{shayan_fig_11}
\end{figure}

An innovative joint design for tag waveforms and reader detectors is presented in \cite{Gang_Yang}, which eliminates direct-link interference without adding complexity to the hardware. Additionally, the optimal detection threshold is derived in a closed-form expression based on the ML detector. Compared to the conventional design, the AmBC system demonstrates superior performance over OFDM signals. Further, it has been demonstrated that increasing the number of subcarriers and the length of CP leads to better BER performance. It is important to note that there is a trade-off between data rate and BER, as smaller CP lengths and more subcarriers result in higher BER rates and higher data rates, respectively. More specifically, Fig. \ref{shayan_fig_10} shows the BER versus the average SNR with the averaging detector \cite{Vincent_Liu} as the baseline scheme. By increasing the average SNR, the proposed ED outperforms the averaging detector. Due to the strong direct-link interference, the averaging detector maintains a high BER near $0.18$. In Fig. \ref{shayan_fig_11}, the effect of different $N_{\text{cp}}$'s is shown on BER performance. As $N_{cp}$ increases, the BER decreases, and the SNR gain becomes smaller for a larger $N_{\text{cp}}$. One interesting result is that there is not any BER floor for different $N_{\text{cp}}$'s since the direct interference is canceled out in the proposed scheme. By choosing a smaller $N$, a higher data rate is achieved at the expense of the higher BER. Accordingly, there is a trade-off between the BER and the data rate.

Next in \cite{Yoan_shin}, an improved ED is designed by proposing an arbitrary positive power operation on the amplitude of the signal rather than a squaring operation. This detection method is based on NP, in which the probability of detection is maximized subject to the probability of phase alarm. In comparison with \cite{Gang_Yang}, the proposed improved ED shows better performance in terms of BER. The authors in \cite{Mohamed_ElMossallamy_2} study the detection performance of AmBC over OFDM signals using null subcarriers. In OFDM systems, all subcarriers are not used, resulting in null edge subcarriers. An ED can detect the information of the tag by shifting the backscattered energy into null subcarriers using this structure. In contrast to \cite{Gang_Yang}, the proposed scheme has better performance in a variety of scenarios since the usable section of the CP in \cite{Gang_Yang} diminishes with the maximum channel delay spread, resulting in worse BER. An ML detector for \bc over ambient OFDM signals using a general $Q$-ary signal constellation is proposed in \cite{Donatella_Darsena_2}. With moderate SNR values, the proposed non-coherent 
detector can obtain low BER at small distances of the tag from an OFDM-based RF source. Next, the authors in \cite{Donatella_Darsena} examine the joint problem of CE, interference cancellation, and information detection in AmBC systems with OFDM-based AP. In this work, it is assumed that the receiver of the \bc is an AP. Due to the rank-deficient linear LS problem caused by traditional training-based approaches (e.g., ML approach), the authors propose a space alternating generalized expectation-maximization algorithm. In this way, the direct channel between AP and tags and the information-bearing symbol are estimated iteratively based on the information-bearing symbol. Detection of the data is then carried out using coherent detection. Additionally, the corresponding algorithm results in interference cancellation at the reader since the channel parameters are updated simultaneously. In \cite{Chong_Zhang}, the authors propose an interference-free transceiver for the AmBC systems in which OFDM signals are transmitted from the RF source to the legacy receivers. Among the design factors of the interference-free transceiver are the tag signal design, the signal interference cancellation technique, and the DFT operation. Adding the CP to an OFDM source symbol eliminates the ISI in a multipath channel, resulting in higher detection accuracy for the reader. In more detail, the tag design consists of four phases. During phase two, the tag begins to reflect by modulating its binary data onto the received RF source signal, while during the rest of the phases, the tag does not reflect. By using the signals arriving in phases two and four of the OFDM signal, the reader can cancel the interference. Each legacy receiver can also remove the CP from the received OFDM signal, indicating there is no interference experienced by the legacy receivers. The ML detector at the reader is proposed with two detection thresholds, and the interference cancellation in the proposed OFDM-based AmBC system is demonstrated to result in a low BER. Using a general $Q$-ary signal constellation at the BS, \cite{Donatella_Darsena_3} investigates non-coherent ML detection by treating direct link interference as undesirable. A comparison is made between the proposed detector and the CP-based ED \cite{Ying-Chang} and the classical ED. This study illustrates that the proposed non-coherent detector performs better than other baseline schemes in terms of BER. An innovative approach is described in \cite{Jinho_Choi} that performs carrier estimation at the reader through joint estimation and semi-coherent detection without any use of the repeating structure of the OFDM signal. In comparison to \cite{Ying-Chang}, the proposed approach is more appropriate for short-length CPs. Moreover, unlike \cite {Jing_Qian}, the performance of the proposed scheme is examined under frequency-selective fading channels.

\begin{table*}[]
\caption{Summary of OFDM-based AmBC systems. }\label{table_ofdm}	
\centering
\begin{tabular}{|l|l|l|l|l|}
\hline
Works   & Tag modulation type     & Detection design    & Setup parameters& Achieved BER     \\ \hline
\rowcolor[HTML]{E0FFFF} 

\cite{Donatella_Darsena}       & OOK & ML detector& \begin{tabular}[c]{@{}l@{}}$F_c = 2.4$ GHz\\ $T_s = 3.6$ $\mu$s\\ $N_s = 128$ \\ $N_{\text{cp}} = 16$  \\ $P_s = 20$ dBm\end{tabular} & $\sim 10^{-2}$    \\

\cite{Chong_Zhang}     & OOK & ML detector& \begin{tabular}[c]{@{}l@{}}$N_{\text{cp}} = 256$\\$ N_s = 20$\\  SNR $= 20$ dB\end{tabular}& \cellcolor[HTML]{FFFFFF}$ \sim \hspace{0.3em}< 10^{-2}$ \\
\rowcolor[HTML]{E0FFFF} 

\cite{Donatella_Darsena_2}    & OOK & \begin{tabular}[c]{@{}l@{}}Energy detector\\ Non-coherent ML detector\end{tabular}   & \begin{tabular}[c]{@{}l@{}}$F_c = 500$ MHz\\ $N_s = 512$ \\ $N_{\text{cp}} = 40$\\ SNR $= 20$ dB\end{tabular} & \cellcolor[HTML]{E0FFFF} $\sim10^{-2} $\\

\cite{Gang_Yang}  & OOK & \begin{tabular}[c]{@{}l@{}} CP-based energy detector\\ Energy detector\end{tabular}  & \begin{tabular}[c]{@{}l@{}}$F_c = 20$ MHz\\ $N_s = 512$ \\$ N_{\text{cp}} = 64$\\ SNR $= 20$ dB\end{tabular}   & \cellcolor[HTML]{FFFFFF} $\sim \hspace{0.3em}< 10^{-2}$ \\
\rowcolor[HTML]{E0FFFF} 

\cite{Yoan_shin} & OOK & Neyman–Pearson & \begin{tabular}[c]{@{}l@{}}$F_c = 20$ MHz\\ $N_s = 512$ \\$ N_{\text{cp}} = 64$\\ SNR $= 10$ dB\end{tabular}&$ \sim \hspace{0.3em}< 10^{-2} $\\

\cite{Mohamed_ElMossallamy_2} & \cellcolor[HTML]{FFFFFF}OOK & Energy detector   & \begin{tabular}[c]{@{}l@{}}$F_c = 10$ MHz\\$ N_s = 1024$\\ $N_{\text{cp}} = 72$\\ SNR $= 20$ dB\end{tabular}& \cellcolor[HTML]{FFFFFF}$\sim \hspace{0.3em}< 10^{-2}$ \\
\rowcolor[HTML]{E0FFFF} 

\cite{Donatella_Darsena_3}    & OOK & \cellcolor[HTML]{E0FFFF}\begin{tabular}[c]{@{}l@{}}Non-coherent ML Detector\\ CP-based energy detector\\ Energy detector\end{tabular} & \begin{tabular}[c]{@{}l@{}}$F_c = 10$ MHz\\ $N_s = 512$\\$ N_{\text{cp}} = 64$\\ SNR $= 20$ dB\end{tabular} & $\sim \hspace{0.3em}< 10^{-2} $\\

\cite{Mohamed_ElMossallamy}    & BFSK& Energy detector &  \begin{tabular}[c]{@{}l@{}}$F_c = 10$ MHz\\ $N_s = 512$\\ $N_{\text{cp}} = 64$\\ SNR $= 20$ dB\end{tabular} & \cellcolor[HTML]{FFFFFF}$\sim \hspace{0.3em}< 10^{-2}$ \\
\rowcolor[HTML]{E0FFFF} 

\cite{Takanori_Hara}   & PSK, DSK      & ML detector& \begin{tabular}[c]{@{}l@{}}$N_s = 64$\\ $N_{\text{cp}} = 16$\\ SNR $= 4$ dB\end{tabular}       & $\sim 10^{-2}-10^{-3}$ \\
\cite{Jinho_Choi}& OOK & ML Detector& \begin{tabular}[c]{@{}l@{}}$N_s = 32$\\ $N_{\text{cp}} = 4$\\ SNR $= 4$ dB\end{tabular}  & \cellcolor[HTML]{FFFFFF} $\sim \hspace{0.3em}< 10^{-3}$  \\ \hline
\end{tabular}
\end{table*}

The authors of \cite{Takanori_Hara} consider PSK and delay-shift keying (DSK) modulation schemes for an OFDM-based \abc system. Using a pilot-based approach, an optimal ML detector is proposed for the PSK modulation scheme. In the case of the DSK modulation scheme, the tag is composed of $M$ delay circuits that propagate reflections in different ways. Based on the estimated channel, the optimal ML detector is used to detect the transmitted bits at the reader's end. PSK modulation achieves the lowest SER at bandwidth efficiencies of less than $3$ (bits/channel use), whereas DSK achieves the lowest SER at bandwidth efficiencies exceeding $2$ (bits/channel use). A more general modulation scheme, called BFSK, is considered in \cite{Mohamed_ElMossallamy}, which includes BPSK as a special case. They propose an optimal non-coherent detector with symmetric error probabilities for ``$0$"s and ``$1$"s that do not require SNR-dependent threshold estimation but require knowledge of high and low guard subcarriers. In addition, the exact expression for the BER is derived. As a result of symmetric error probabilities, the proposed detection method shows a $3$ dB improvement over the energy detection method \cite{Gang_Yang,Mohamed_ElMossallamy_2}.

\section{Open Issues and Future Research Directions}\label{Sec:open_issue}
Several open research problems have not been fully addressed in the literature, which is discussed in the following section.

\subsection{Channel Modeling and Hardware Impairments}
While the various contributions reviewed above have comprehensively considered the problem of signal detection in AmBC systems, the analysis has been typically restricted to Rayleigh. Further research is needed to deal with more general fading distributions, including Rician, Nakagami, and shadowed fading using slow and fast fading models. In addition, when a BS is located above the rooftops, the angular spread is small, making the assumption of a spatially correlated channel at the receiver valid. Future studies should also consider handling the case of multiple angular paths at the reader. As for the hardware impairments, AmBC systems do not necessarily have stable power sources, and incorporating this instability into BER analysis would be an interesting future direction. Second, AmBC transceivers suffer from various RF front-end imperfections as a result of non-ideal manufacturing processes and mismatched components. The CFO and the in-phase/quadrature (I/Q) imbalances are both serious concerns in AmBC since they can introduce errors that negatively affect the demodulation's performance. As a result, AmBC receivers should be able to estimate and compensate for impairments of this nature.

\subsection{Channel Estimation}
CE is a crucial component of AmBC systems, and there are several open issues and future research directions that need to be addressed to improve the performance and efficiency of these systems. One of the main challenges in AmBC systems is interference from other wireless communication systems, which affects the accuracy of CE. Future research should focus on developing effective interference mitigation techniques to improve the accuracy of CE. Another challenge is the low SNR, which can negatively impact the CE accuracy. Future research should focus on developing CE techniques that are robust to low SNR. Additionally, the channel in AmBC is often non-stationary, meaning that it changes over time, which makes it challenging to estimate. Future research should focus on developing CE techniques that are robust to non-stationary channels. Furthermore, CE requires a significant amount of energy, which can negatively impact the energy efficiency of the system. Developing novel energy-efficient CE techniques is required. In many cases, CE and data estimation are performed independently, which can adversely affect system performance. Future research should focus on developing joint channel and data estimation techniques to improve overall performance. Finally, deep learning has shown promising results for CE in various wireless communication systems, and future research should focus on developing deep learning-based CE techniques specifically for AmBC systems.

 \subsection{Channel Coding}
In essence, coding is used to ensure the secure transmission of messages by guarding against interference, collisions, and any intentional changes made to the signal. There are several established codes available for \bc systems, such as non-return to zero (NRZ), Manchester, Miller, and FM0. However, the NRZ code has limitations and doesn't work well with data consisting of a long sequence of bits, while the Manchester code requires transmitting more bits than in the original signal. Thus, Miller and FM0 coding techniques are commonly used in \bc systems to enhance signal reliability, reduce noise, and simplify the system. However, as \bc is rapidly evolving and growing in terms of applications, conventional channel coding may not be enough to meet the demands for high data rates, long-range communication, and robustness. Thus, designing training sequences and joint estimation and detection is a promising approach for improving the data rate and reliability of \bc systems.

\subsection{Constellation Design}
High-order modulation is suitable for situations that require high-speed IoT communication. As a consequence, it is desirable to use higher-order modulation at the tag by using more than two different impedance values to achieve a high data rate. To achieve optimal constellation design with $M$-PAM schemes, each symbol magnitude should be separated as widely as possible, and magnitude degeneracy should be eliminated. However, this may complicate the reader and tag circuit design for other modulation schemes since the reader must have access to the full CSI. The reason for this is that non-coherent detection eliminates the phase of the received signals, which makes high-order modulation schemes such as $M$-PSK and $M$-QAM ineffective.

\subsection{Integrating AmBC with 5G and Subsequent Networks} The potential application of AmBC systems within future 5G and succeeding wireless networks remains a relatively untapped research area. A critical understanding of how these systems could be seamlessly incorporated into existing network architectures is essential, particularly concerning their interoperability with non-terrestrial networks (NTNs) and heterogeneous networks (HetNets). Challenges to be surmounted include interference management, security assurance, and maintaining an optimal quality of service (QoS). Furthermore, the development of innovative protocols and schemes is necessitated due to the amplified complexity and heterogeneity of these networks, particularly in the context of advanced signal detection techniques.

\subsection{Promoting Energy Efficiency and Green AmBC} Energy consumption poses a formidable challenge for AmBC systems, particularly for IoT devices, where the crux lies in prolonging battery life and ensuring energy efficiency. Hence, further studies are warranted to design energy-conserving algorithms and efficient transmission techniques. Additionally, adopting a green communication approach, which focuses on minimizing the overall carbon footprint of these systems, could prove beneficial in terms of environmental sustainability and efficient signal propagation.

\subsection{Ensuring Security and Privacy in AmBC Systems} 
While AmBC proffers a plethora of benefits, its susceptibility to various security threats necessitates urgent attention. These threats may encompass eavesdropping, signal jamming, and replay attacks, all of which could potentially disrupt signal detection. Therefore, the development of robust and secure protocols is imperative to fortify against these threats. Furthermore, research into privacy-preserving mechanisms is essential to safeguard user data within these systems, particularly in the context of secure signal transmission.

\subsection{Exploiting Machine Learning in AmBC Systems} 
Incorporating MaL has the potential to significantly amplify system performance by automating and optimizing numerous facets of the AmBC communication process, including advanced signal detection techniques. MaL can play a critical role in CE, as discussed before.  By leveraging data-driven predictive models, MaL can enhance CE accuracy, leading to more precise signal detection. Additionally, the dynamic nature of MaL can be utilized to optimally select modulation and coding schemes based on current network conditions, leading to more efficient spectrum usage and improved reliability. However, despite the substantial promise of integrating MaL into AmBC, more research is required to fully unlock its potential, particularly in the context of signal detection. 
As a potential avenue for future work, DRL can also be explored. DRL's ability to learn and make decisions in complex environments might provide innovative solutions for AmBC signal detection.

\section{Conclusion}\label{Sec:conclude}
This paper underscores the pivotal role that future low-power wireless networks are set to play in enriching our everyday lives. A key facilitator of this revolution is the AmBC system. By harnessing and modulating existing signals, AmBC brings us closer to a future of ubiquitous, scalable, and efficient connectivity, enabled by the strategic integration of various groundbreaking technologies. 

This paper presents a comprehensive review of the literature on signal detection in AmBC systems, a cornerstone for enabling large-scale, low-power, and cost-effective wireless communication. Beginning with an in-depth overview of signal detection techniques, their relevant parameters, and practical implementations, the paper transitions into a detailed exploration of receiver design in AmBC systems. It further delves into diverse applications of AmBC, emphasizing the transformative potential of signal detection. Lastly, it puts forth pertinent questions about future research directions and underlines the open problems within this domain. Even though the adoption and optimization of 6G networks bring significant challenges, they hold unprecedented potential benefits to global communication systems. Progress in this shift requires rigorous understanding and research in areas like AmBC.

\bibliographystyle{ieeetr}
\bibliography{ref}

\begin{IEEEbiography}	[{\includegraphics[width=1in,height=1.25in,clip,keepaspectratio]{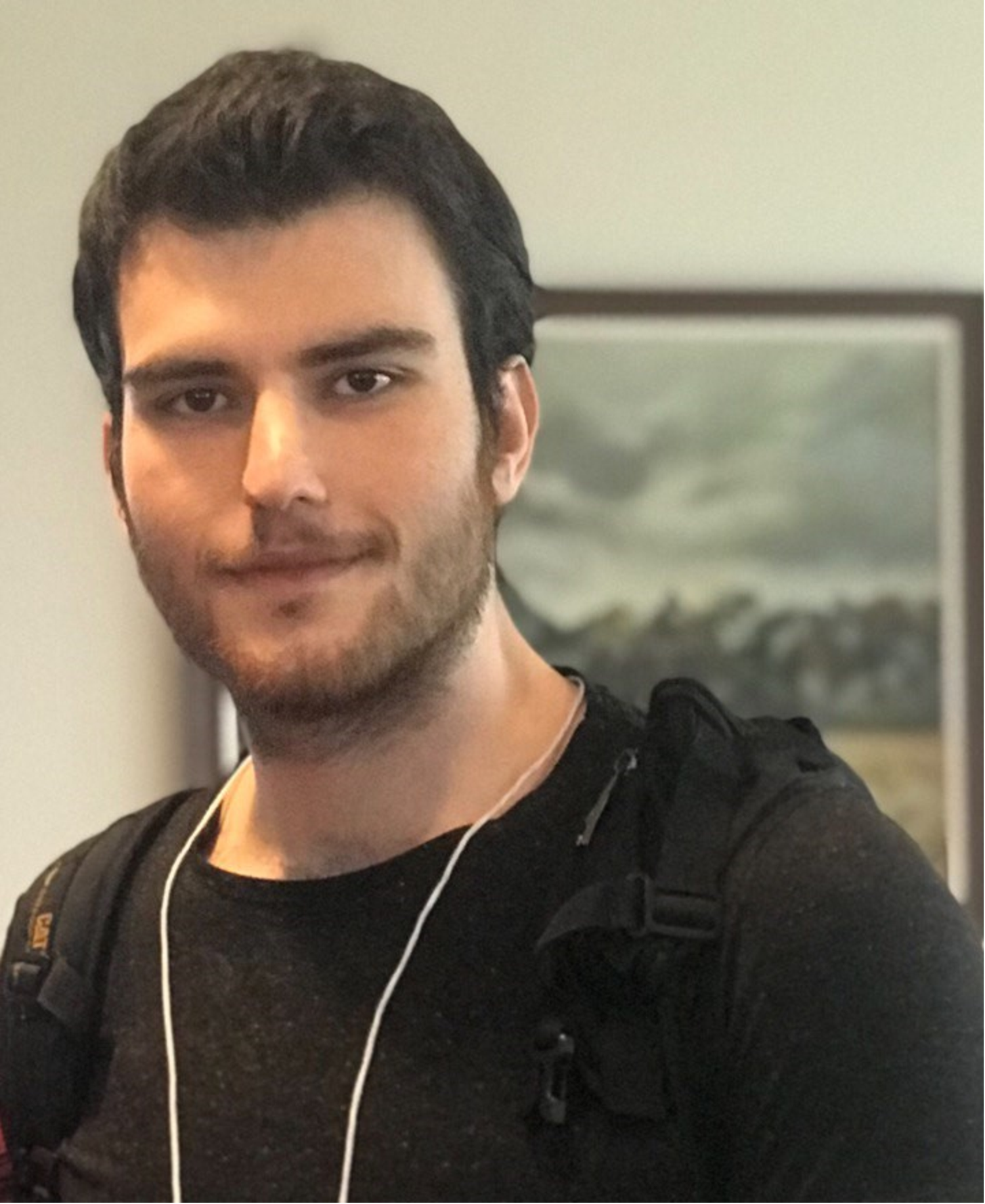}}] 
	{Shayan Zargari} received the B.Sc. degree and M.Sc. degree in Electronic Engineering and Telecommunication Engineering from the Azad University South Tehran Branch and Iran University of Science and Technology in 2018 and 2020, respectively. He is currently pursuing a Ph.D. degree in the field of communications and signal processing at the University of Alberta, Edmonton, Canada. From 2019 until 2020, he was a visiting researcher at the Electronics Research Institute, Sharif University of Technology, Tehran, Iran. From 2020 until 2021, he was a visiting researcher at the Department of Electrical and Computer Engineering, Tarbiat Modares University, Tehran, Iran. His research interests include optimization theory, intelligent reflecting surface (IRS), unmanned aerial vehicle (UAV) communications, resource allocation in wireless communication, and green communication. He is a Reviewer of several IEEE journals, such as IEEE JOURNAL ON SELECTED AREAS IN COMMUNICATIONS, IEEE TRANSACTIONS ON COMMUNICATIONS, and IEEE TRANSACTIONS ON WIRELESS COMMUNICATIONS.
\end{IEEEbiography}

\begin{IEEEbiography}[{\includegraphics[width=1in,height=1.25in,clip,keepaspectratio]{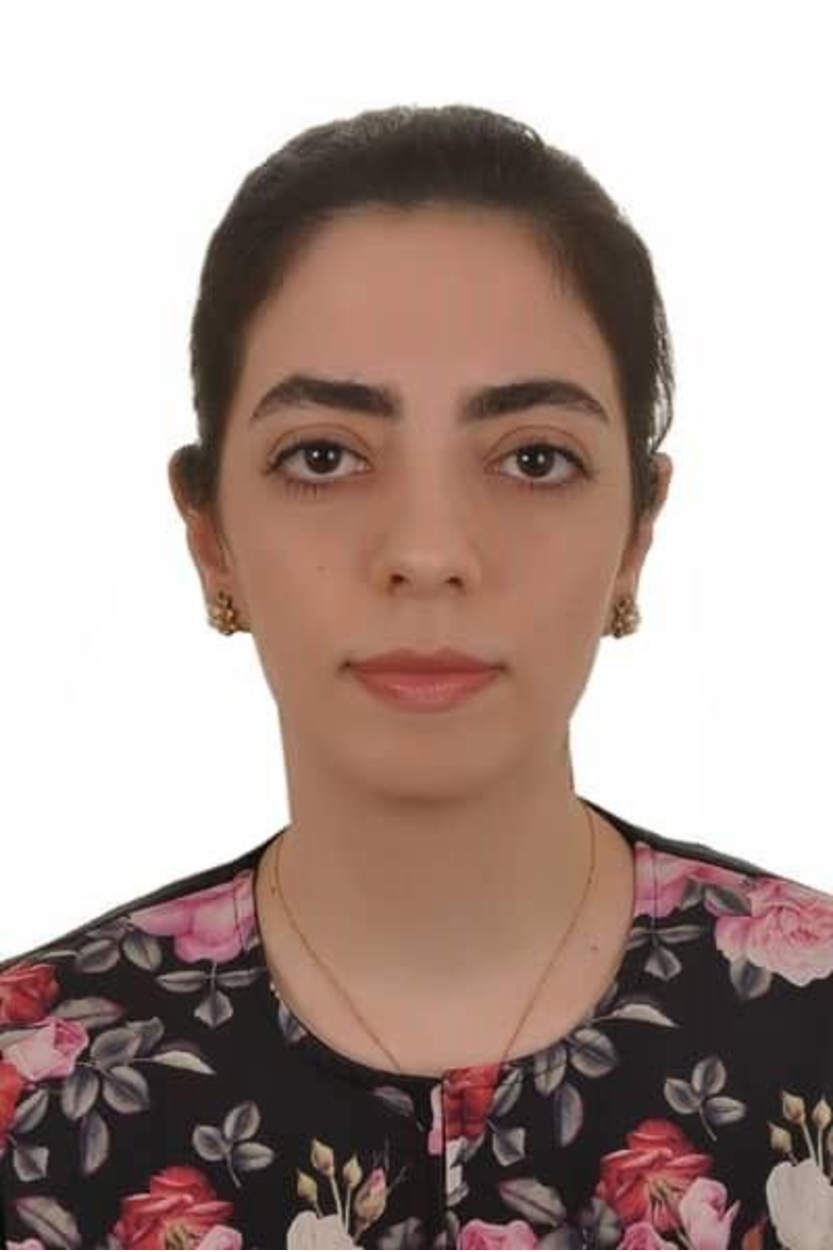}}]
{Azar Hakimi} received the B.Sc. and the M.Sc. degree in Electrical Engineering and communication systems from Shahrekord University, Iran,  in 2015 and 2018, respectively. She is currently a Ph.D. student at the University of Alberta. Her research interests include wireless communication systems, Internet of Things systems, backscatter communication, wireless information and power transfer, non-orthogonal multiple
access communications, MIMO communication, and intelligent reflecting surfaces.
\end{IEEEbiography}

\begin{IEEEbiography}[{\includegraphics[width=1in,height=1.25in,clip,keepaspectratio]{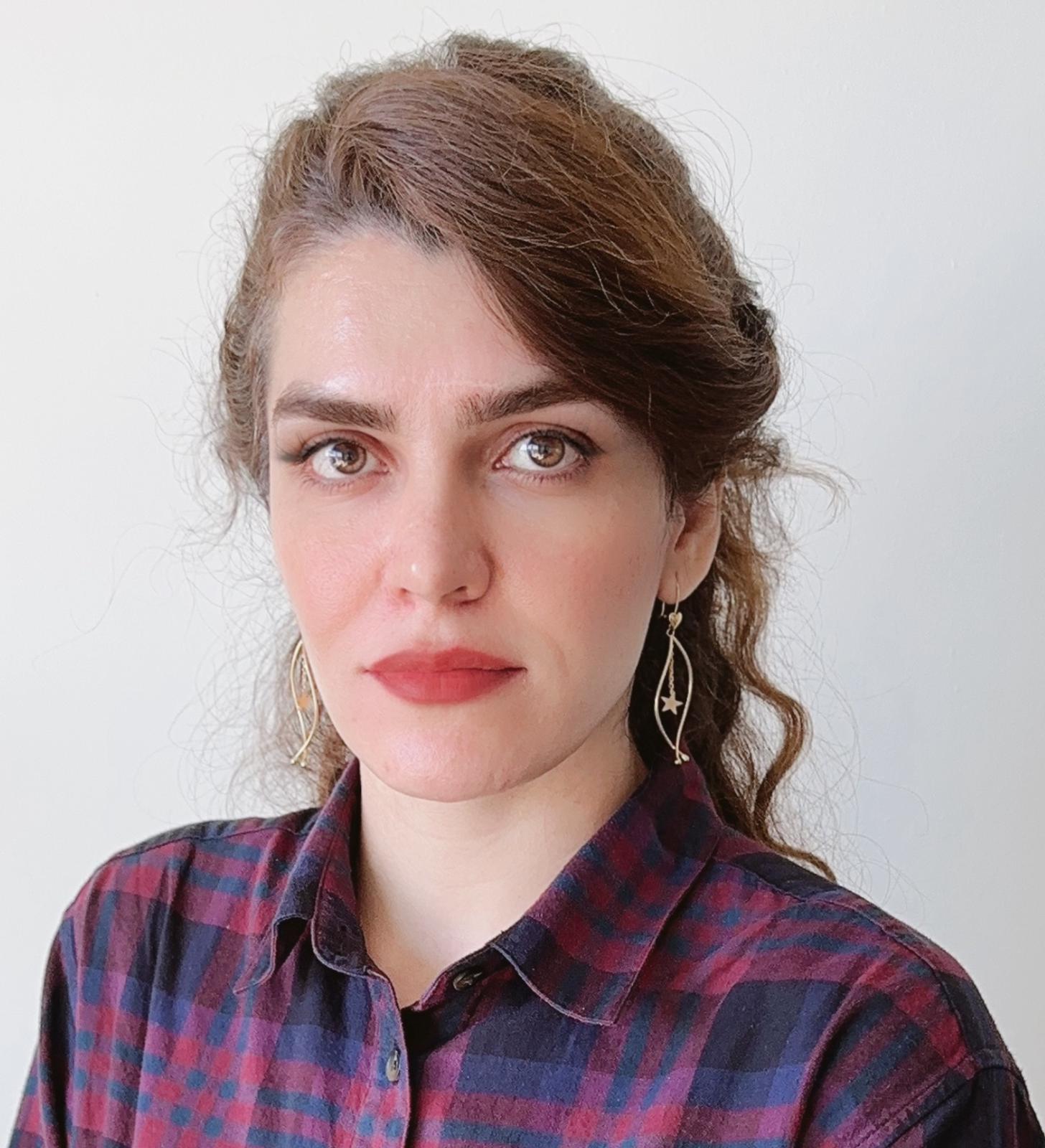}}]
{Fatemeh Rezaei}(S'12--M'22) received her Ph.D. degree (with distinction) in Electrical Engineering from Yazd University (YU), Iran, in 2021. She is currently a Postdoctoral Fellow at the Department of Electrical and Computer Engineering, University of Alberta, Canada, where she was a Visiting Researcher from February 2019 to September 2020. Her current research interests are in wireless communications and signal processing for 5G and beyond cellular systems, including backscatter communications, internet of things, energy harvesting systems, MIMO and massive MIMO, intelligent reflective surfaces, non-orthogonal multiple access, interference alignment, cognitive radio networks, and machine learning for wireless networks. 
\end{IEEEbiography}

\begin{IEEEbiography}[{\includegraphics[width=1in,height=1.25in,clip,keepaspectratio]{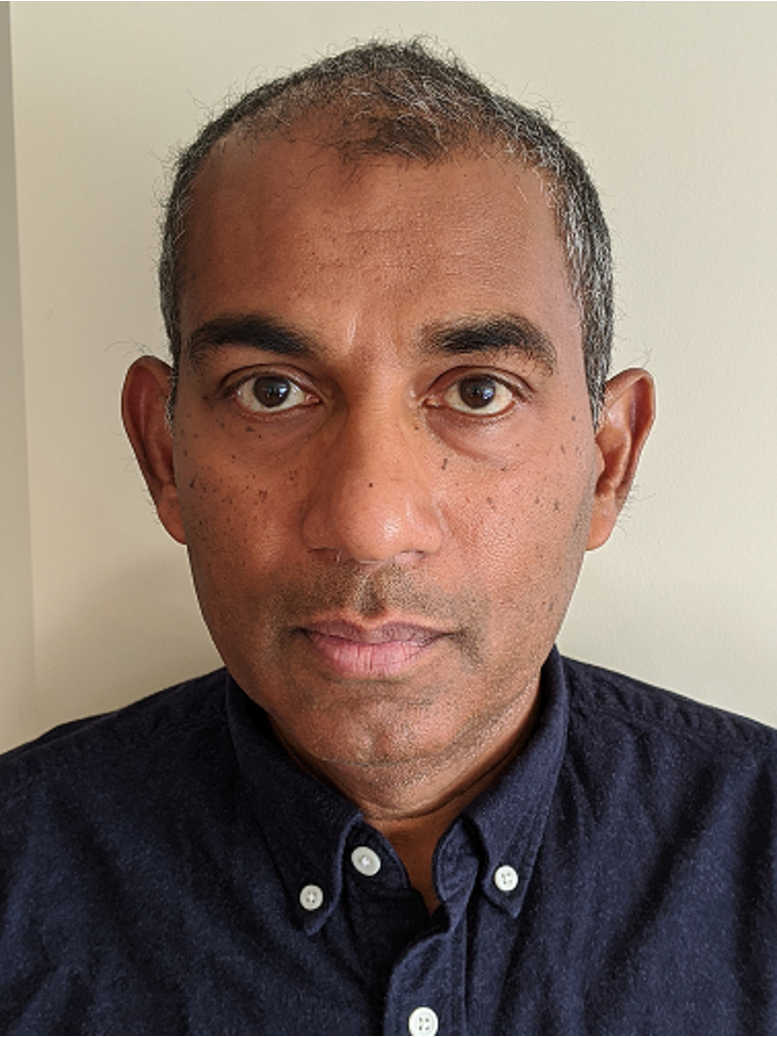}}]
{Chintha Tellambura}(Fellow, IEEE) received the B.Sc. degree in electronics and telecommunications from the University of Moratuwa, Sri Lanka, the M.Sc. degree in electronics from the Kings College, University of London, and the Ph.D. degree in electrical engineering from the University of Victoria, Canada.

He was with Monash University, Australia, from 1997 to 2002. Since 2002, he has been with the Department of Electrical and Computer Engineering, University of Alberta, where he is currently a Full Professor. He has authored or co-authored over 560 journal and conference papers, with an h-index of 83 (Google Scholar). He has supervised or co-supervised over 70 M.Sc., Ph.D., and PDF trainees. His current research interests include future wireless networks and machine learning algorithms. He was elected as a fellow of The Canadian Academy of Engineering in 2017. He received the Best Paper Awards from the IEEE International Conference on Communications (ICC) in 2012 and 2017. He is the winner of the prestigious McCalla Professorship and the Killam Annual Professorship from the University of Alberta. He served as an Editor for the IEEE TRANSACTIONS ON COMMUNICATIONS from 1999 to 2012 and IEEE TRANSACTIONS ON WIRELESS COMMUNICATIONS from 2001 to 2007. He was an Area Editor of Wireless Communications Systems and Theory from 2007 to 2012.
\end{IEEEbiography}

\begin{IEEEbiography}[{\includegraphics[width=1in,height=1.25in,clip,keepaspectratio]{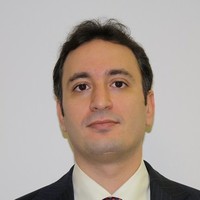}}]
{Amine Maaref}(Senior Member, IEEE) has been with the Huawei Technologies Canada Research Center (CRC) in Ottawa, ON, since 2011, where he currently holds the position of Senior Expert and Team Leader, focusing on 5G/6G radio access design. 

As a technical leader in LTE and NR physical layer air interface, he has represented Huawei Technologies in 3GPP RAN1 standards and was actively involved in the standardization of the foundational 5G NR releases. Prior to joining Huawei, he was with Mitsubishi Electric Research Labs (MERL), in Cambridge, MA, USA, as a Research Scientist, where he conducted advanced research in broadband mobile communications and was actively involved in 3GPP LTE/LTE-Advanced and WiMAX IEEE 802.16m standardization. 
 
Over the last 20 years, he has co-authored more than 100 international peer-reviewed publications and standard contributions in the relevant fora, filed more than 500 worldwide awarded patents and pending patent applications. He holds many standard essential patents and received numerous prestigious awards for his outstanding research and scholarly achievements, including the 5G Outstanding Fundamental Research and Standard Award for the recognition of outstanding contribution to the 5G fundamental research and standardization and the top 10 invention award in recognition of inventions that have become important commercial features of products and have the potential to generate huge commercial value in the future, from Huawei Technologies. He was the Guest Editor for several special issues of IEEE journals and magazines and was the founding Managing Editor of IEEE 5G/Future Networks Tech Focus. From 2014 to 2020, he served as an Editor of the IEEE Transactions on Wireless Communications and currently serves as an Editor of IEEE Transactions on Communications.

\end{IEEEbiography}

\EOD

\end{document}